\documentclass[aps,pra,reprint,superscriptaddress,floatfix,amsfonts,amsmath]{revtex4-1}
\usepackage{array}
\usepackage{amssymb}
\usepackage{amsmath}
\usepackage{graphicx}
\usepackage{contour}
\usepackage[bookmarks=false,colorlinks=true,urlcolor=blue,citecolor=blue,linkcolor=blue]{hyperref}

\usepackage{tikz}
\usetikzlibrary{calc,arrows}

\usepackage[compat=1.1.0]{tikz-feynman}
\tikzfeynmanset{/tikzfeynman/warn luatex=false}

\newcommand{\be}{\begin{equation}}
\newcommand{\ee}{\end{equation}}
\newcommand{\bea}{\begin{eqnarray}}
\newcommand{\eea}{\end{eqnarray}}
\newcommand{\beal}{\begin{align}}
\newcommand{\eeal}{\end{align}}
\newcommand{\bes}{\begin{equation} \begin{split}}
\newcommand{\ees}{\end{split} \end{equation}}

\newcommand{\f}{\frac}

\newcommand{\qv}{\vec{q}}
\newcommand{\pv}{\vec{p}}
\newcommand{\qvp}{\vec{q}^{\,\prime}}

\newcommand{\Lambdamat}{\mathbf{\Lambda}}

\newcommand{\Kmat}{\mathbf{K}}

\newcommand{\Kinv}{K^{-1}}

\newcommand{\Keffinv}{K^{-1}_{\mathrm{eff}}{}}
\newcommand{\tr}{\mbox{Tr}}

\newcommand{\rv}{\vec{r}}

\newcommand{\Sv}{\vec{S}}

\newcommand{\Qv}{\vec{Q}}

\newcommand{\xhat}{\hat{x}}
\newcommand{\yhat}{\hat{y}}
\newcommand{\I}{I}

\newcommand{\Ic}{Ic}
\newcommand{\Mc}{Mc}

\newcommand{\Msc}{Mc}
\newcommand{\Id}{Id}

\newcommand{\Msd}{Md}

\newcommand{\IdMd}{Id/Md}

\newcommand{\IdMsd}{Id/Md}
\newcommand{\IfdMsd}{Ifd/Md}
\newcommand{\Jonetwo}{$J_1$-$J_2$}

\begin{document}

\title{Interplay between Magnetic and Vestigial Nematic Orders in the Layered {\Jonetwo} Classical Heisenberg Model}
\author{Olav  F.~\surname{Sylju{\aa}sen}}
\affiliation{Department of Physics, University of Oslo, P.~O.~Box 1048 Blindern, N-0316 Oslo, Norway}

\author{Jens Paaske}
\email{paaske@nbi.ku.dk}
\affiliation{Center for Quantum Devices, Niels Bohr Institute, University of Copenhagen, 2100 Copenhagen, Denmark}

\author{Michael Schecter}
\email{schecter@umd.edu}
\affiliation{Condensed Matter Theory Center and Joint Quantum Institute, Department of Physics, University of Maryland, College Park, Maryland 20742, USA}

\date{\today}


\begin{abstract}
We study the layered {\Jonetwo} classical Heisenberg model on the square lattice using a self-consistent bond theory. We derive the phase diagram for fixed $J_1$ as a function of temperature $T$, $J_2$ and interplane coupling $J_z$. Broad regions of (anti)ferromagnetic and stripe order are found, and are separated by a first-order transition near $J_2\approx 0.5$ (in units of $|J_1|$). Within the stripe phase the magnetic and vestigial nematic transitions occur simultaneously in first-order fashion for strong $J_z$. For weaker $J_z$ there is in addition, for $J_2^*<J_2 < J_2^{**}$, an intermediate regime of split transitions implying a finite temperature region with nematic order but no long-range stripe magnetic order. In this split regime, the order of the transitions depends sensitively on the deviation from $J_2^*$ and $J_2^{**}$, with split second-order transitions predominating for $J_2^* \ll J_2 \ll J_2^{**}$. We find that the value of $J_2^*$ depends weakly on the interplane coupling and is just slightly larger than $0.5$ for $|J_z| \lesssim 0.01$. In contrast the value of $J_2^{**}$ increases quickly from $J_2^*$ at $|J_z| \lesssim 0.01$ as the interplane coupling is further reduced. In addition, the magnetic correlation length is shown to directly depend on the nematic order parameter and thus exhibits a sharp increase (or jump) upon entering the nematic phase. Our results are broadly consistent with predictions based on itinerant electron models of the iron-based superconductors in the normal-state, and thus help substantiate a classical spin framework for providing a phenomenological description of their magnetic properties.

\end{abstract}

\maketitle

\section{Introduction}\label{sec:intro}

The iron-based superconductors (FeSCs) represent an intriguing class of materials exhibiting both unconventional superconductivity as well as peculiar normal-state properties (see Refs.~\onlinecite{PaglioneGreene2010,Stewart2011,Dai2012} for reviews).
In particular, in the normal-state there are strong indications that the stripe spin-density wave order with wavevector $\Qv=(0,\pi)\,{\rm or}\,(\pi,0)$, that occurs below a critical temperature, is intimately related to the structural distortion of the lattice that occurs at a higher critical temperature. This structural distortion breaks the tetragonal symmetry, leading to long-ranged lattice-nematic order. Several proposals have thus suggested that spin fluctuations of the stripe order drive the nematic transition\cite{Fang2008,CenkeXu2008,Fernandes2014}, while the lattice distortion arises secondarily via coupling to the spin-driven nematic order parameter.

This scenario motivates a purely magnetic approach that captures the interplay between the vestigial nematic order and the fluctuations of the magnetic stripe order that drive it. Previous proposals have focused on the {\Jonetwo} model of localized spins~\cite{Yildrim2008,SiAbrahams2008}, which is known, as we discuss below, to have nematic order in the strictly two-dimensional (2D) limit at finite temperatures~\cite{CCL1990,Weber2003}. This type of order has become known as Ising-nematic order~\cite{Fernandes2014} because the order parameter space (the choice between $\Qv=(\pi,0)$ or $\Qv=(0,\pi)$) is effectively $Z_2$ due to the Mermin-Wagner theorem~\cite{MerminWagner1966} that precludes spontaneous spin-rotation symmetry breaking in 2D. This Ising-nematic phase transition extends to the {\em layered} {\Jonetwo} model where there is an additional phase transition to an ordered magnetic stripe phase~\cite{Fang2008,CenkeXu2008}. Thus the layered {\Jonetwo} model may serve as the simplest phenomenological model capable of describing the putative spin-fluctuation triggered nematic phase in the FeSCs.


Nevertheless, the fact that the FeSCs are not insulators indicate that the normal-state of the FeSCs is likely best described as an itinerant magnet at low doping~\cite{Singh2009}, and a rich phase diagram based on itinerant magnetism has been predicted~\cite{Fernandes2012}.
It is unclear whether the phase diagram of the layered {\Jonetwo} model is as rich, or if and where the {\Jonetwo} model consists of regions of split nematic and magnetic transitions.
Moreover, the nature of the phase transitions is an important aspect of the problem that is known in the itinerant models to depend sensitively on the microscopic parameters near the bifurcation point where the simultaneous (first order) transition splits into two separate transitions. In particular, as the transitions split, there appears to be a narrow intermediate regime where the order of the transitions may be different for the magnetic and nematic order parameters. This can lead, e.g., to a metanematic transition at which the finite nematic order parameter jumps to a higher value as the magnetic order sets in. In addition, the itinerant models predict a sharp increase of the spin correlation length upon entering the nematic phase due to its direct dependence on the nematic order parameter~\cite{Fernandes2012}.

It is the purpose of this paper to determine whether such phenomena arise also within a specific microscopic model based on localized spins: the layered square lattice {\Jonetwo} classical Heisenberg model.  We study this model using the nematic bond theory developed in Ref.~\onlinecite{Schecter2017}, which can detect nematic and magnetic orders independently and determine the order of the transitions. A dual purpose of this paper is also to explain this method in depth. The nematic bond theory allows us to construct (for the first time, to the best of the authors' knowledge) the phase diagram of the layered {\Jonetwo} model for fixed $J_1$ as a function of temperature $T$, $J_2$ and interlayer coupling $J_z$.
The nematic bond theory can be employed to investigate temperature-dependent properties of any classical Heisenberg hamiltonian, and requires considerably less numerical efforts than Monte Carlo simulations.
We show that practically all of the phenomena discovered in the itinerant electron models arise also in the Heisenberg model, including an intermediate regime of transition types near the bifurcation points, and a sharp increase of the spin correlation length upon entering the nematic phase.

\begin{figure}[t]
\begin{center}
\includegraphics[width=\columnwidth]{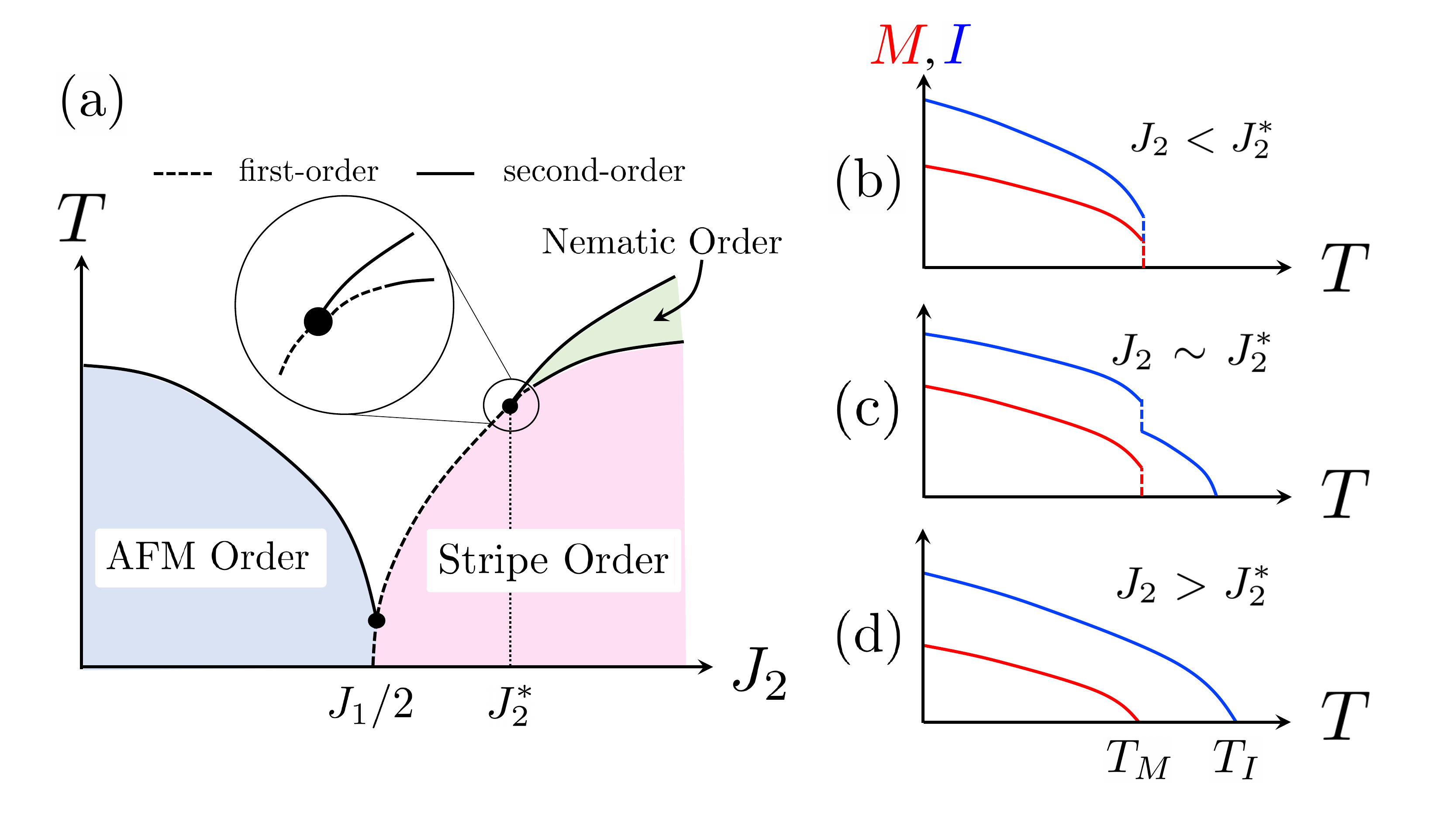}
\caption{(a) Schematic phase diagram of Eq.~\eqref{Hamiltonian} for a fixed value of $J_z>0$.  The phase diagram is invariant under $J_1\to-J_1$ and antiferromagnetic order (AFM) $\to$ ferromagnetic order (FM). For $J_2/J_1>1/2$ the system enters the stripe phase via a simultaneous (magnetic and nematic) first-order transition when $J_2<J_2^*$, depicted in (b). For $J_2>J_2^*$ the magnetic and nematic order parameters, $M$ and $I$, develop at different temperatures, $T_M$ and $T_I$, respectively. Near the branching point, $J_2\sim J_2^*$, one (or both) of the split transitions may remain first-order, while for $J_2\gtrsim J_2^*$ the transitions are predominantly second-order, as shown in (d). In the case where the magnetic transition is first-order, as shown in (c), one has a metanematic transition whereby the nematic order parameter jumps from a finite value to a higher one. }
\label{fig:schematic}
\end{center}
\end{figure}

A schematic phase-diagram illustrating the behavior of the order parameters in the various regimes is presented in Fig.~\ref{fig:schematic}. One sees that for $J_1/2\lesssim J_2<J_2^*$ the transition into the stripe phase is simultaneous and first-order, while for $J_2>J_2^*$ the transitions are split and predominantly second-order. Near the bifurcation point, $J_2\sim J_2^*$, one of the transitions may remain first-order. This leads (in the example shown in Fig.~\ref{fig:schematic}) to a metanematic transition, similar to what happens in the itinerant models for various parameter regimes.  Although we will not attempt to make a direct connection to any FeSC compound in particular, our results show that at a phenomonological level the physics of the {\Jonetwo} model indeed appears to capture the essential aspects of the normal-state magnetic properties of the FeSCs and therefore may be of use as a simplifying alternative to the itinerant approach.

While some previous studies have attempted to address the issues discussed above within a classical spin framework, it is important to emphasize that here we study \emph{directly} the microscopic layered {\Jonetwo} model. An effective model of the layered {\Jonetwo} model, the Ising O(3)-model~\cite{CCL1990}, has been studied previously using Monte Carlo simulations~\cite{Kamiya2011} which yields a simultaneous first order phase transtion for large interlayer couplings and two split transitions for weaker interlayer couplings. Other studies have included a phenomenological biquadratic coupling to the Hamiltonian to mimic the longwavelength effective action that arises upon coarse graining~\cite{CCL1990,Fang2008,CenkeXu2008}. Our treatment is different in that we do not explicitly assume a separate Ising degree of freedom from the outset and do not insert a biquadratic coupling by hand. Instead, we shall utilize and develop a tractable technique, the nematic bond theory, that can tackle the {\Jonetwo} model head-on and provide its phase diagram in terms of microscopic parameters.

The rest of the paper is organized as follows. We first define the Hamiltonian of the \Jonetwo--model, Sec.~\ref{sec:model}, and then describe in details the method we use to solve it in Sec.~\ref{sec:method}.
Then we discuss in Sec.~\ref{sec:orderpar} the order parameters relevant for the \Jonetwo--model and  how to compute them using our method. The results for the \Jonetwo--model are then described in Sec.~\ref{sec:results}, followed by Sec.~\ref{sec:correlations} that describes the spin correlations. We end by a general discussion in Sec.~\ref{sec:discussion}.

\section{\Jonetwo--model \label{sec:model}}
\noindent
The Hamiltonian of the layered classical \Jonetwo--model is
\be \label{Hamiltonian}
H = J_1 \sum_{\langle ij \rangle} \Sv_{i} \cdot \Sv_{j}  + J_2 \sum_{\langle \langle ij \rangle \rangle} \Sv_{i} \cdot \Sv_{j}
+J_z \sum_{\{ij\}} \Sv_{i} \cdot \Sv_{j}
\ee
where $\langle ij \rangle$,$\langle \langle ij \rangle \rangle$ denotes nearest and next-nearest in-plane neighbors respectively and $\{ij \}$ denotes nearest neighbors in adjacent layers on a cubic lattice. The classical spin degrees of freedom $\Sv$ are unit length vectors with $N_s=3$ components.
We will focus on the frustrated case $J_2 >0$ (AF). Without loss of generality we also take $J_1,J_z<0$ (FM).
Due to the bipartite structure of the lattice, our results are equally valid for the corresponding AF cases (up to a corresponding $\pi$ shift in the $\Qv$ vector), obtained by simply reversing the spins on one sublattice ($J_1 \to -J_1$), or by reversing the spins on adjacent layers ($J_z \to -J_z$).

The idea that a spin model with continuous symmetry has a separate Ising-nematic phase was first predicted by Henley~\cite{Henley1989} as an example of the order from disorder scenario~\cite{Villain1980}. This was extended to the Heisenberg \Jonetwo--model by Chandra,Coleman and Larkin\cite{CCL1990} and can be explained in the following way.
In the large $J_2$ limit the spins on each sublattice are strongly coupled. This causes them to align antiferromagnetically on each of the two interpenetrating sublattices. The effective field on a spin mediated through the nearest-neighbor couplings $J_1$ is thus zero. So the sublattices are effectively decoupled resulting in a zero energy cost to rotate all the (anti)aligned spins on one sublattice relative to those on the other sublattice. However, the entropy of thermal spin fluctuations depends on this relative orientation, and selects collinear spins. This can be achieved either by forming stripes of spins along the coordinate x-axis or along the y-axis. This axial orientation of spin correlations along one of the two crystal directions is essentially a discrete Ising degree of freedom that can order at a finite temperature even in a two dimensional system where long-ranged magnetic order is prohibited by the Mermin-Wagner theorem~\cite{MerminWagner1966}. We will refer to this ordering as lattice Ising-nematic order with order parameter denoted by $I$. The presence of long-range magnetic order will be indicated by the order parameter $M$.

In the {\em layered} \Jonetwo--model it is believed that the order-from-disorder scenario still holds for weakly coupled layers, so that there is a region of Ising-nematic order that exists before the long-range stripe order sets in at low enough temperatures, and for strongly coupled layers the nematicity and stripe order occur simultaneously in a first order transition~\cite{Fang2008,CenkeXu2008}.  While these prior expectations are based on an effective description that utilizes the biquadratic coupling, here we confirm this scenario for the layered \Jonetwo--model directly and provide its phase diagram for the first time.

\section{Method \label{sec:method}}
\noindent
In order to make the \Jonetwo--model tractable we write the Hamiltonian in $\qv$-space
\be
H = \sum_{\qv} J_{\qv} \Sv_{\qv} \cdot \Sv_{-\qv},
\label{qHamiltonian}
\ee
where the sum goes over the first Brillouin zone and
\be
\label{jq}
J_{\qv} = J_1 \left( \cos{q_x} + \cos{q_y} \right)
+ 2J_2 \cos{q_x} \cos{q_y} +J_z \cos{q_z} -C.
\ee
We set $J_1=-1$ which defines our unit of energy. For convenience we have subtracted a parameter dependent constant $C$ so that the energy of the minimum of $J_{\qv}$ is zero.
The momentum vectors giving these minimal energies are $\Qv=(0,0,0)$ for $J_2<1/2$, and $\Qv=(\pm\pi,0,0),(0,\pm\pi,0)$ for $J_2>1/2$. 
Thus, the ground state is a FM for $J_2 < 1/2$, and stripe-ordered (with broken lattice rotation symmetry) for $J_2>1/2$.


The spins on all sites are unit length vectors: $\Sv_{\rv} \cdot \Sv_{\rv} =1$. These constraints are enforced in the partition function by writing them as integral representations of $\delta$-functions
\be
\delta \left( \Sv_{\rv} \cdot \Sv_{\rv} -1 \right) = \int \f{\beta d \lambda_{\rv}}{2\pi} \; e^{-i \beta \lambda_{\rv} \left( \Sv_{\rv} \cdot \Sv_{\rv} - 1 \right)}
\ee
where we have scaled the integration variable by the inverse temperature, $\beta=1/T$.
This gives the partition function
\be
Z = \int D\Sv d\Delta D\lambda e^{-\beta\sum_{\qv,\qvp}\left( \Kmat_{\qv,\qvp} - \Lambdamat_{\qv,\qvp} \right) \Sv_{\qv}^* \cdot \Sv_{\qvp}  + V \beta \Delta}
\ee
where we have introduced a matrix $\Lambdamat_{\qv,\qvp} = -i \lambda_{\qv-\qvp} (1-\delta_{\qv,\qvp})$, where $\lambda_{\qv}$ is the Fourier-transformed constraint integration variable. We have separated out its $\qv =0$ component and written it as  $\Delta = i \lambda_{\qv=0}$ and put it into the diagonal momentum space matrix $\Kmat_{\qv,\qvp} \equiv K_{\qv} \; \delta_{\qv,\qvp}$, where $K_{\qv} \equiv J_{\qv} + \Delta$. Factors of $\beta$ have been absorbed into the integration measures.

The enforcement of the unit length constraints as $\delta$-functions allows
us to treat the integrals over the spin components as independent Gaussian integrals. We generalize the spins to have $N_s$ vector components, but will set $N_s=3$ at the end of the calculation. We then scale the spin components by a factor $1/\sqrt{\beta}$ and perform the Gaussian integrals to get
\be
Z = \int d\Delta D\lambda e^{-S_{\lambda}}
\ee
where we have redefined the integration measure with appropriate factors of $\beta$ and
\be
S_\lambda = \f{N_s}{2}  \tr  \ln{\left( \Kmat - \Lambdamat \right)} - \beta V \Delta.
\ee
The thermal average of any spin correlation function can be obtained by adding sources to the action and performing the appropriate derivatives. This yields the momentum dependent susceptibility
\be
\chi_{\qv} \equiv \langle \Sv^*_{\qv} \cdot \Sv_{\qv} \rangle = \f{N_s T}{2}
\langle \left( \Kmat - \Lambdamat \right)^{-1}_{\qv,\qv} \rangle_{S_\lambda} \label{first_chi_eq}
\ee
where the brackets denote the average with respect to $S_\lambda$.

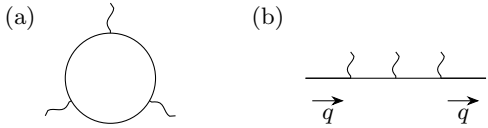
\begin{figure}[h]
\begin{tikzpicture}
\node[] at (-1.2,0.8) {(a)};
\begin{feynman}
\draw (0,0) circle (0.6cm);
\foreach \i in {0,...,2}{
\vertex (a\i) at (120*\i+90:0.6);
}

\foreach \i in {0,...,2}{
\vertex (b\i) at (120*\i+90:1.0);
}

\diagram*{
(a0) -- [photon] (b0);
(a1) -- [photon] (b1);
(a2) -- [photon] (b2);
};
\end{feynman}

\begin{scope}[xshift=0.3\linewidth]
\node[] at (-0.5,0.8) {(b)};
\begin{feynman}
\foreach \i in {0,...,4}{
\vertex (a\i) at (0.6*\i,0);
\vertex (b\i) at (0.6*\i,0.35);
\vertex (c\i) at (0.6*\i,1);
}

\diagram*{
(a0) --[momentum'=$q$] (a1);
(a0) -- (a4);
(a1) -- [photon] (b1);
(a2) -- [photon] (b2);
(a3) -- [photon] (b3);
(a3) --[momentum'=$q$] (a4);
(a0) --[draw=none] (c0);
};
\end{feynman}
\end{scope}
\end{tikzpicture}

\caption{(a) The $n=3$ term coming from expanding $S_\lambda$. (b) The $m=3$ diagram in the expansion of the integrand for the momentum dependent susceptibility.\label{diagrams}}
\end{figure}

In order to calculate this average we first expand the action $S_\lambda$ and the integrand $\left( \Kmat - \Lambdamat \right)^{-1}$ in powers of $\lambda$ and treat everything except the quadratic term as perturbations. Expanding the action gives rise to ring diagrams with $n \geq 3$ wavy lines, one for each $\lambda$-factor, separated by solid lines that each contributes a factor $\Kinv_{\pv}$, see Fig.~\ref{diagrams}(a) for the $n=3$ diagram. The expansion of the integrand gives a sum of diagrams each with two external lines that carry a momentum $\qv$ and $m \geq 0$ wavy lines, see Fig.~\ref{diagrams}(b).

The diagram rules are: The (bare) spin propagator is drawn as a solid line and gives a factor $K^{-1}_{\qv}$. A wavy line, which we will refer to as the (bare) constraint propagator, gives a factor
\be
D_{0\qv} = \f{2}{N_s} \left[ \sum_{\pv} \Kinv_{\pv+\qv} \Kinv_{\pv} \right]^{-1} \label{bareD}
\ee
which originates from the quadratic part of the action $S_\lambda$. The $\qv=0$ component of $D_{0\qv}$ is explicitly set to 0. Every line, solid or wavy, carries a momentum. External lines have a fixed momentum, and there is momentum conservation at each vertex. Undetermined momenta are integrated over. The numerical factors associated to a diagram are: a factor $-i$ for each vertex in a diagram, a factor $N_s/2m$ for each ring with $m$ wavy lines, and an overall combinatorial factor $1/(k_3!k_4!k_5!\ldots)$ where $k_m$ is the number of rings with $m$ wavy lines.

\begin{figure}[h]
\begin{tikzpicture}
\begin{scope}[scale=0.5]
\begin{feynman}
\foreach \i in {0,...,3}{
\vertex (a\i) at (0.5*\i,0);
}
\vertex[left=0.5cm of a0] (i);
\vertex[right=0.5cm of a3] (o);
\diagram*{
(i) --[very thick] (o);
};
\end{feynman}
\end{scope}
\end{tikzpicture}
=
\begin{tikzpicture}
\begin{scope}[scale=0.5]
\begin{feynman}
\foreach \i in {0,...,3}{
\vertex (a\i) at (0.5*\i,0);
}
\vertex[left=0.5cm of a0] (i);
\vertex[right=0.5cm of a3] (o);
\diagram*{
(i) --[] (o);
};
\end{feynman}
\end{scope}
\end{tikzpicture}
+
\begin{tikzpicture}[baseline=(i)]
\begin{scope}[scale=0.5]
\begin{feynman}[inline=(i)]
\foreach \i in {0,...,3}{
\vertex (a\i) at (0.5*\i,0);
}
\vertex[left=0.5cm of a0] (i);
\vertex[left=0.1cm of a3,blob] (b){$\Sigma$};
\vertex[right=0.5cm of a3] (o);
\diagram*{
(i) -- (b);
(b) --[very thick] (o);
};
\end{feynman}
\end{scope}
\end{tikzpicture}
\caption{
Dyson equation for the renormalized spin propagator. \label{Dysonselfenergy}
}
\end{figure}
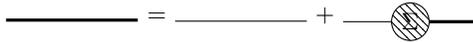

Performing the average over $\lambda$ amounts to writing down all diagrams and connecting the wavy lines. As usual, only connected diagrams with two external legs having the same momentum $\qv$ will contribute to the momentum dependent susceptibility. We approximate the averaging over $\lambda$ by summing just a selected infinite subset of diagrams in the following way. First we define a proper self-energy $\Sigma_{\qv}$ which renormalizes the spin propagator according to the Dyson equation shown in Fig.~\ref{Dysonselfenergy}. The renormalized spin propagator is drawn as a bold solid line.
Solving this equation gives the renormalized spin propagator
\be
\Keffinv_{\qv} = \f{1}{J_{\qv} + \Delta - \Sigma_{\qv}} \label{Keffinv}.
\ee

Similarly a proper polarization $\Pi_{\qv}$ is introduced
\begin{figure}[ht]
\def\outerlinelengths{0.5cm}
\begin{tikzpicture}
\begin{scope}[scale=0.5]
\begin{feynman}
\foreach \i in {0,...,3}{
\vertex (a\i) at (0.5*\i,0);
}
\vertex[left=\outerlinelengths of a0] (i);
\vertex[right=\outerlinelengths of a3] (o);
\diagram*{
(i) --[photon,very thick] (o);
};
\end{feynman}
\end{scope}
\end{tikzpicture}
=
\begin{tikzpicture}
\begin{scope}[scale=0.5]
\begin{feynman}
\foreach \i in {0,...,3}{
\vertex (a\i) at (0.5*\i,0);
}
\vertex[left=\outerlinelengths of a0] (i);
\vertex[right=\outerlinelengths of a3] (o);
\diagram*{
(i) --[photon] (o);
};
\end{feynman}
\end{scope}
\end{tikzpicture}
+
\begin{tikzpicture}[baseline=(i)]
\begin{scope}[scale=0.5]
\begin{feynman}[inline=(i)]
\foreach \i in {0,...,3}{
\vertex (a\i) at (0.5*\i,0);
}
\vertex[left=\outerlinelengths of a0] (i);
\vertex[left=0.02cm of a3,blob] (b){$\Pi$};
\vertex[right=\outerlinelengths of a3] (o);
\diagram*{
(i) --[photon] (b);
(a3) --[photon,very thick] (o);
};
\end{feynman}
\end{scope}
\end{tikzpicture}
\caption{
Dyson equation for the effective constraint-field propagator. \label{Dysonwavy}
}
\end{figure}
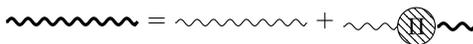
to make a Dyson equation for the renormalized constraint propagator (bold wavy line), see Fig.~\ref{Dysonwavy}. Solving this gives
\be
D^{-1}_{\qv} = D^{-1}_{0\qv} - \Pi_{\qv}. \label{Deff}
\ee

\begin{figure}[h]
\def\outerlinelengths{0.1cm}
\begin{tikzpicture}[baseline=(i)]
\begin{scope}[xshift =-0.22\textwidth]
\begin{feynman}[inline=(i)]

\pgfmathsetmacro{\radius}{0.3};
\pgfmathsetmacro{\x}{0.5};
\pgfmathsetmacro{\y}{0};
\vertex (m)[blob] at ({\x,\y}){\contour{white}{$\Sigma$}};
\foreach \i/\angle in {0/0,1/90,2/180,3/270}{
\vertex (s\i) at ({\x+\radius*cos(\angle))},{\y+\radius*sin(\angle)});
};

\vertex[left=\outerlinelengths of s2] (i);
\vertex[right=\outerlinelengths of s0] (o);

\diagram*{
};
\end{feynman}
\end{scope}
\end{tikzpicture}
=
\begin{tikzpicture}[baseline=(i)]
\begin{feynman}[inline=(i)]
\foreach \i in {0,...,1}{
\vertex (a\i) at (1*\i,0);
}

\foreach \i in {0,...,1}{
\vertex (b\i) at (1*\i,1);
\vertex [above=0.7cm of b\i] (c\i);
}

\vertex[left=\outerlinelengths of a0] (i);
\vertex[right=\outerlinelengths of a1] (o);

\diagram*{
(i) -- (a0);
(a1)-- (o)
(a0) -- [very thick] (a1);
(a0) -- [photon, very thick, half left] (a1);
};
\end{feynman}
\end{tikzpicture}

\vspace{0.5cm}
\def\outerlinelengths{0.3cm}
\begin{tikzpicture}[baseline=(i)]
\begin{feynman}[inline=(i)]

\pgfmathsetmacro{\radius}{0.3};
\pgfmathsetmacro{\x}{0.5};
\pgfmathsetmacro{\y}{0};
\vertex (m)[blob] at ({\x,\y}){\contour{white}{$\Pi$}};
\foreach \i/\angle in {0/0,1/90,2/180,3/270}{
\vertex (s\i) at ({\x+\radius*cos(\angle))},{\y+\radius*sin(\angle)});
};

\vertex[left=\outerlinelengths of s2] (i);
\vertex[right=\outerlinelengths of s0] (o);

\diagram*{
};
\end{feynman}
\end{tikzpicture}
=
\begin{tikzpicture}[baseline=(i)]
\begin{feynman}[inline=(i)]

\pgfmathsetmacro{\radius}{0.3};
\pgfmathsetmacro{\x}{0.5};
\pgfmathsetmacro{\y}{0};
\draw[very thick] (\x,\y) circle (\radius);
\foreach \i/\angle in {0/0,1/60,2/120,3/180}{
\vertex (s\i) at ({\x+\radius*cos(\angle))},{\y+\radius*sin(\angle)});
};

\vertex[left=\outerlinelengths of s3] (i);
\vertex[right=\outerlinelengths of s0] (o);

\diagram*{
(i) --[photon] (s3);
(s0) --[photon] (o);
};
\end{feynman}
\end{tikzpicture}
-
\begin{tikzpicture}[baseline=(i)]
\begin{feynman}[inline=(i)]

\pgfmathsetmacro{\radius}{0.3};
\pgfmathsetmacro{\x}{0.5};
\pgfmathsetmacro{\y}{0};
\draw (\x,\y) circle (\radius);
\foreach \i/\angle in {0/0,1/60,2/120,3/180}{
\vertex (s\i) at ({\x+\radius*cos(\angle))},{\y+\radius*sin(\angle)});
};
\vertex[left=\outerlinelengths of s3] (i);
\vertex[right=\outerlinelengths of s0] (o);

\diagram*{
(i) --[photon] (s3);
(s0) --[photon] (o);
};
\end{feynman}
\end{tikzpicture}
\caption{Self-consistent equations for the self-energy and the polarization. Note that the bold lines on the right hand sides also include the self-energy and the polarization.\label{attempt}}
\end{figure}
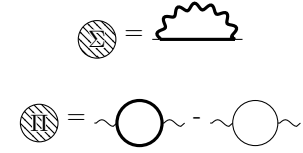

Next we approximate the self-energy and the polarization by the self-consistent diagrams in Fig.~\ref{attempt}
which is equivalent to writing
\begin{align}
\Sigma_{\qv} &= (-i)^2 \sum_{\pv} \Keffinv_{\qv-\pv} D_{\pv}. \label{selfenergy_eq} \\
\Pi_{\qv} &= (-i)^2 \f{N_s}{2} \sum_{\pv} \Keffinv_{\pv+\qv} \Keffinv_{\pv} -(-i)^2 \f{N_s}{2} \sum_{\pv} \Kinv_{\pv+\qv} \Kinv_{\pv}
\label{constraintprop_eq_a}
\end{align}

The expression for the proper polarization Eq.~(\ref{constraintprop_eq_a}) is finally converted, using Eqs.~(\ref{bareD}) and (\ref{Deff}), into an equation for the renormalized constraint propagator
\be
D_{\qv}^{-1} = \f{N_s}{2} \sum_{\pv} \Keffinv_{\pv+\qv} \Keffinv_{\pv}. \label{constraintprop_eq}
\ee

Equations~(\ref{Keffinv}), (\ref{selfenergy_eq}) and (\ref{constraintprop_eq}) represent the averaging over the non-zero momentum modes of the constraint field and define a system of self-consistent equations that can be solved iteratively to give $\Keffinv_{\qv}$ as a function of $\Delta$.
After averaging over these modes the expression for the spin susceptibility, Eq.~(\ref{first_chi_eq}), becomes
\be
\chi_{\qv} = \f{N_s T}{2} \langle \Keffinv_{\qv} \rangle_{S_\Delta}
\ee
where the brackets denote the remaining average over the zero momentum mode(homogeneous component) $\Delta$ of the constraint field taken with respect to the weight $e^{-S_\Delta} = \int D \lambda e^{-S_\lambda}$.
This averaging is carried out by simply replacing it with a single value of $\Delta$ which for self-consistency is the value which ensures that the unit vector constraint is satisfied as an average: $\langle \Sv_{\rv} \cdot \Sv_{\rv} \rangle = 1$ which is equivalent to $\f{1}{V} \sum_{\qv} \chi_{\qv}=1$. Thus the value of $\Delta$ (contained in $\Keffinv$) is chosen so that it satisfies
\be
\f{N_s T}{2V} \sum_{\qv} \Keffinv_{\qv} = 1, \label{sumcond}
\ee
and the spin susceptibility becomes
\be
\chi_{\qv} = \f{N_s T}{2} \Keffinv_{\qv}, \label{chi_eq}
\ee
using the particular value of $\Delta$ that satisfies Eq.~(\ref{sumcond}).

Instead of fixing $T$ from the outset and seeking a value of $\Delta$ that satisfies Eq.~(\ref{sumcond}), we will rewrite Eq.~(\ref{sumcond}) as a way to {\em calculate} the temperature given a fixed value of $\Delta$:
\be
T = \left[ \f{N_s}{2V} \sum_{\qv} \Keffinv_{\qv} \right]^{-1},  \label{T_eq}
\ee
and solve the self-consistent equations keeping the value of $\Delta$ fixed. That is, we introduce an extra ``mass renormalization'' step where $\Delta$ is restored to its original value after each iteration.

Thus our procedure for solving the equations is as follows. First pick a value of $\Delta$ and an initial guess for $\Sigma_{\qv}$. Set $N_s=3$. Then iterate the following steps:
\begin{enumerate}
\item Subtract a constant from $\Sigma_{\qv}$ so that its minimum value becomes zero.
\item Make $\Keffinv_{\qv}$ according to Eq.~(\ref{Keffinv}), and calculate $T$ according to Eq.~(\ref{T_eq}). If $T$ has converged then exit and use the obtained $\Keffinv_{\qv}$ to compute $\chi_{\qv}$, Eq.~(\ref{chi_eq}).
\item Calculate $D_{\qv}$ using the new $\Keffinv_{\qv}$, Eq.~(\ref{constraintprop_eq}).
\item Calculate the new self-energy $\Sigma_{\qv}$, Eq.~(\ref{selfenergy_eq}).
\item Go to step 1.
\end{enumerate}

This whole procedure is repeated for a range of $\Delta$ values, typically two hundred values ranging from $10^{-9}$ to $1$ evenly spaced on a log-scale. We use the convergence criterion that $T$ has converged when the relative difference between the temperatures at succesive iterations, $|T_{n+1}-T_{n}|/T_{n} < 10^{-8}$. Convergence is typically reached after 10-20 iterations.

For each value of $\Delta$ the iteration procedure yields a value of $T$ and a corresponding susceptibility $\chi_{\qv}$. While $T$ is usually a single-valued function of $\Delta$, there are temperature regions near phase transitions where different values of $\Delta$ correspond to the same value of $T$ but to different values of $\chi_{\qv}$. Thus in these regions the equations give several different solutions $\chi_q$ for the same value of $T$.  We will deal with this by selecting the solution with the lowest free energy, which we take to be the one smoothly connected to the unique solution on the low-$T$ side of the multi-valued region.

\section{Order parameters \label{sec:orderpar}}
The calculation of the momentum-dependent susceptibility $\chi_{\qv}$ at different temperatures allows us to make inferences about different phases and order parameters. We focus on the planar nematic order parameter
\be\label{eq:I}
\I
= \f{1}{V} \sum_{\rv} \langle \vec{S}_{\rv} \cdot \vec{S}_{\rv+\xhat} - \vec{S}_{\rv} \cdot \vec{S}_{\rv+\yhat} \rangle
\ee
where $\xhat,\yhat$ are planar unit lattice vectors and $\langle...\rangle$ denotes the thermal expectation value. In terms of the momentum space susceptibility, the order parameter is
\be
I = \f{1}{2V} \sum_{\qv} \left( \cos{q_x}-\cos{q_y} \right) \chi_{\qv}
\ee
where the sum over $\qv$ is taken over the first Brillouin zone.
The planar nematic order parameter detects anisotropy in bond ordering on the planes, and does not break spin rotational symmetry.

We also calculate the magnetization order parameter which breaks spin rotation symmetry. We infer this from the coefficient of the diverging susceptibility as the system size goes to infinity.
The $\qv$-dependent susceptibility at the ordering vector is
\be
\chi_{\Qv} = \vec{m}_{\Qv} \cdot \vec{m}_{-\Qv} V + \sum_{\rv} \delta f(\rv) e^{-i \Qv \cdot \rv}
\ee
where $\delta f(\rv)$ is the spin fluctuation correlation function characterized by a correlation length $\xi$. For finite $\xi$ less than the linear system size, the last term will be independent of $V$. Therefore for large system sizes we can keep only the first term, which diverges with increasing $V$, and arrive at
\be
M^2 \equiv \vec{m}_{\Qv} \cdot \vec{m}_{-\Qv} = \f{N_s T}{2V} \Keffinv_{\Qv}
=\f{N_s T}{2V \Delta}. \label{magnetization}
\ee
In the last equality we have used the fact that the maximum value $\Keffinv_{\Qv}$ is always $1/\Delta$ because both the self-energy $\Sigma_{\Qv}$ and $J_{\Qv}$ is zero for the maximum value of $\Keffinv$.
Extracting the magnetization this way gives the dominating magnetic order corresponding to the wave vector $\Qv$ where $\Keffinv$ is maximal.


\section{Results \label{sec:results}}
For simplicitly, we begin by studying the $J_2=0$ Hamiltonian where our results can be compared to Monte Carlo data. In this case the Hamiltonian reduces to an unfrustrated layered ferromagnet (FM) where we expect a finite temperature phase transition to an ordered state with a FM magnetic moment at wavevector $\Qv=(0,0,0)$. We first consider the isotropic FM $(J_z=-1)$ and solve the self-consistent equations numerically for a range of $\Delta$'s and obtain results for the magnetization using Eq.~(\ref{magnetization}). We always start the iterations of the self-consistent equations with an initial guess for $\Sigma_{\qv}$ which breaks the nematic symmetry so that the initial $I$ is slightly negative.

\begin{figure}[t]
\begin{center}
\begin{tikzpicture}
\begin{scope}[]
\node[anchor=south west,inner sep=0] (image1) at (0\textwidth,0) {\includegraphics[width=0.22\textwidth]{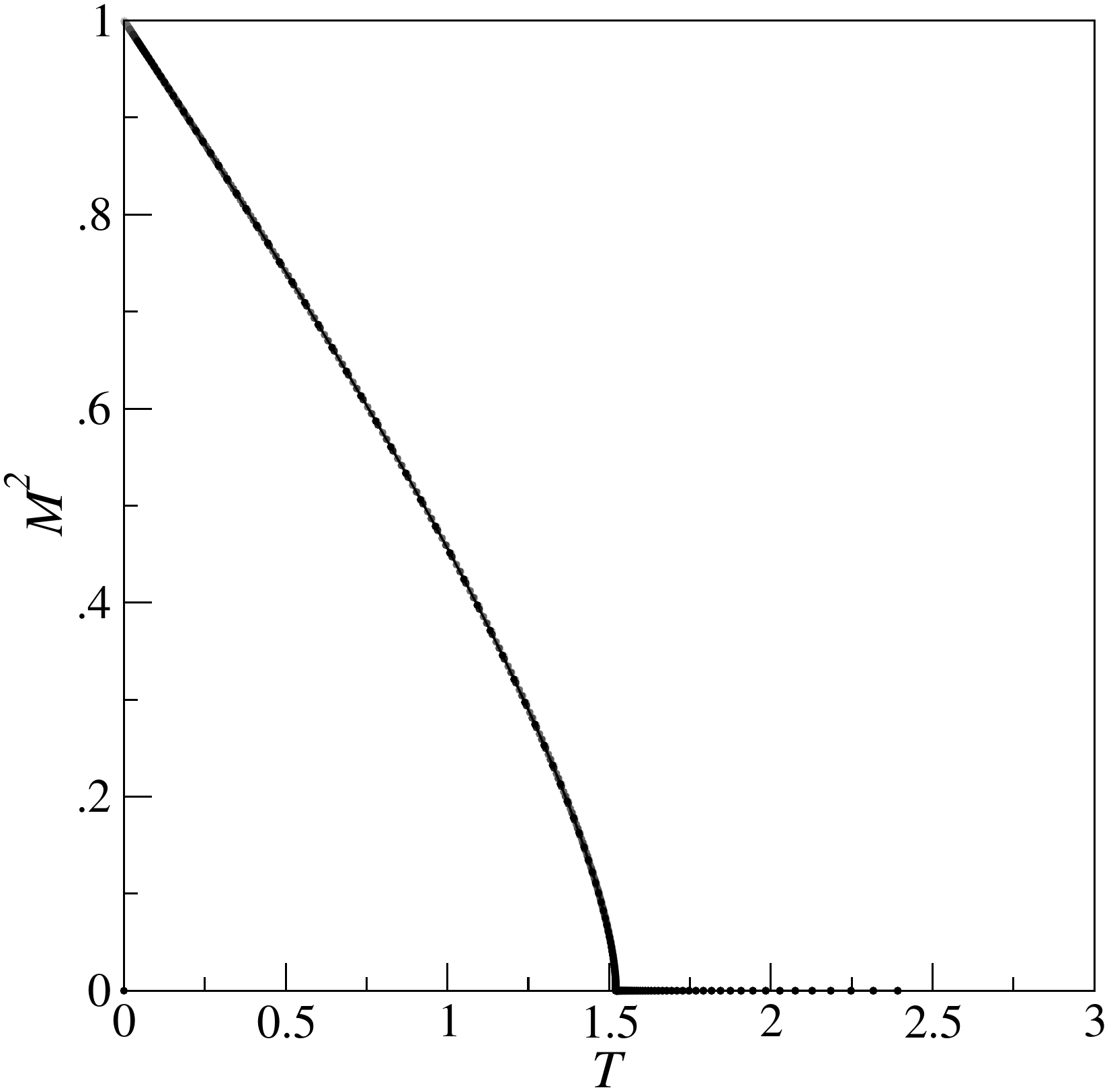}};
\node[] at (3.66,3.61) {(a)};
\end{scope}
\begin{scope}[xshift =0.23\textwidth]
\node[anchor=south west,inner sep=0] (image2) at (0\textwidth,0) {\includegraphics[width=0.2372\textwidth]{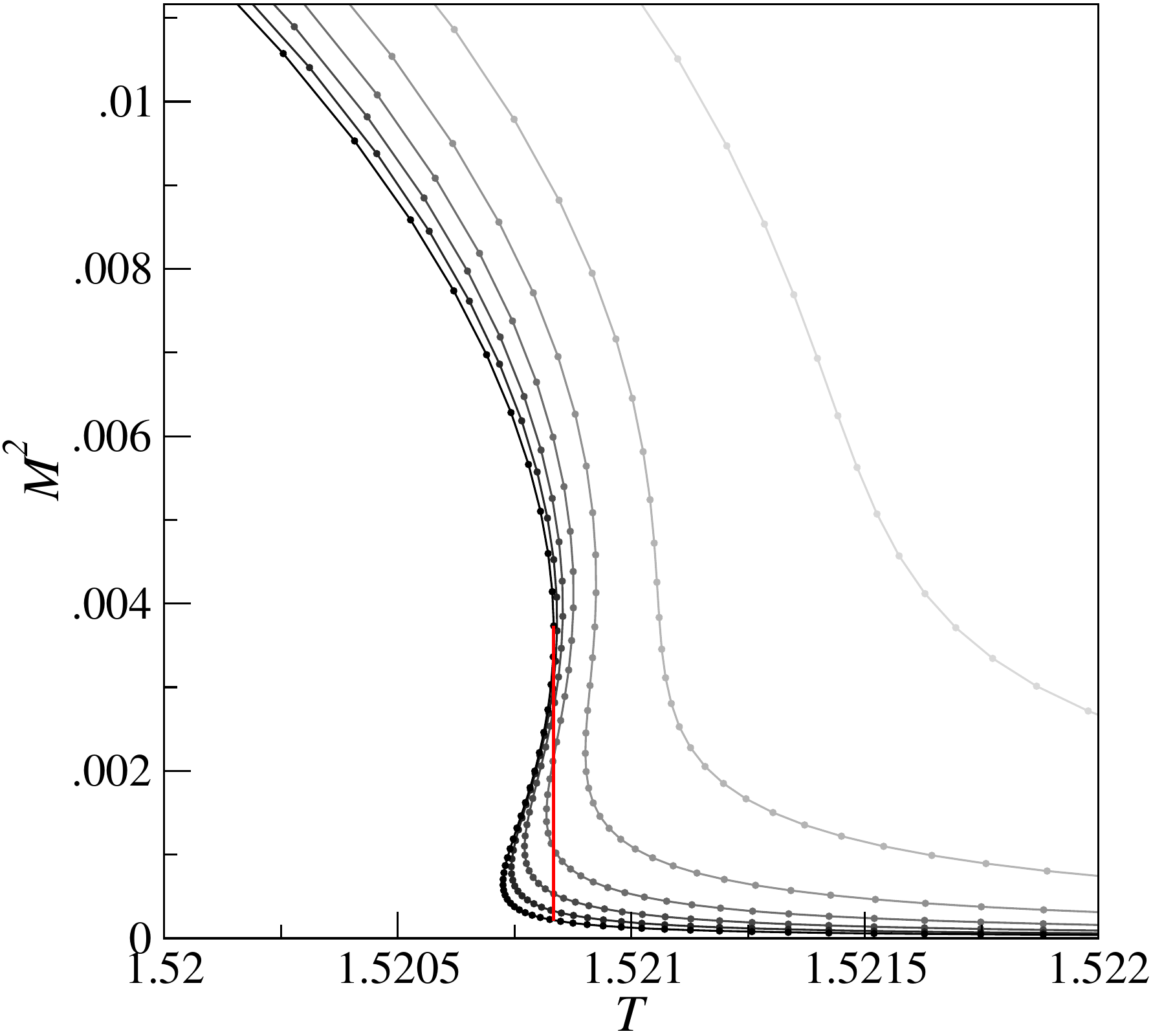}};
\node[] at (3.83,3.6) {(b)};
\end{scope}
\end{tikzpicture}
\caption{Magnetic order parameter squared for the unfrustrated 3D FM: $J_z=-1.0$, $J_2=0$. Both panels show finite-size curves with $L=100-400$ where darker curves indicate larger $L$. Panel (b) shows a zoom-in on panel (a) near the phase transition point. A vertical red line indicating a jump in the order parameter is drawn for the largest $L$ curve.
\label{3DFM}
}
\end{center}
\end{figure}

Figure~\ref{3DFM}(a) shows the magnetization squared as a function of temperature for cubic systems of different linear sizes $L=100-400$. The finite-size curves fall almost on top of each other and do not cross. On magnifying the behavior close to $T_c$ where the magnetization vanishes, finite-size effects become apparent, see Fig.~\ref{3DFM}(b), and as $T$ increases there is a slight overshoot of the magnetization curves for the largest system sizes. These magnetization curves are clearly unphysical as the overshoot leads to a temperature region near $T_c$ where the magnetization is a multivalued function of $T$. We interpret this behavior as the system is having multiple possible solutions to the self-consistent equations with different free energies at the same temperature. The free energy increases with temperature, thus in order to choose the solution with lowest free energy we pick the branch in the multivalued region that is connected smoothly to the low temperature single-valued region. When following this branch upon increasing $T$, the curve will turn around at some temperature and will, if continued, give other solutions with higher free energies. We omit these by drawing a vertical line towards zero order parameter at the first turning point  (infinite slope of $M^2(T)$) upon increasing $T$. This is drawn as a red line for the largest system size in Fig.\ref{3DFM}(b). Then the order parameter evolution with $T$ continues from where the line hits the order parameter curve again. This results in a discontinuous jump in the order parameter at the critical temperature which we interpret as a discontinuous phase transition. However, in the case of Fig.~\ref{3DFM}(b) the discontinuous jump decreases as the system size is increased, which when extrapolated to infinite size results in a \emph{continuous} transition, as is expected for a 3D FM.

In order to determine $T_c$ of the magnetic phase transition we use different methods based on how the finite-size curves behave. If they cross, we pick the crossing-points between successive linear system sizes $L$ and $L+50$ and extrapolate these crossings to infinite size using a cubic polynomial in $1/L$. For finite-size curves that do not cross but has an overshoot we identify the temperature of each finite-size curve at the point where the magnetization curve turns back, and then extrapolate these points to the infinite $L$ limit using a cubic polynomial in $1/L$. A third method we use, which also works when the magnetization curve is single-valued, is to use the expected behaviour of the magnetization near a continuous phase transition $M = A(T_c-T)^{\beta}$. To do so we make a Kouvel-Fisher plot\cite{KouvelFisher1964}, i.e. we plot $\left[-\f{d \ln M}{dT} \right]^{-1}$ vs. $T$ and find for each finite-size curve the temperature where this quantity crosses the temperature axis. These are finally extrapolated to the infinite-size limit using a cubic polynomial in $1/L$. For $J_z=-1.0$ and $J_2=0$ these methods give results that are very close to each other: $T_c=1.52083$ (disc.) and $T_c = 1.52077$ (Kouvel-Fisher).
This is to be compared with the most accurate Monte Carlo result\cite{ChenFerrenbergLandau1993} which is $T_c=1.442928(77)$, a relative difference of $\sim 5\%$. We attribute this difference to the lack of vertex corrections in our self-consistent equations. Thus we estimate that our critical temperatures are approximate with a relative accuracy of roughly $5\%$.
For very weakly coupled layers where the system is almost two dimensional we use a fourth method where the magnetic $T_c$ is determined as the temperature at which the spin-spin correlation length diverges, see Sec.~\ref{sec:correlations}.

We now turn to the weakly frustrated FM regime $0 < J_2 <0.5$ with isotropic interlayer couplings $J_z=-1$. We find that
for increasing $J_2$, $T_c$ goes down, and the magnetization overshoot seen for finite sizes at $J_2=0$ quickly becomes smaller. For $J_2>0.5$ there is no longer a low temperature finite FM magnetization. Instead the dominating divergence of the susceptibility is at $\Qv=(\pi,0,0)$ or $\Qv=(0,\pi,0)$ which indicates stripe magnetic order which we will denote by $M$. In addition there is also a finite value of the nematic order parameter $I$, see Fig.~\ref{JZ-1.0_J20.51}(a). In this figure (and in following figures), the stripe magnetization squared is shown as positive values, while the nematic order parameter is shown as negative values. Finite size effects are small for $J_2=0.51$, and those present indicate that the jump in the order parameters increases slightly with system size. Thus we conclude that for $J_2=0.51$ there are simultaneous discontinuous phase transitions in both the nematic order parameter $I$ and the stripe magnetization $M$, which we write in short-form {\IdMsd}, where the d means discontinuous.
On further increasing $J_2$ this simultaneous {\IdMsd} character of the transition persists up to the largest $J_2$ value studied, $J_2=2$, see Fig.~\ref{JZ-1.0_J20.51}(b). The quantitative changes upon increasing $J_2$ include: larger finite size effects, increasing $T_c$, and smaller jumps in the order parameters.
\begin{figure}[t]
\begin{center}
\begin{tikzpicture}
\begin{scope}[]
\node[anchor=south west,inner sep=0] (image1) at (0\textwidth,0) {
\includegraphics[width=0.22\textwidth]{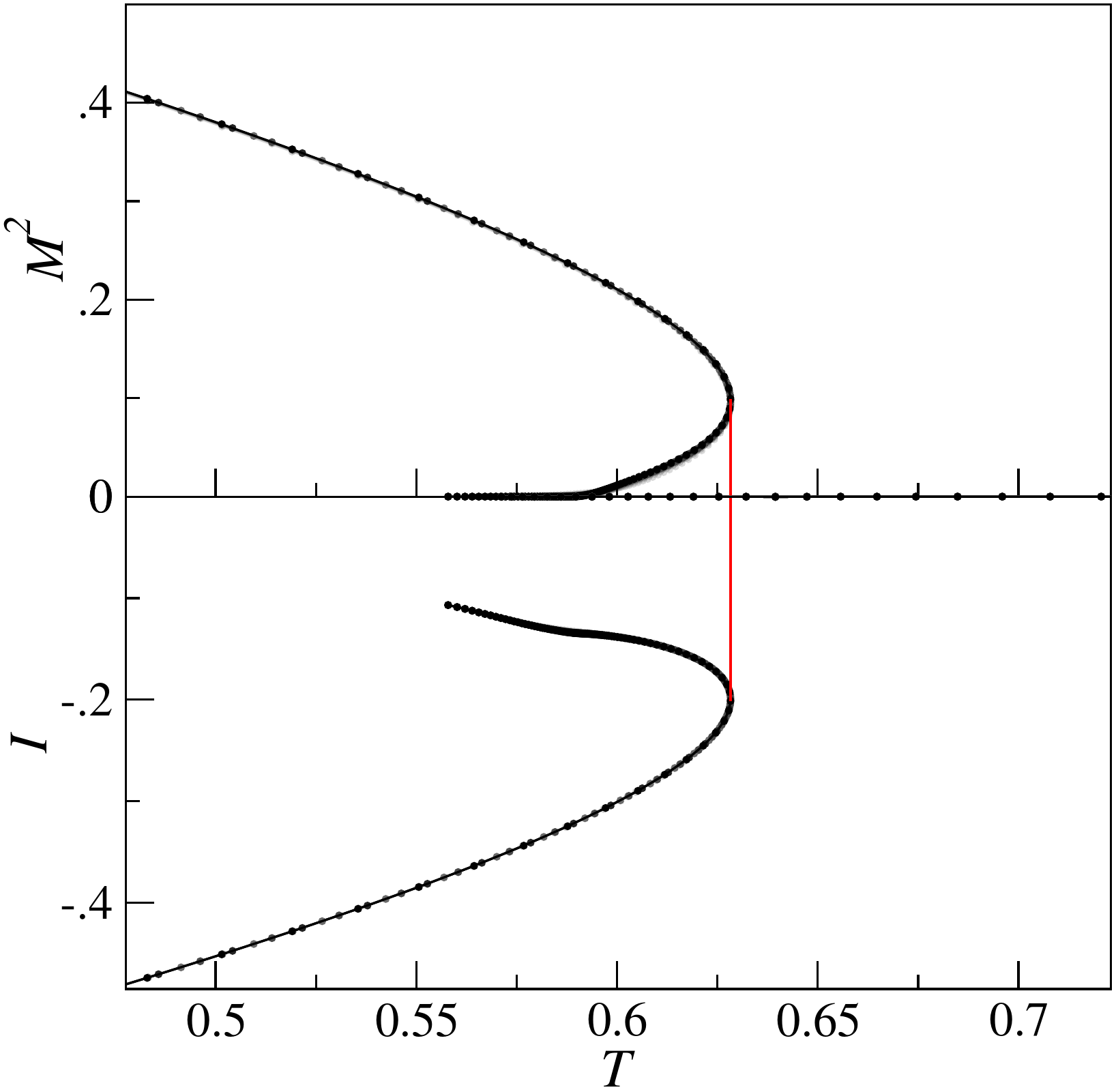}};
\node[] at (3.7,3.65) {(a)};
\end{scope}
\begin{scope}[xshift=0.23\textwidth]
\node[anchor=south west,inner sep=0] (image2) at (0\textwidth,0) {
\includegraphics[width=0.232\textwidth]{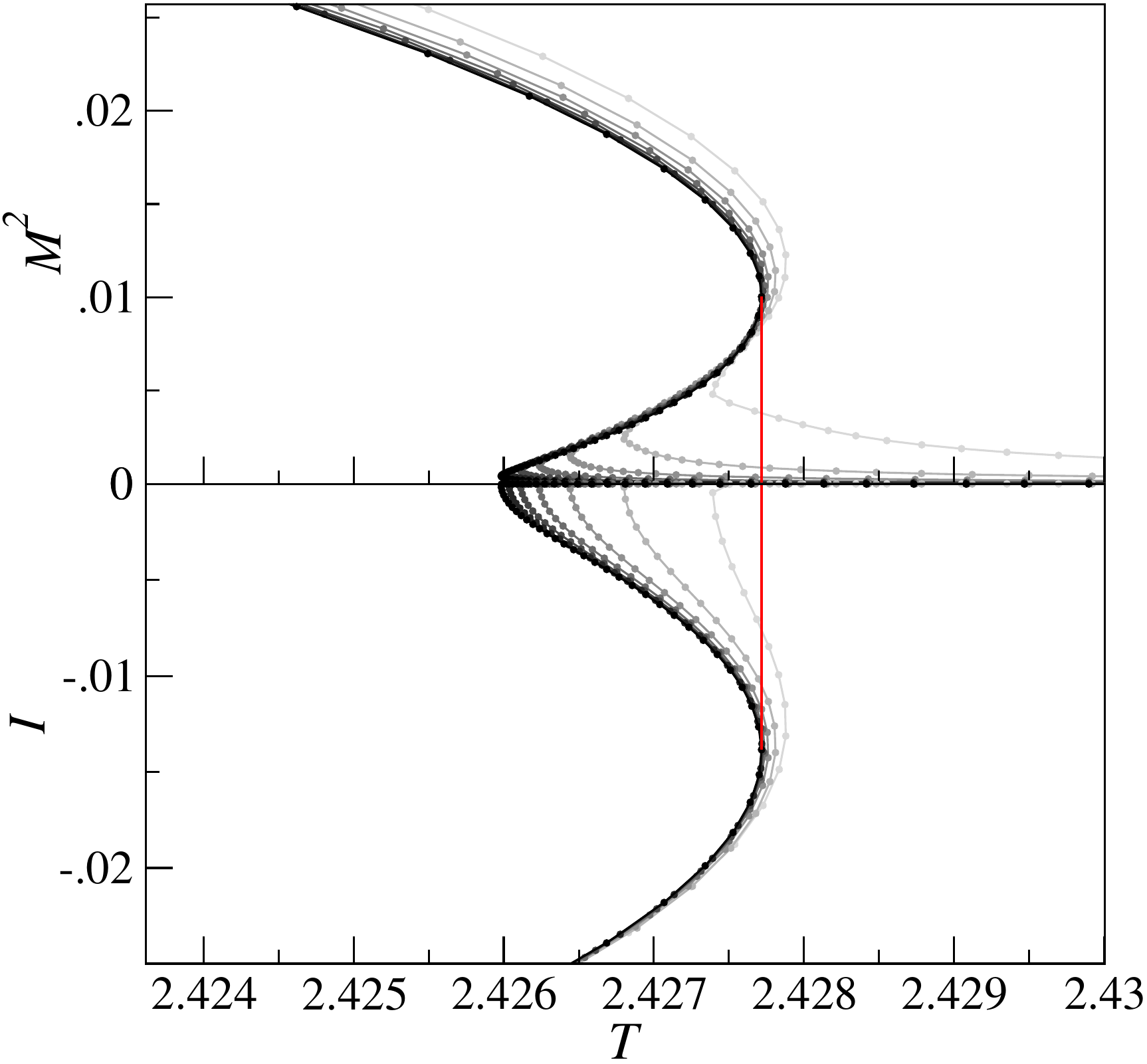}};
\node[] at (3.79,3.65) {(b)};
\end{scope}
\end{tikzpicture}
\caption{Order parameters for $J_z=-1$, and (a) $J_2=0.51$, (b) $J_2=2.0$ for various system sizes $L=100-400$. Darker curves indicate larger sizes. The stripe magnetization squared is shown as positive values, while the nematic order parameter $I$ is shown as negative values.
\label{JZ-1.0_J20.51}
}
\end{center}
\end{figure}

\begin{figure}[t]
\begin{center}
\begin{tikzpicture}
\begin{scope}[]
\node[anchor=south west,inner sep=0] (image1) at (0\textwidth,0) {
\includegraphics[width=0.45\textwidth]{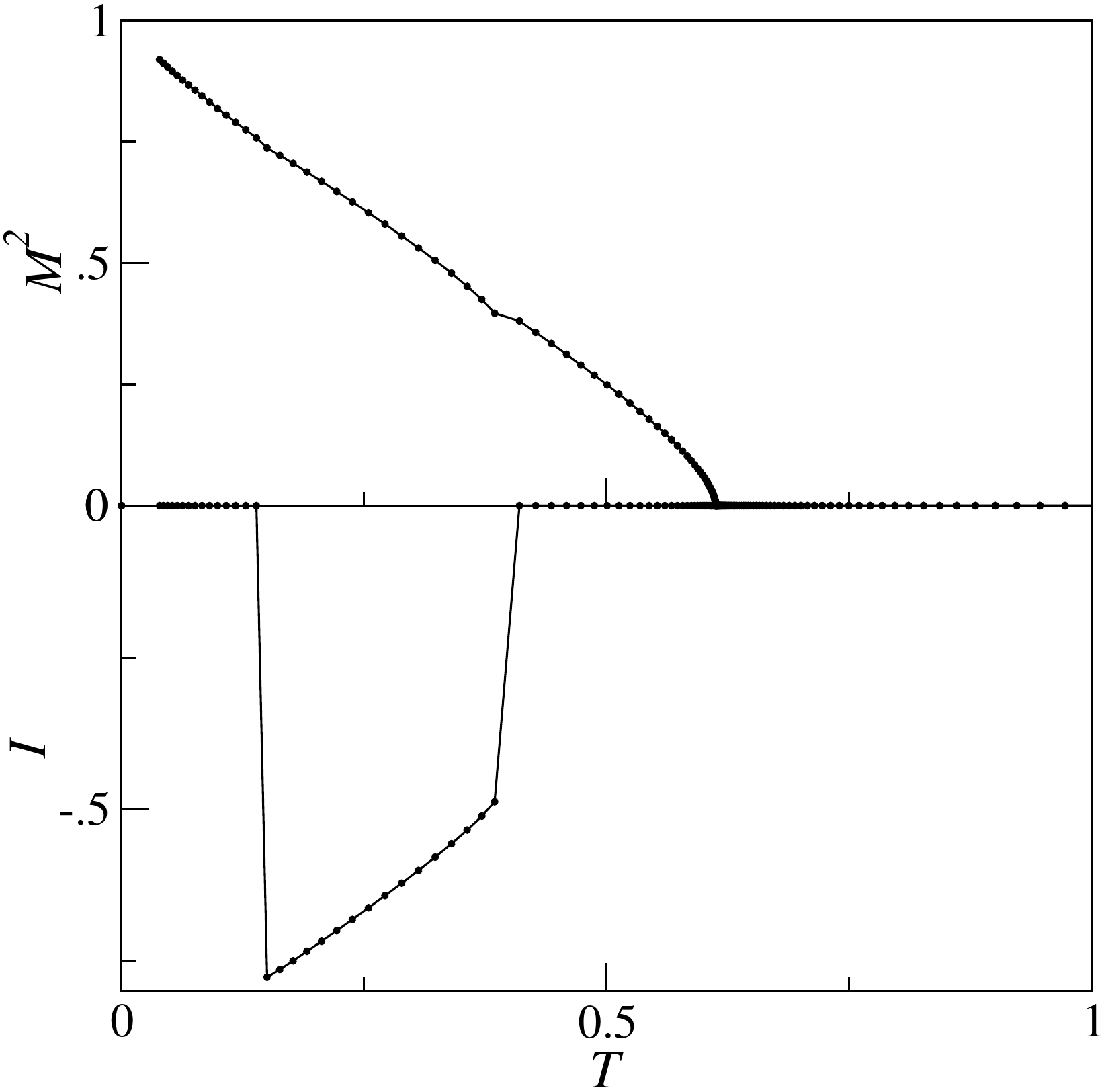}};
\end{scope}
\end{tikzpicture}
\caption{Order parameters for $J_z=-1.0$, $J_2=0.495$. The nematic order parameter $I$ is shown as negative values. The dominant magnetization is shown as positive values. An intermediate regime is clearly visible with non-zero nematic order parameter $I$.
\label{lowTphase}
}
\end{center}
\end{figure}

For $J_2=0.495$, i.e slightly less than the boundary between the FM and the stripe phase, the peak magnetic ordering wave vector changes with temperature from being FM at high temperatures, to stripe order, and then back to FM again at the lowest temperatures. The intermediate stripe order is also indicated by the finite temperature region with nonzero $I$ in Fig.~\ref{lowTphase}. Thus, at finite temperatures the nematic phase extends slightly into the region $J_2 < 0.5$.

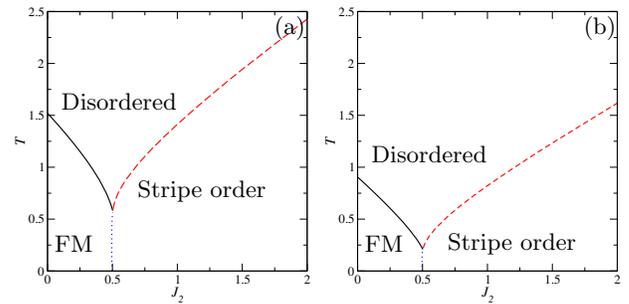
\begin{figure}[t]
\begin{center}
\begin{tikzpicture}
\begin{scope}[]
\node[anchor=south west,inner sep=0] (image1) at (0\textwidth,0) {
\includegraphics[width=0.22 \textwidth]{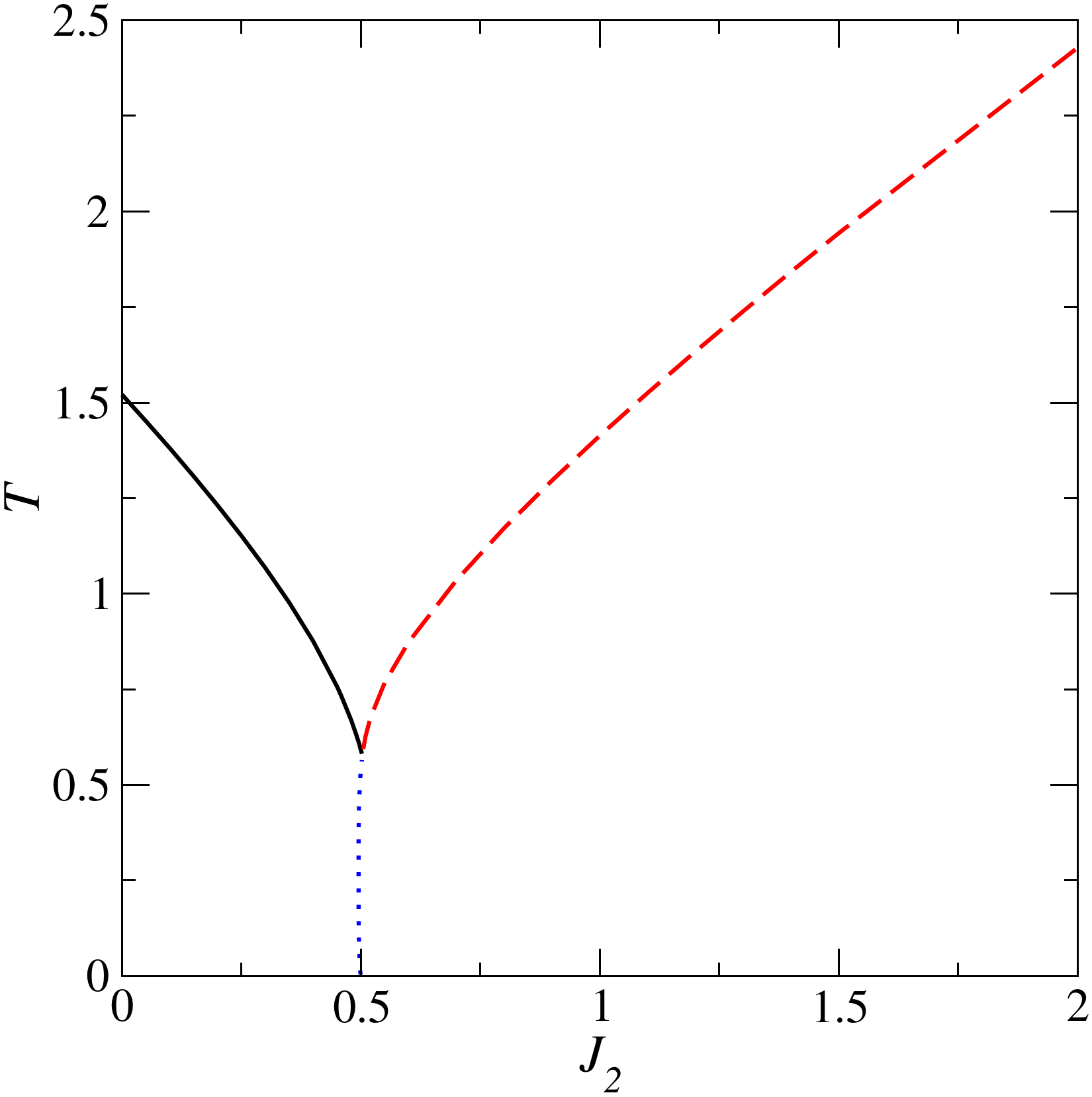}};
\node[] at (3.67,3.68) {(a)};
\node[] at (0.8,0.8) {FM};
\node[] at (2.5,1.5) {Stripe order};
\node[] at (1.4,2.7) {Disordered};
\end{scope}
\begin{scope}[xshift =0.23\textwidth]
\node[anchor=south west,inner sep=0] (image2) at (0\textwidth,0) {
\includegraphics[width=0.22 \textwidth]{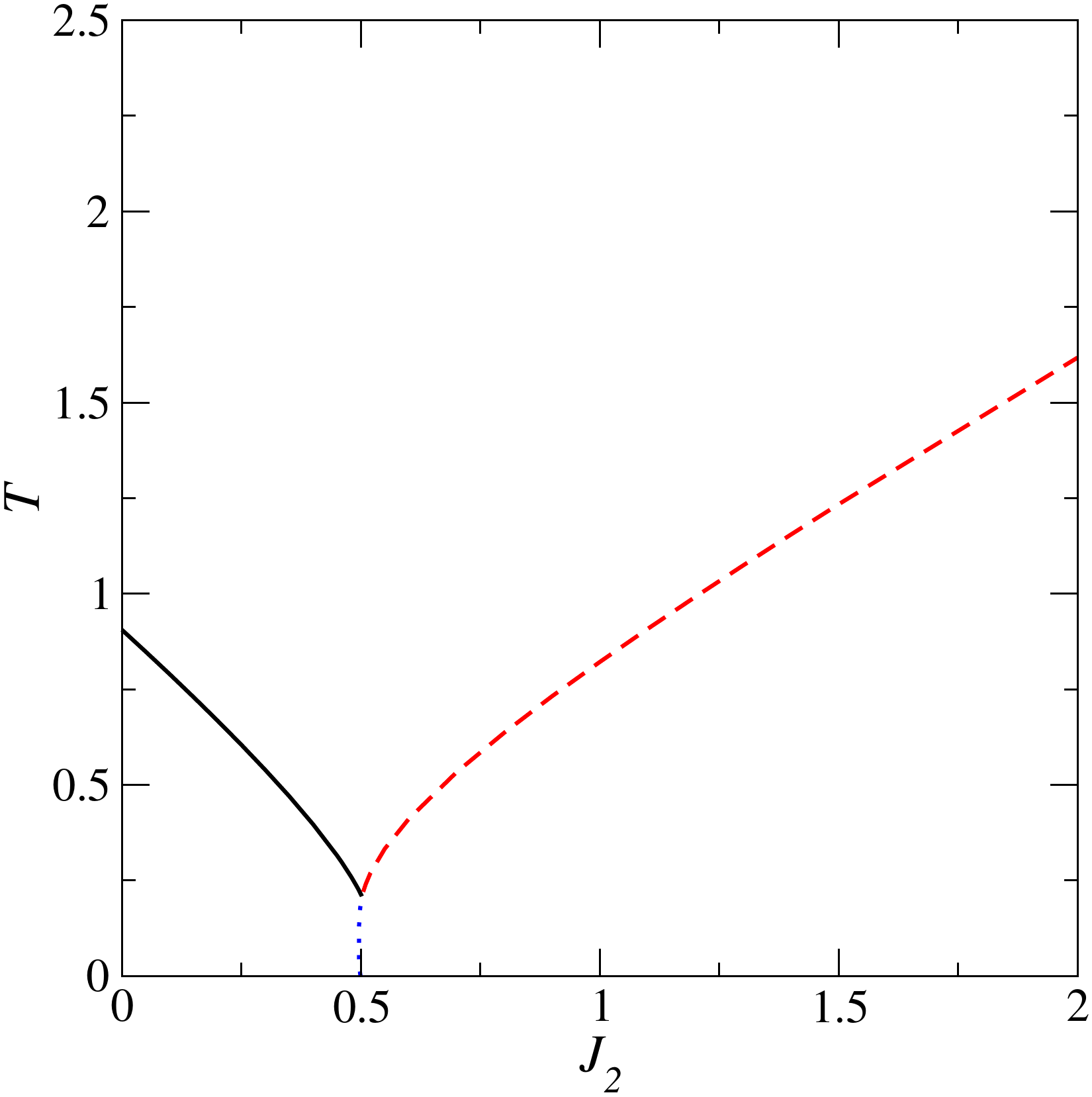}};
\node[] at (3.67,3.68) {(b)};
\node[] at (0.8,0.8) {FM};
\node[] at (2.5,0.8) {Stripe order};
\node[] at (1.4,2.) {Disordered};
\end{scope}
\end{tikzpicture}
\caption{Phase diagrams for (a) $J_z=-1.0$ and (b) $J_z=-0.1$. Disorder-FM phase boundary (solid black), disorder-stripe phase boundary(dashed red). FM-stripe phase boundary(dotted blue).
\label{Phasediagram_JZ-1.0}
}
\end{center}
\end{figure}

Summarizing these results for $J_z=-1$ gives the phase diagram shown in Fig.~\ref{Phasediagram_JZ-1.0}(a) where the black curve shows the continuous FM phase transition while the red curve marks the discontinuous simultaneous nematic and stripe phase transitions.  The blue curve which is slightly bent towards the FM phase defines the boundary between the ordered FM and stripe phases.  The phase diagram for $J_z=-0.1$ similarly obtained is shown in Fig.~\ref{Phasediagram_JZ-1.0}(b).

\begin{figure}[t]
\begin{center}
\begin{tikzpicture}
\begin{scope}[]
\node[anchor=south west,inner sep=0] (image1) at (0\textwidth,0) {
\includegraphics[width=0.23 \textwidth]{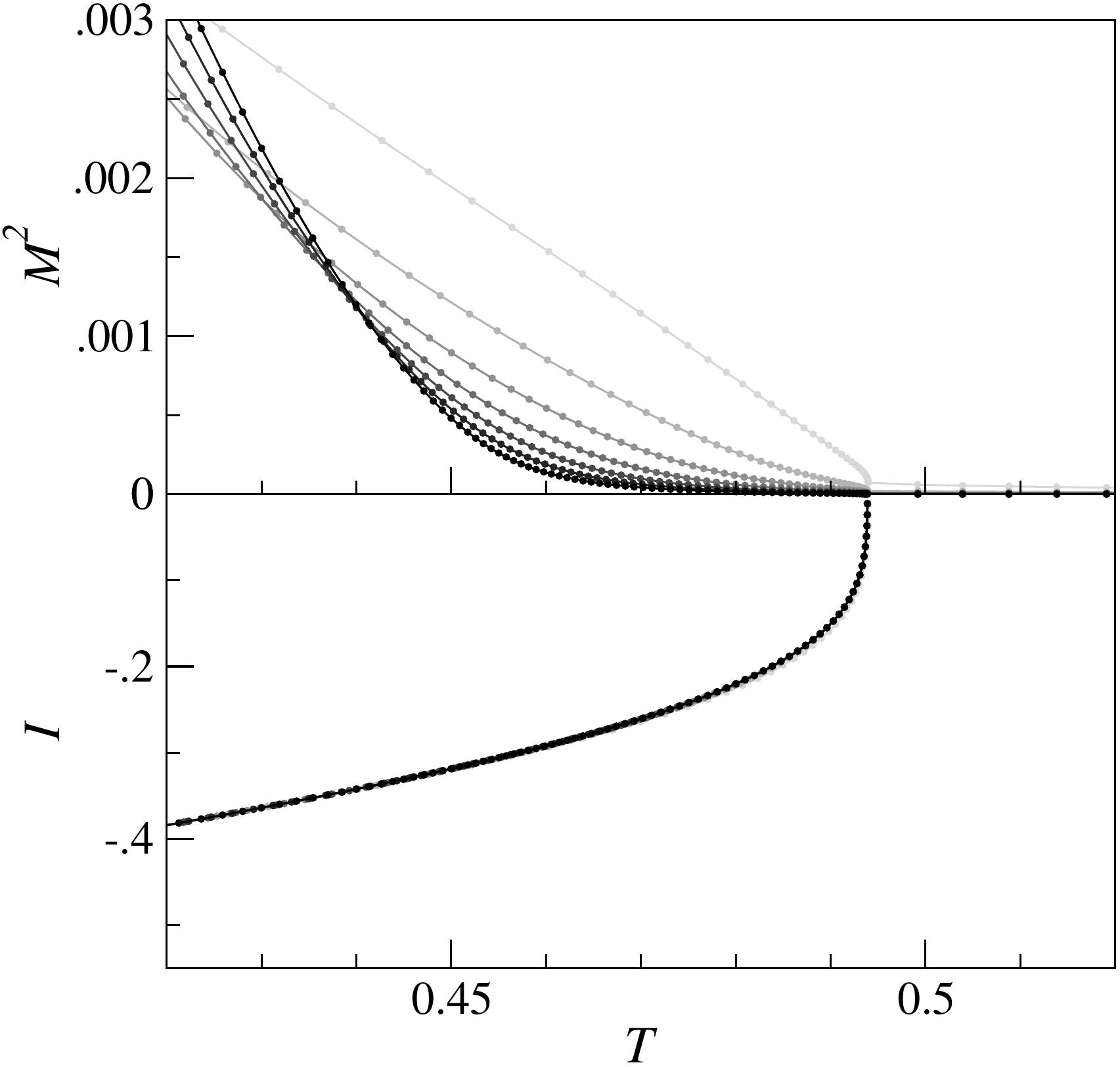}};
\node[] at (3.88,3.65) {(a)};
\end{scope}
\begin{scope}[xshift =0.24\textwidth]
\node[anchor=south west,inner sep=0] (image2) at (0\textwidth,0) {
\includegraphics[width=0.23 \textwidth]{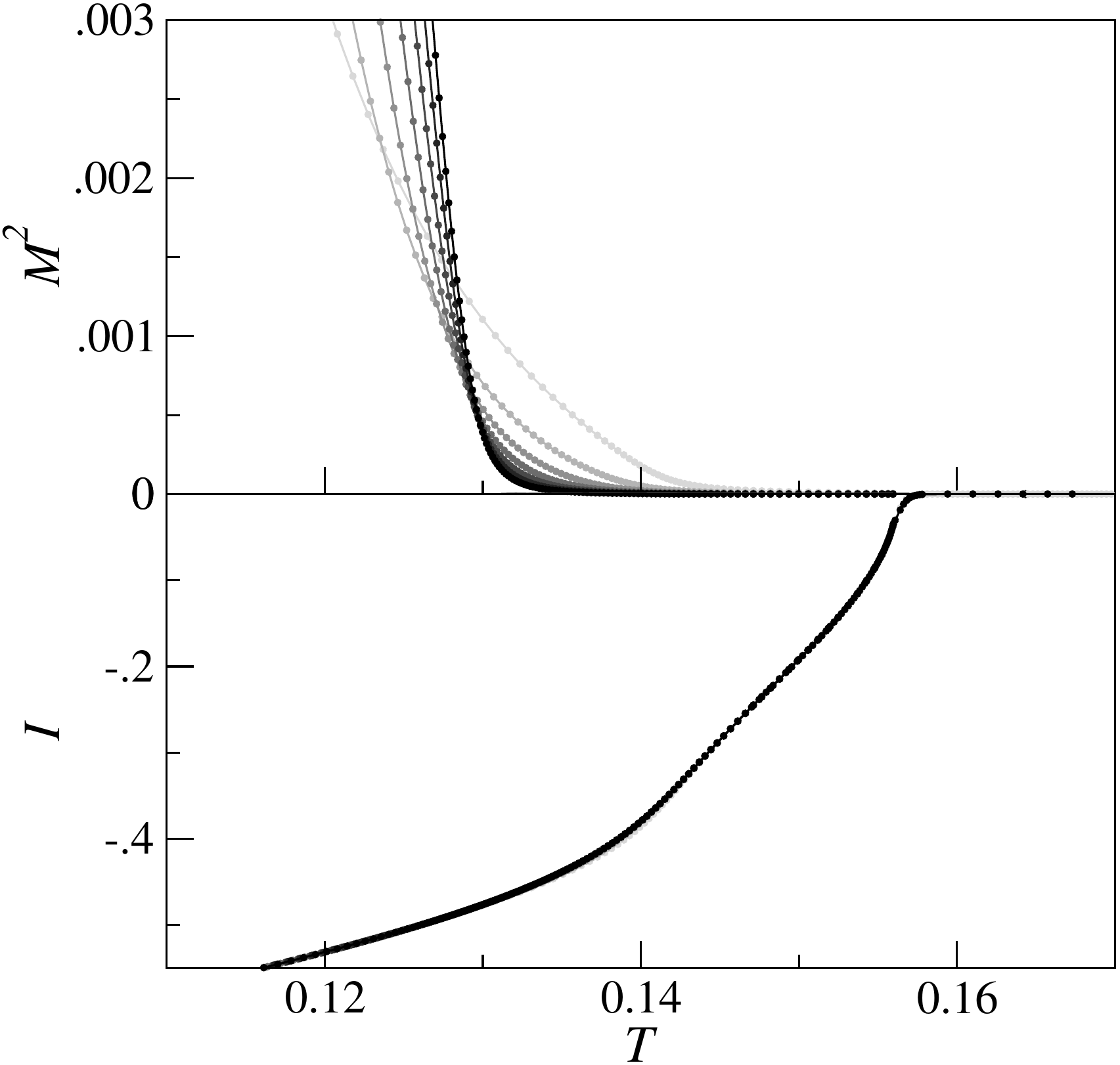}};
\node[] at (3.88,3.65) {(b)};
\end{scope}
\end{tikzpicture}

\begin{tikzpicture}
\begin{scope}[xshift=0.01\textwidth]
\node[anchor=south west,inner sep=0] (image2) at (0\textwidth,0) {
\includegraphics[width=0.235 \textwidth]{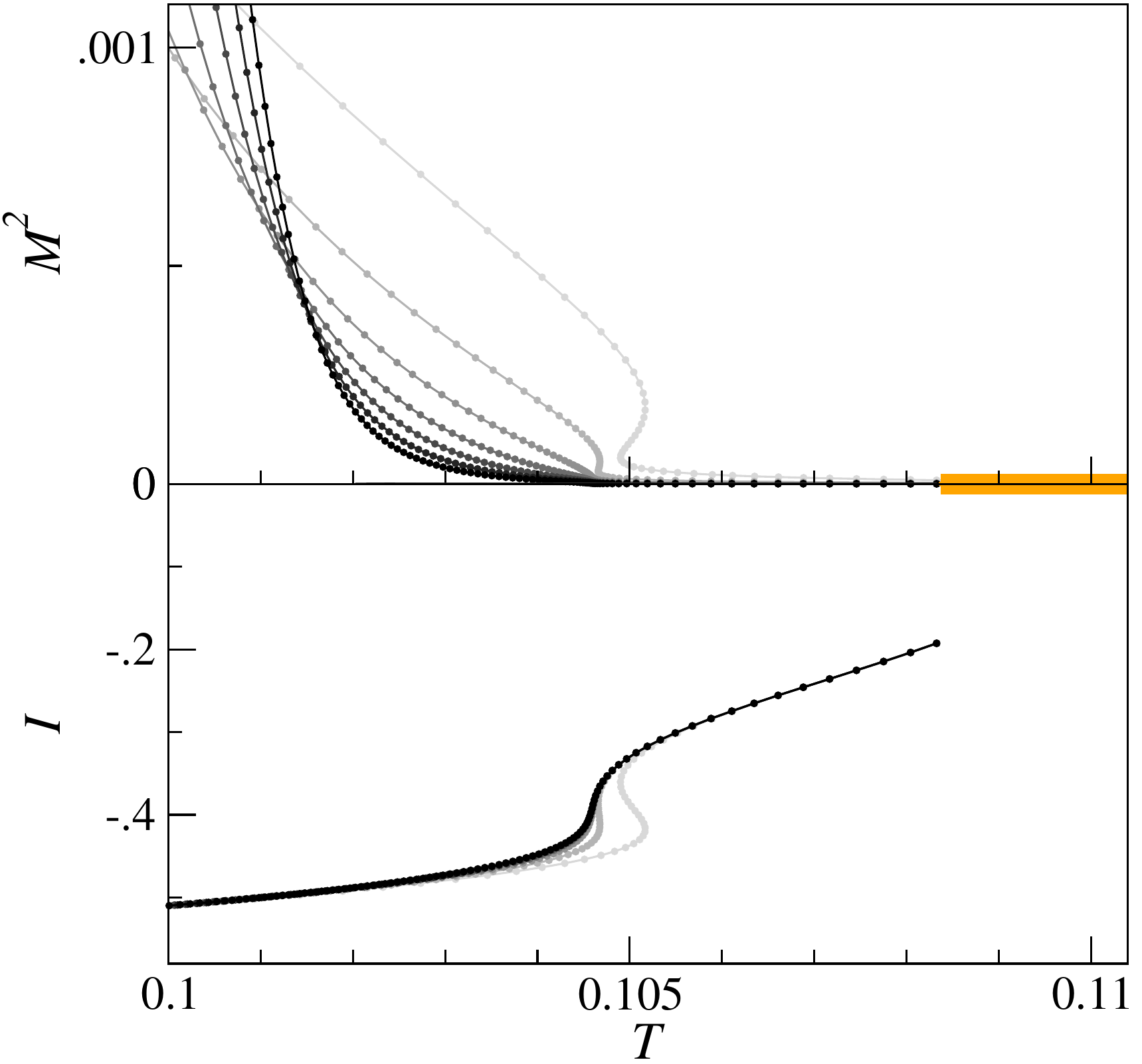}};
\node[] at (3.94,3.72) {(c)};
\end{scope}
\begin{scope}[xshift =0.25\textwidth]
\node[anchor=south west,inner sep=0] (image2) at (0\textwidth,0) {
\includegraphics[width=0.235 \textwidth]{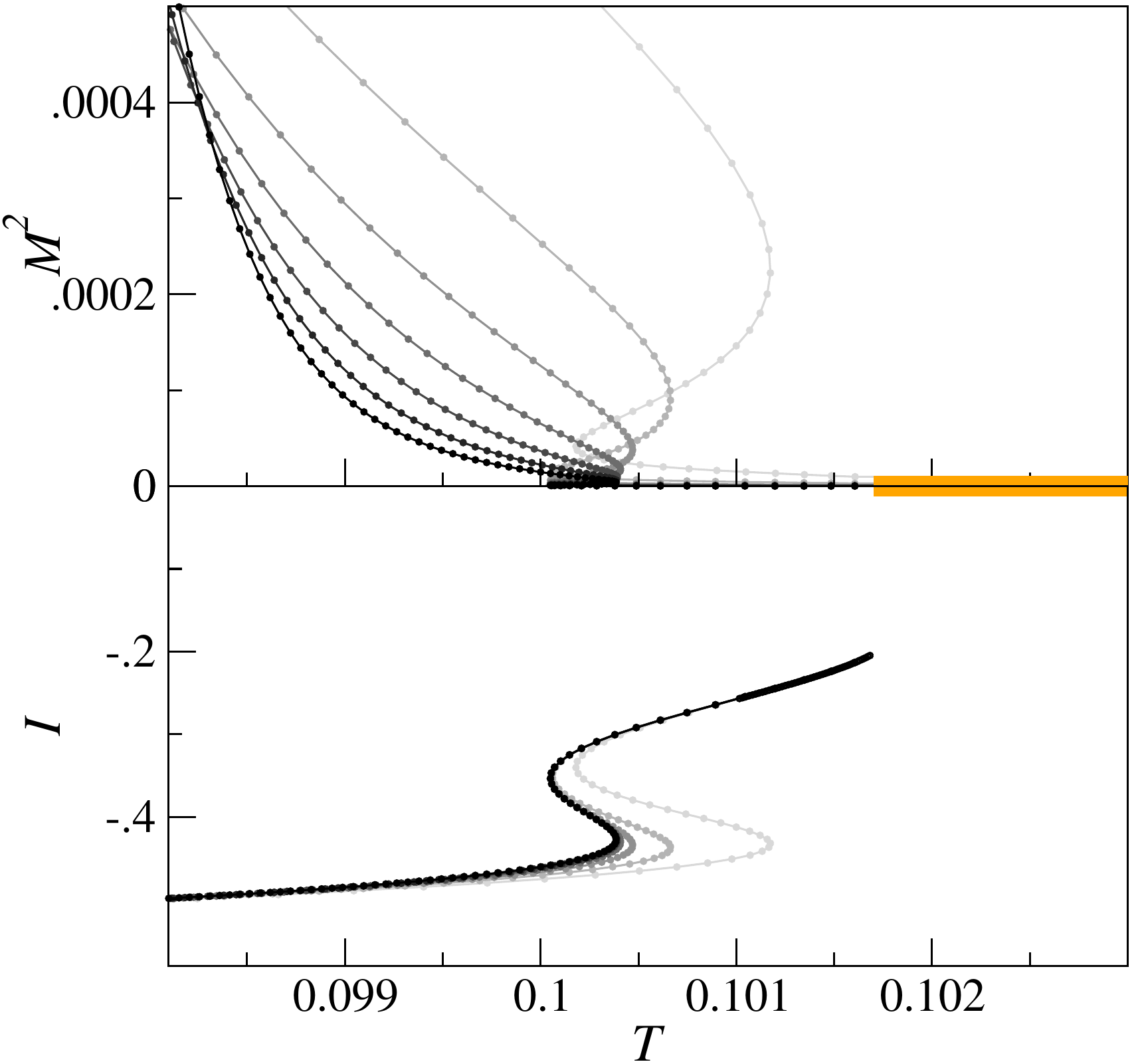}};
\node[] at (3.92,3.72) {(d)};
\end{scope}
\end{tikzpicture}

\begin{tikzpicture}
\begin{scope}[xshift=-0.007\textwidth]
\node[anchor=south west,inner sep=0] (image2) at (0\textwidth,0) {
\includegraphics[width=0.235 \textwidth]{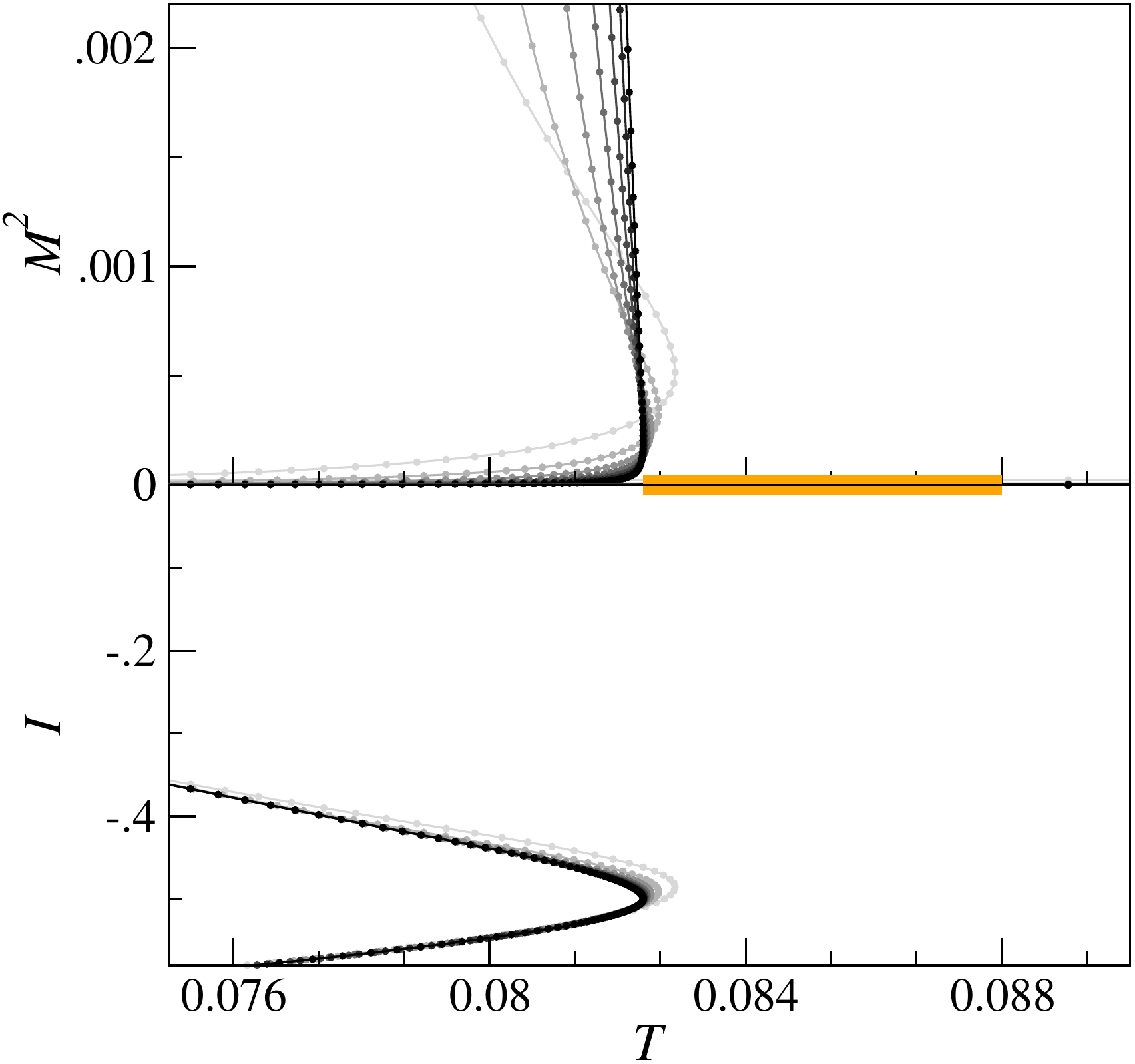}};
\node[] at (3.97,3.74) {(e)};
\end{scope}
\begin{scope}[xshift =0.235\textwidth]
\node[anchor=south west,inner sep=0] (image2) at (0\textwidth,0) {
\includegraphics[width=0.235 \textwidth]{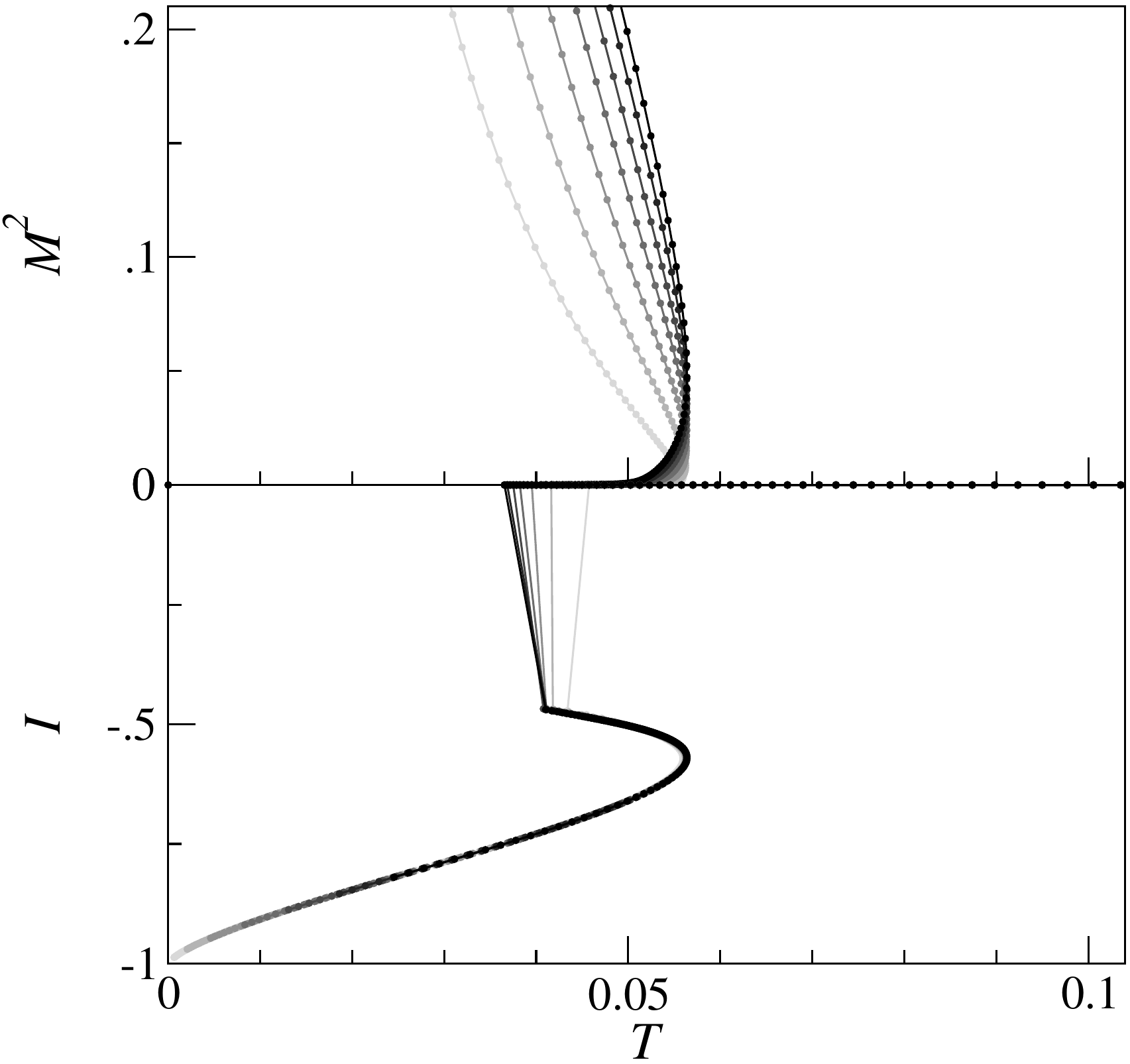}};
\node[] at (3.97,3.74) {(f)};
\end{scope}

\end{tikzpicture}

\caption{Order parameters for $J_z=-0.0001$.
(a) $J_2=0.8$. (b) $J_2=0.513$. (c) $J_2=0.506$. (d) $J_2=0.5054$. (e) $J_2=0.503$. (f) $J_2=0.5005$. The different curves are for different system sizes, the darkest being the largest system size. The nematic order parameter $I$ is shown as negative values while the stripe magnetic order parameter squared is shown as positive values. The orange thick lines indicate temperature intervals of very slow convergence in solving the nematic bond equations.
\label{cubic_m2x_JZ-0.0001_J20.8}
}
\end{center}
\end{figure}
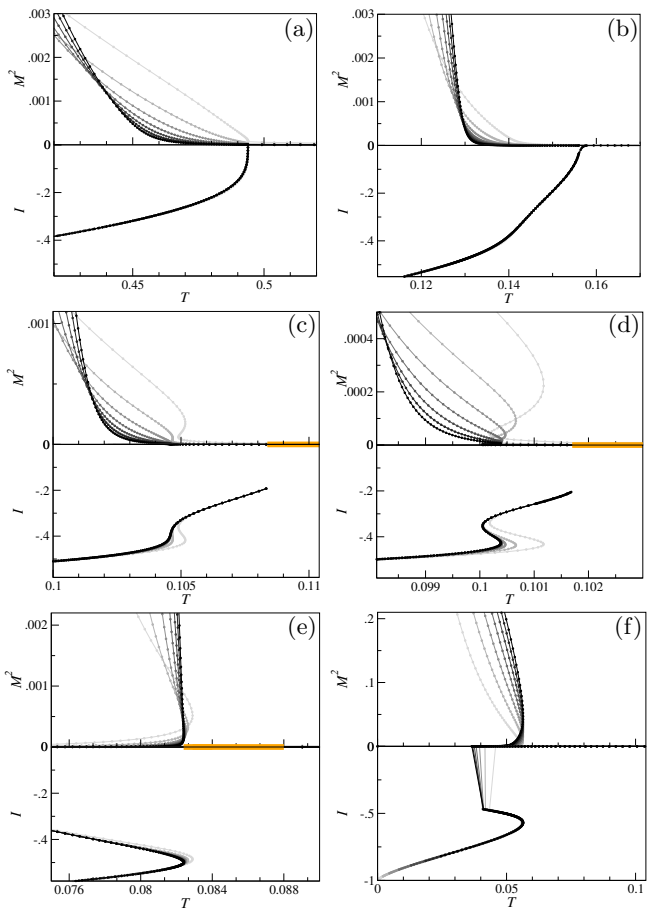

For weaker coupling $J_z$ between the layers the phase diagram is richer. In particular, the nematic and stripe phase transitions cease to occur simultaneously, and there is a finite temperature window where nematic long-range order exists without magnetic stripe order. This last fact can be seen from Fig.~\ref{cubic_m2x_JZ-0.0001_J20.8}(a) at $J_z=-0.0001$ and $J_2=0.8$. There the magnetization curves $M$ for different system sizes cross each other and the nematic order parameter $I$ becomes non-zero, with no overshoot, at a higher temperature than the magnetization crossings occur.
The $T_c$ of the stripe magnetic order is determined by extrapolating the crossings of successive finite-size magnetization curves using a cubic polynomial in $1/L$. These curves are single-valued so we conclude that these transitions are continuous. In what follows we indicate these split phase transitions by writing ({\Ic},{\Msc}), where the phase transition with the highest $T_c$ is written first. This intermediate temperature region with nematic but no stripe order exists also for smaller $J_2$ almost all the way to the FM phase as can be seen from Fig.~\ref{cubic_m2x_JZ-0.0001_J20.8}(b) for $J_2=0.513$.

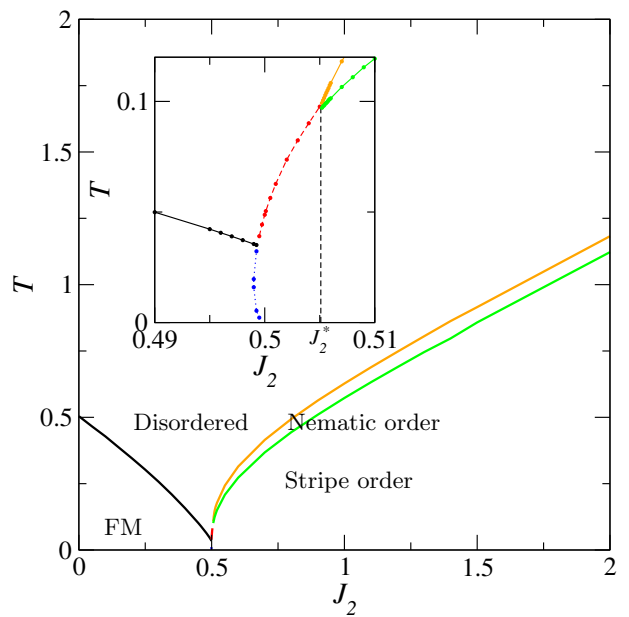
\begin{figure}[t]
\begin{center}
\begin{tikzpicture}
\begin{scope}[]
\node[anchor=south west,inner sep=0] (image2) at (0\textwidth,0) {
\includegraphics[width=0.45\textwidth]{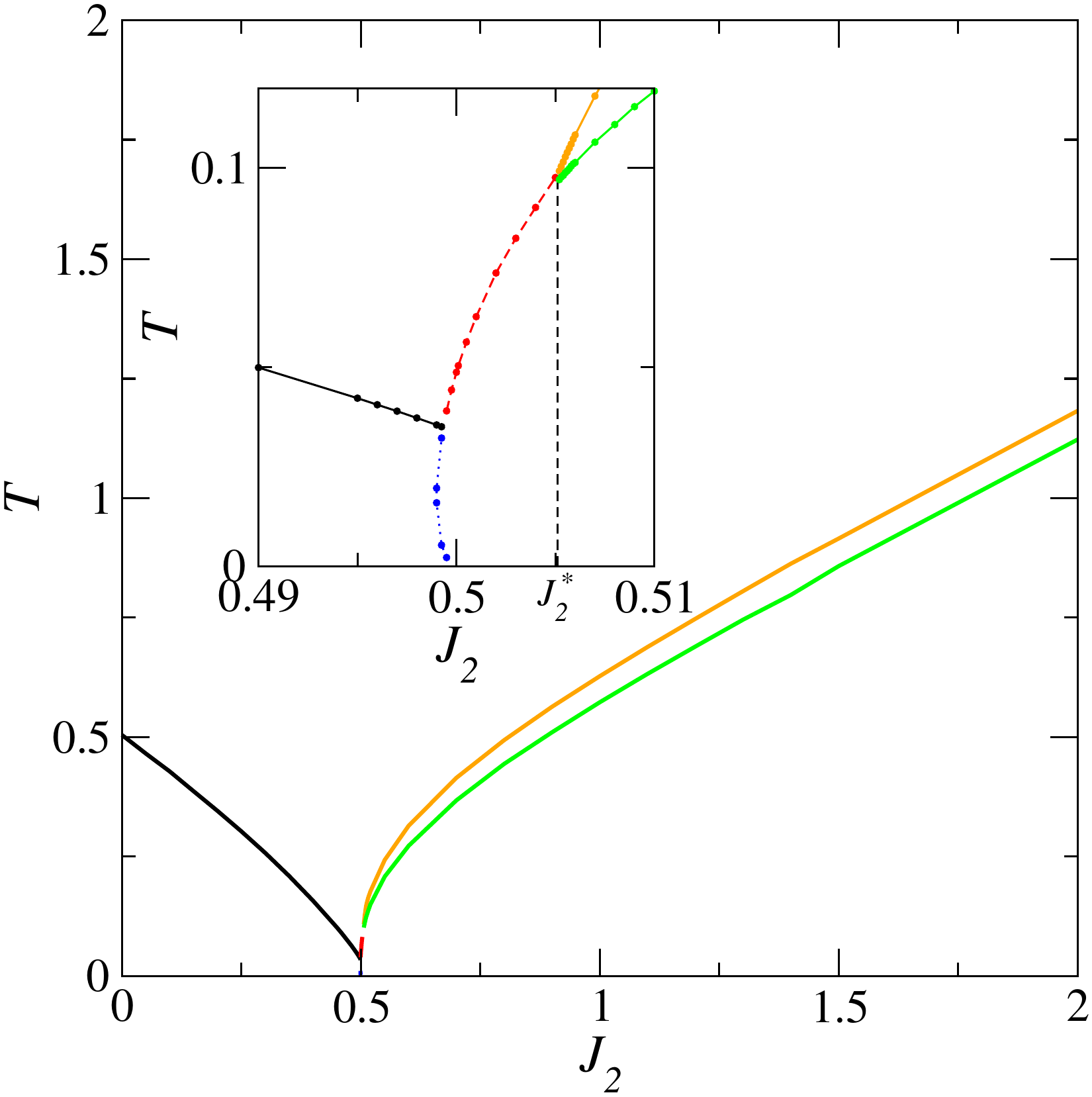}};
\node[] at (1.5,1.2) {FM};
\node[] at (4.7,2.6) {Nematic order};
\node[] at (4.5,1.8) {Stripe order};
\node[] at (2.4,2.6) {Disordered};
\end{scope}
\end{tikzpicture}
\caption{
Phase diagram for $J_z=-0.0001$.
The inset shows a blow-up of the region where all lines meet. The black line is the disorder-FM phase transition. The orange line is the disorder-nematic phase transition where the nematic phase has no long-range stripe magnetic order. The green line is where the magnetic order sets in. The dashed red line is the simultaneous {\IdMd} phase transition. The dotted blue line is the phase boundary between the ordered FM and the ordered stripe magnetic phases.
\label{phasediagram_JZ-0.0001}
}
\end{center}
\end{figure}

Very close to the FM phase, $J_2 \approx 0.506$, the nematic order parameter develops an additional kink feature at a finite value of $I$ which rapidly becomes an overshoot, Figs.~\ref{cubic_m2x_JZ-0.0001_J20.8}(c)-(d).
For these values of $J_2$ the $M$ curves also overshoot at the same $T$ as the kink feature in $I$. However, their magnitudes decrease with increasing system size. In each of the panels Fig.~\ref{cubic_m2x_JZ-0.0001_J20.8}(c)-(e) there is a temperature region close to $T_c$ of the nematic phase transition where the procedure of iterating the self-consistent equations converges very slowly. In fact there is a small temperature region (typically a little less than 10\% of $T_c$), indicated by the orange thick line, where we are unable to find solutions to the self-consistent equations for the largest system size.
For $J_2$ even smaller, the branch of $I$ which indicates the region of nematic order without magnetic order quickly moves down in temperature, and concomittantly the crossings of the magnetizations move up in temperature and end up as overshoots which again indicate an {\IdMsd} transition, see Figs.~\ref{cubic_m2x_JZ-0.0001_J20.8}(e)-(f). Thus, close to $J_2^* \approx 0.505$ there is a rapid change from a split regime with two continuous phase transitions to a regime with simultaneous discontinuous phase transitions. The corresponding phase diagram for $J_z=-0.0001$ is shown in
Fig.~\ref{phasediagram_JZ-0.0001}.

\begin{figure}[t]
\begin{center}
\begin{tikzpicture}
\begin{scope}[]
\node[anchor=south west,inner sep=0] (image1) at (0\textwidth,0) {\includegraphics[width=0.45\textwidth]{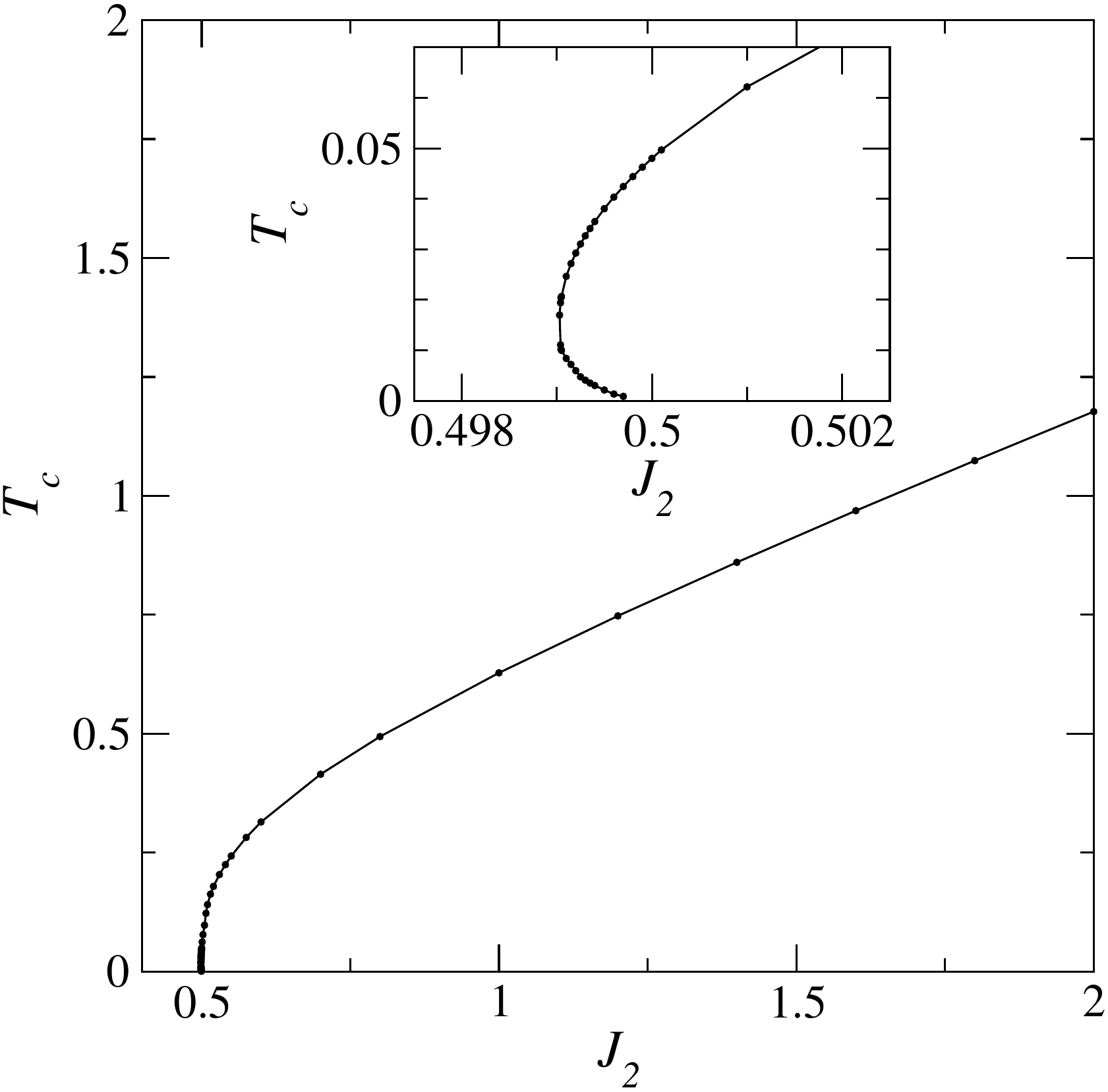}};
\end{scope}
\end{tikzpicture}
\caption{The nematic $T_c$ vs. $J_2$ for the two dimensional model ($|J_z|=0$). The inset shows a zoom in on the region  $J_2 \approx 0.5$. \label{TWOD_1A}}
\end{center}
\end{figure}

For even weaker interplane coupling $|J_z| < 0.0001$ the nematic phase boundary (orange curve in Fig.~\ref{phasediagram_JZ-0.0001})
stays almost unchanged as it becomes equal to the nematic phase transition boundary for a 2D frustrated Heisenberg magnet shown in Fig.~\ref{TWOD_1A}. Note that the nematic phase extends slightly into the region $J_2<0.5$ at finite temperatures also for the strictly 2D case. This is contrary to Ref.\onlinecite{Weber2003} where Monte Carlo data show evidence of an infinite slope at $J_2=0.5$.

\begin{figure}[t]
\begin{center}
\begin{tikzpicture}
\begin{scope}[]
\node[anchor=south west,inner sep=0] (image1) at (0\textwidth,0) {\includegraphics[width=0.45\textwidth]{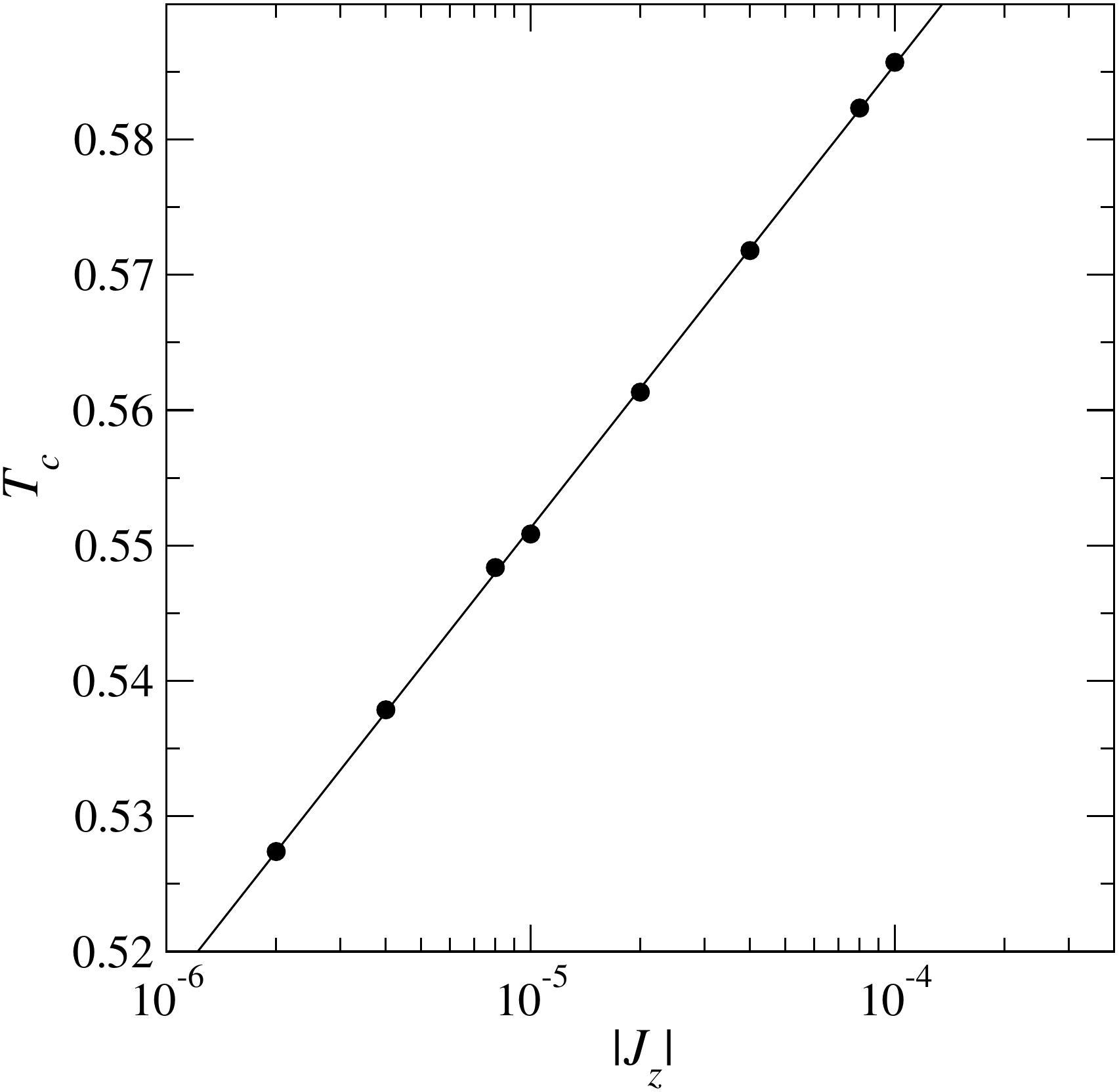}};
\end{scope}
\end{tikzpicture}
\caption{The magnetic $T_c$ vs. $|J_z|$ for $J_2=1$. The values of $T_c$s are here obtained by the Kouvel-Fisher analysis locating the temperature where the spin-spin correlation length diverges as described in conjunction with Fig.~\ref{KF} in Sec.~\ref{sec:correlations}. \label{weak}}
\end{center}
\end{figure}

The other phase boundaries move to lower temperatures, compliant with the expectation that a 2D system cannot sustain long-range magnetic order in accordance with the Mermin-Wagner theorem\cite{MerminWagner1966}. A plot of the magnetic $T_c$ for $J_2=1$ as a function of $|J_z|$ is shown in Fig.~\ref{weak}, and shows a logartihmic behavior $T_c \approx 0.722 + 0.0149 \ln{|J_z|}$ which we interpret to be the first terms in a series expansion of $T_c = a/(1+b\ln{(1/|J_z|)})$, with $a$ and $b$ independent of $J_z$, which is the renormalization group expectation for $N_s=3$ spatially anisotropic nonlinear $\sigma$-models with a small microscopic in-plane coupling\cite{AffleckHalperin1996}.

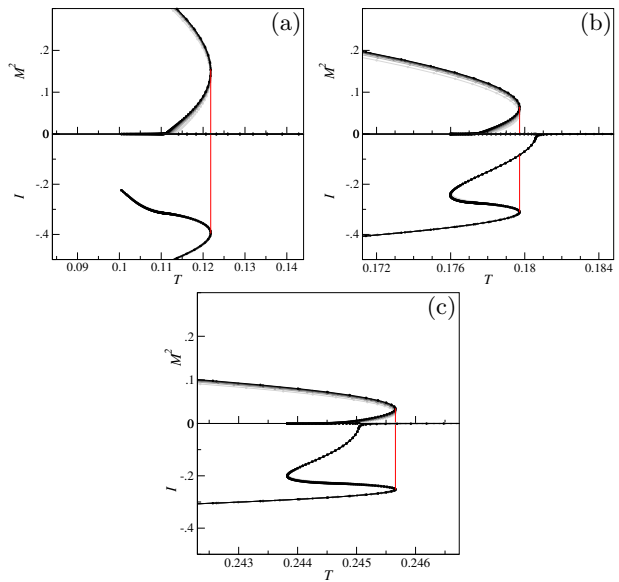
\begin{figure}[t]
\begin{center}
\begin{tikzpicture}
\begin{scope}[]
\node[anchor=south west,inner sep=0] (image1) at (0\textwidth,0) {
\includegraphics[width=0.22\textwidth]{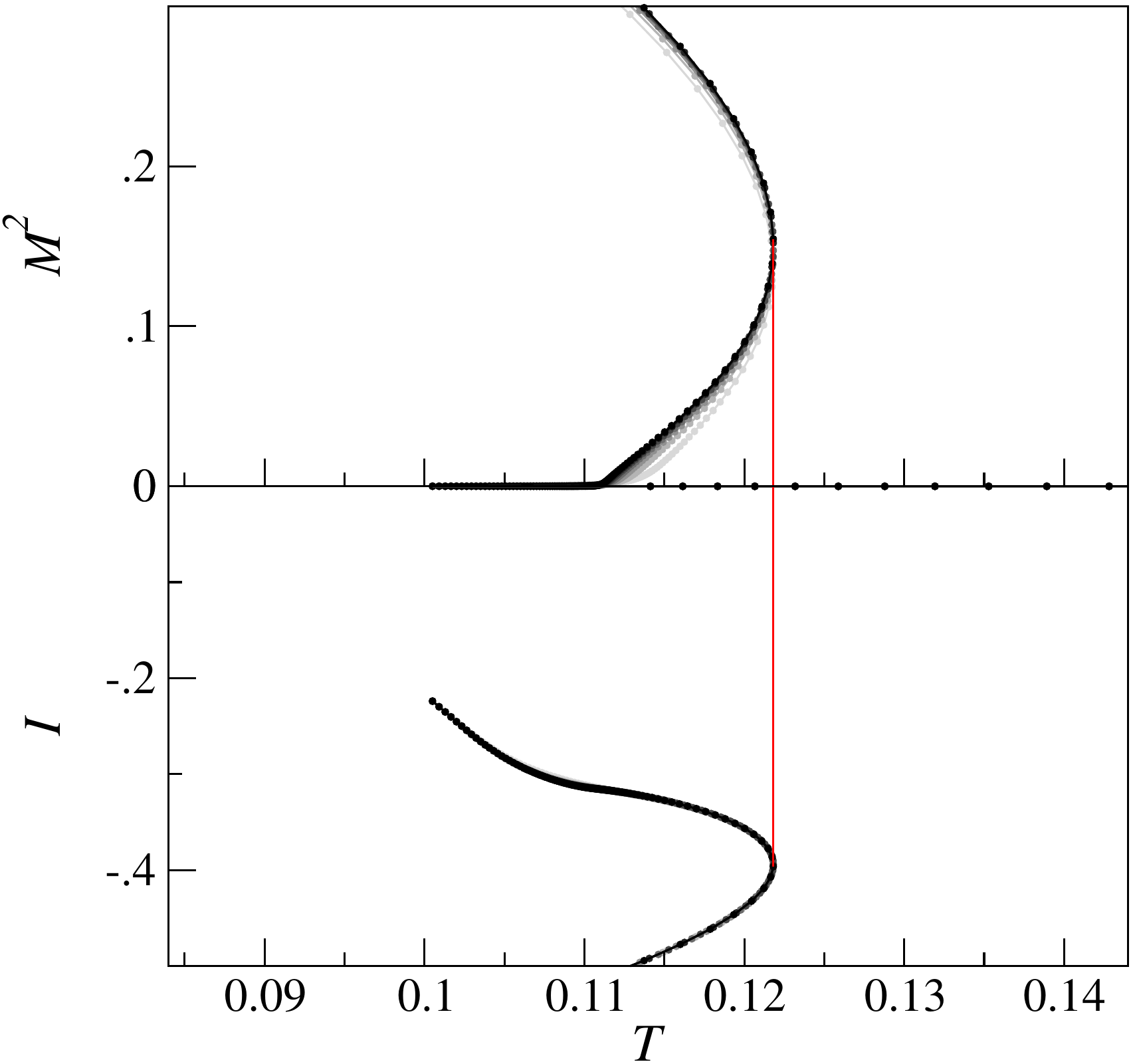}};
\node[] at (3.7,3.48) {(a)};
\end{scope}
\begin{scope}[xshift =0.23\textwidth]
\node[anchor=south west,inner sep=0] (image2) at (0\textwidth,0) {
\includegraphics[width=0.22\textwidth]{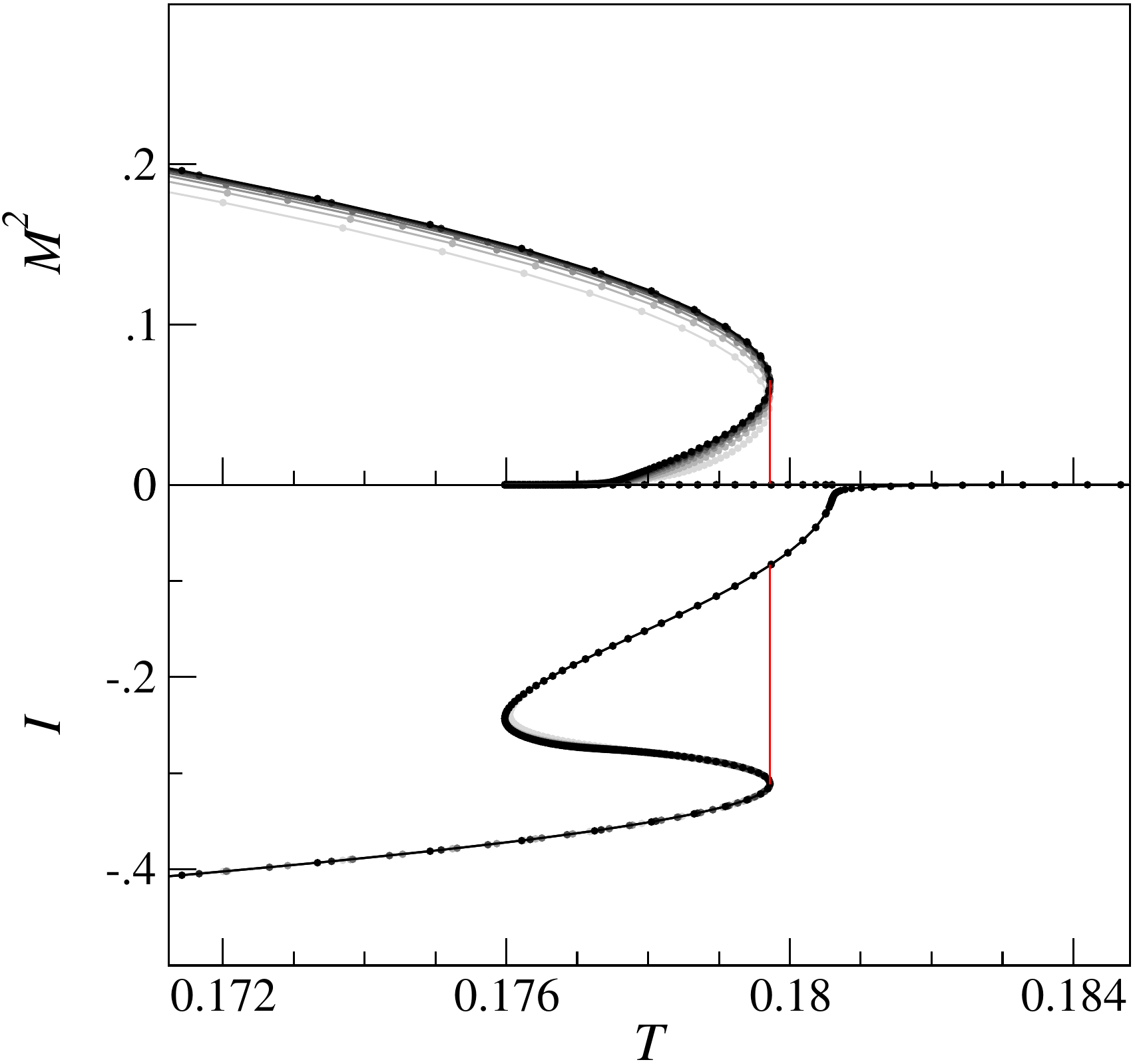}};
\node[] at (3.7,3.48) {(b)};
\end{scope}
\end{tikzpicture}
\begin{tikzpicture}
\begin{scope}[xshift =0\textwidth]
\node[anchor=south west,inner sep=0] (image3) at (0\textwidth,0) {
\includegraphics[width=0.22\textwidth]{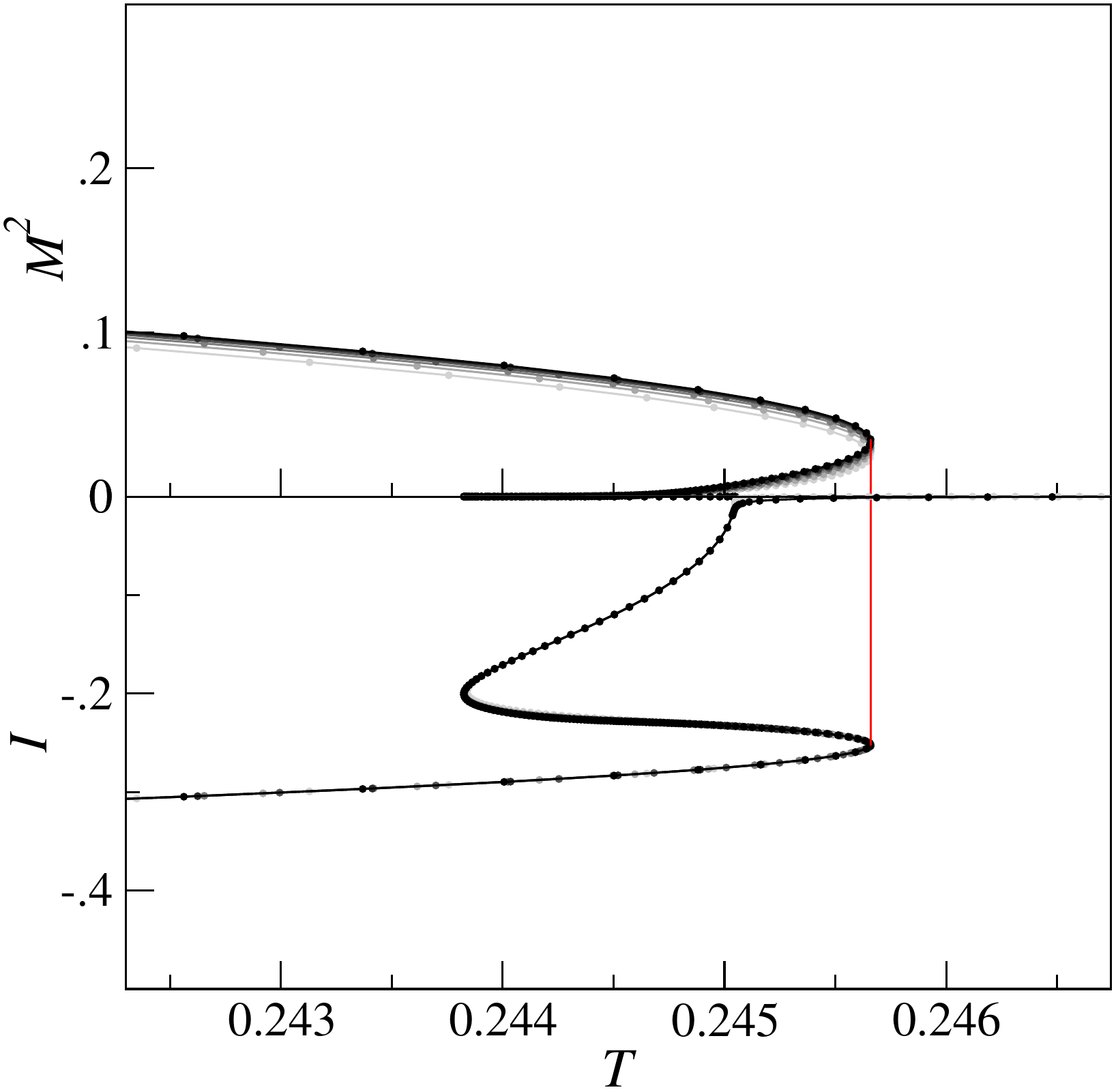}};
\node[] at (3.7,3.65) {(c)};
\end{scope}
\end{tikzpicture}
\caption{Order parameter curves for $J_z=-0.01$,(a) $J_2=0.505$, (b) $J_2=0.52$, and (c) $J_2=0.55$ for various system sizes.
\label{open} }
\end{center}
\end{figure}

It is interesting to explore for what values of $J_2$ the split regime of separate nematic and magnetic stripe phase transitions occurs as the interplane coupling is lowered from $|J_z|=1$. Already at $J_z=-0.01$ there is a hint of the split regime near $J_2=0.5$: Fig.~\ref{open}(a) shows simultaneous {\IdMsd} transitions at $J_2=0.505$, and at $J_2=0.52$ in Fig.~\ref{open}(b), the nematic order parameter becomes non-zero at a higher temperature than the {\IdMsd} occurs. Although there is a region of slow convergence in reaching the solution to the self-consistent equations above this nematic phase transition, it appears to be continuous as there is no visible overshoot. We thus have split phase transitions; an {\Ic} occuring at a higher $T_c$ than the discontinuous stripe magnetic phase transition, that occurs simultaneously with another discontinuous metanematic transition in the nematic order parameter between two finite values, a ({\Ic},{\IfdMsd}) where the letters fd signifies that the transition is a discontinuous jump between two {\em finite} values of the nematic order parameter. For $J_2=0.55$, Fig.~\ref{open}(c), the {\Ic} feature has dropped back below the temperature of the {\IfdMsd} transition, resulting in a single simultaneous {\IdMsd} transition. So for $J_z=-0.01$ the split transitions occur only in a narrow parameter region $J_2^* \approx 0.513 < J_2 < J_2^{**} \approx 0.534$. The differences in critical temperature in the split region for $J_z=-0.01$ are also small; for $J_2=0.52$ we find $\Delta T_c /T_c \sim 0.5\%$.

\begin{figure}[h]
\begin{center}
\begin{tikzpicture}
\begin{scope}[]
\node[anchor=south west,inner sep=0] (image1) at (0\textwidth,0) {\includegraphics[width=0.23 \textwidth]{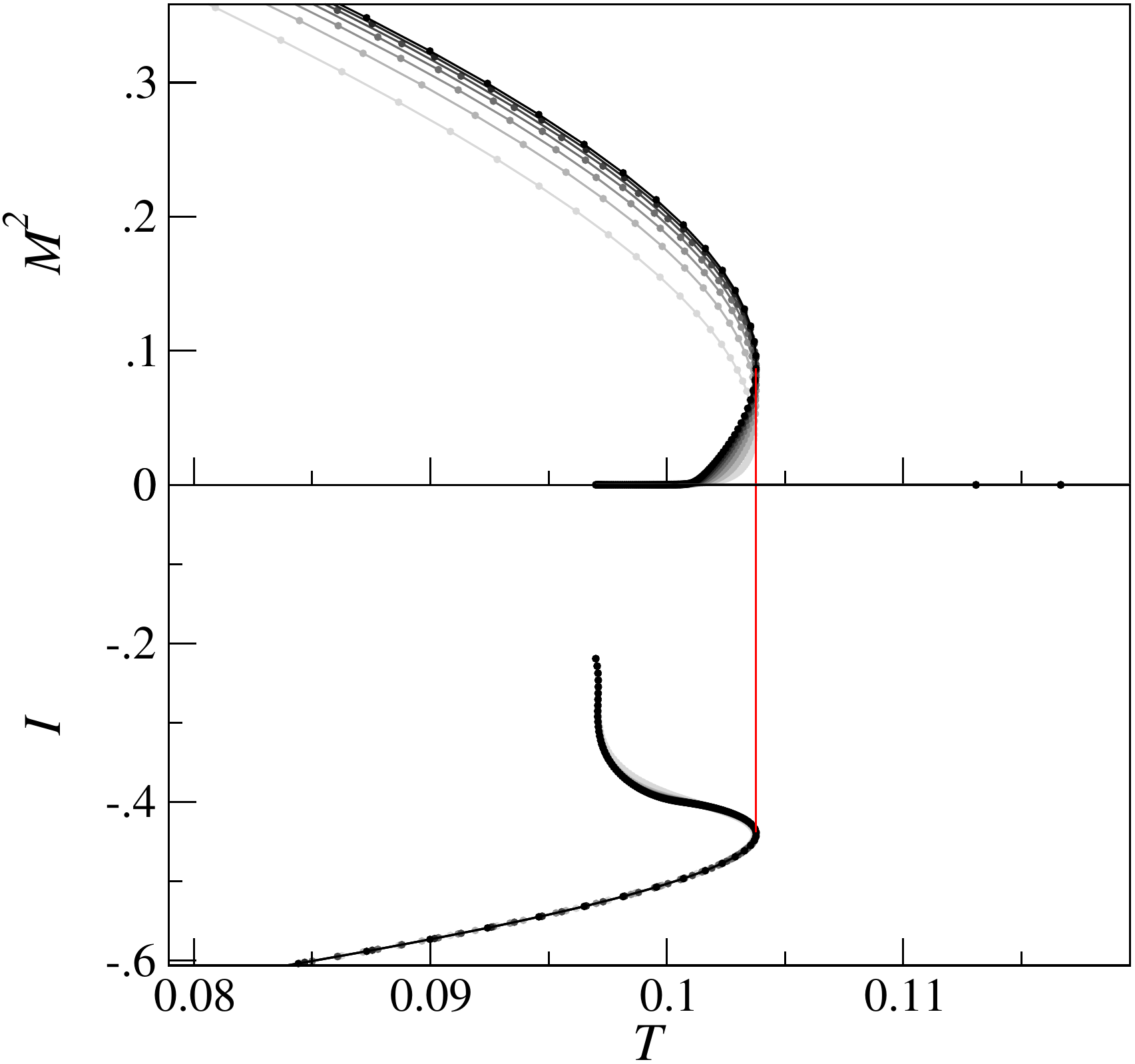}};
\node[] at (3.87,3.63) {(a)};
\end{scope}
\begin{scope}[xshift =0.24\textwidth]
\node[anchor=south west,inner sep=0] (image2) at (0\textwidth,0)
{\includegraphics[width=0.23\textwidth]{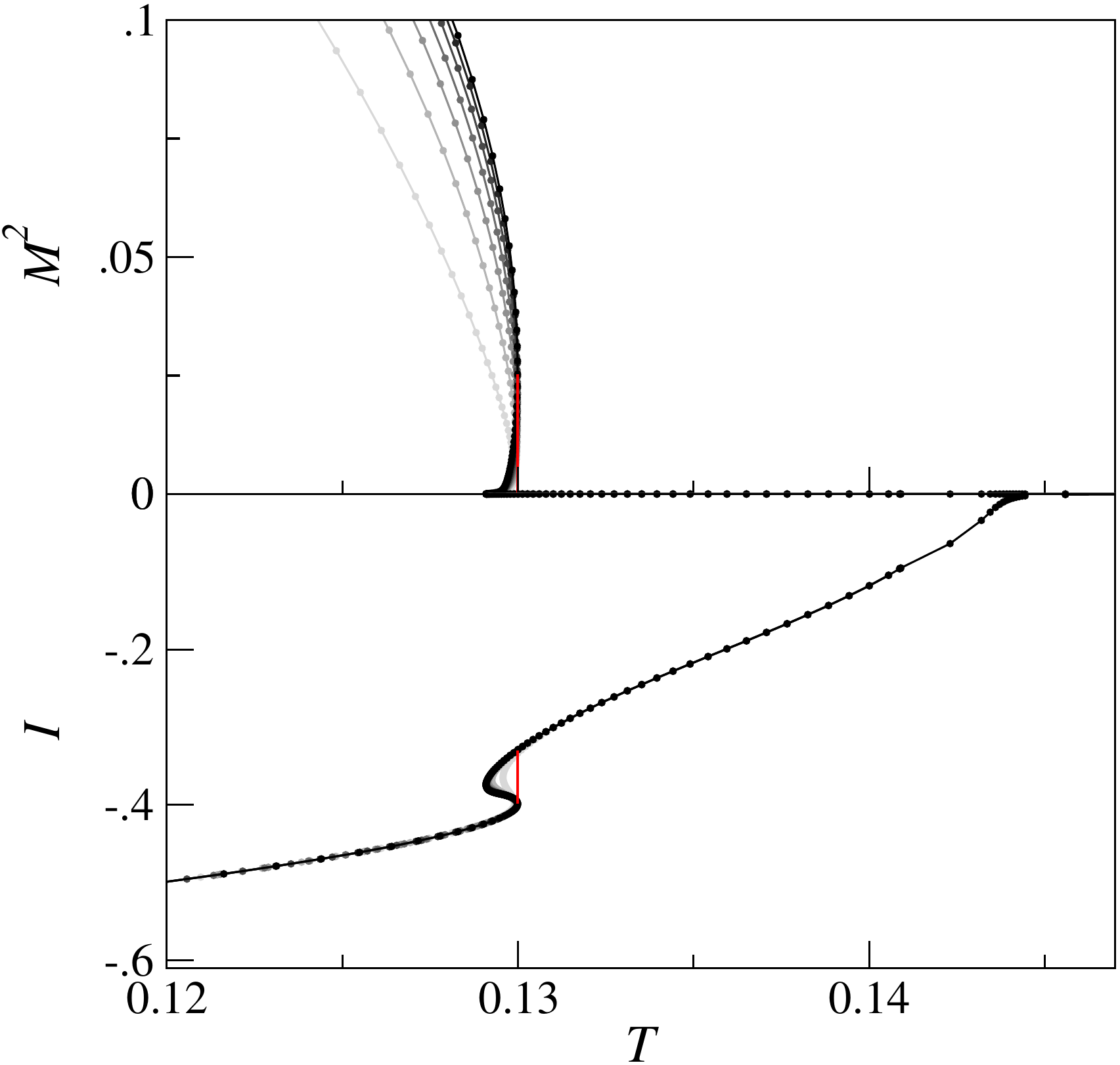}};
\node[] at (3.87,3.63) {(b)};
\end{scope}
\end{tikzpicture}

\begin{tikzpicture}
\begin{scope}[]
\node[anchor=south west,inner sep=0] (image1) at (0\textwidth,0) {\includegraphics[width=0.23 \textwidth]{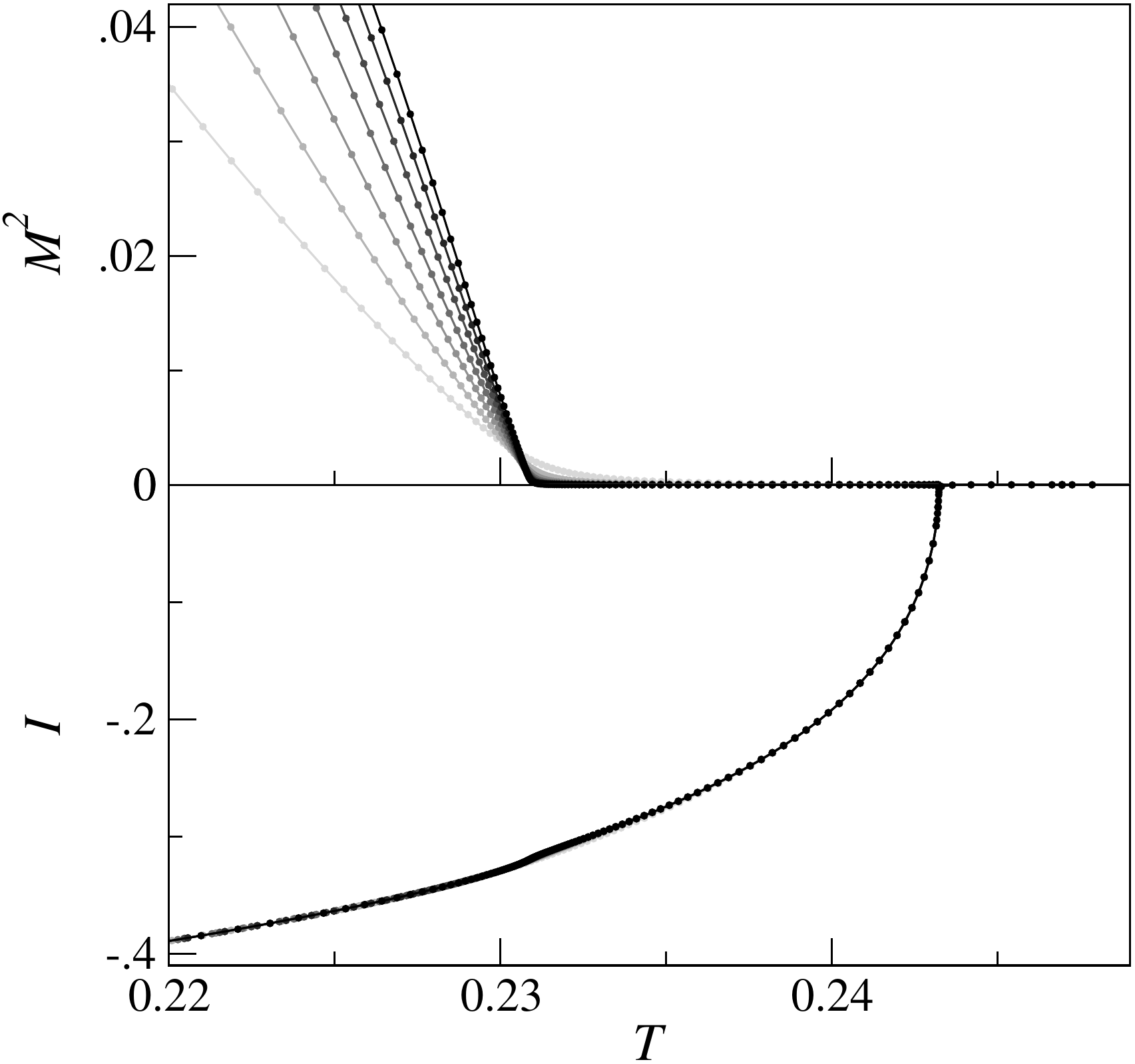}};
\node[] at (3.87,3.63) {(c)};
\end{scope}

\begin{scope}[xshift =0.24\textwidth]
\node[anchor=south west,inner sep=0] (image2) at (0\textwidth,0)
{\includegraphics[width=0.23\textwidth]{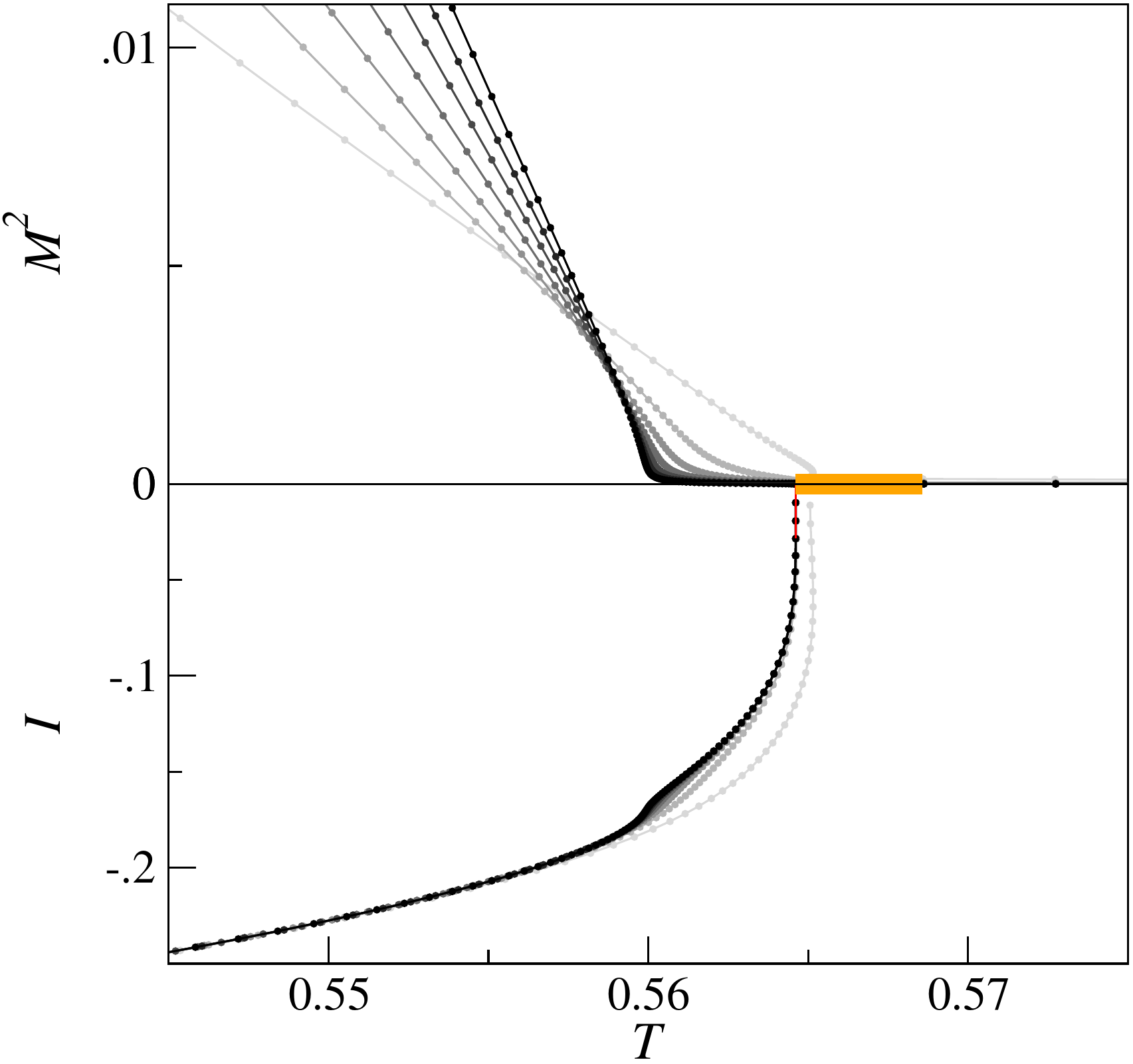}};
\node[] at (3.87,3.63) {(d)};
\end{scope}

\end{tikzpicture}

\begin{tikzpicture}
\begin{scope}[]
\node[anchor=south west,inner sep=0] (image1) at (0\textwidth,0){\includegraphics[width=0.23\textwidth]{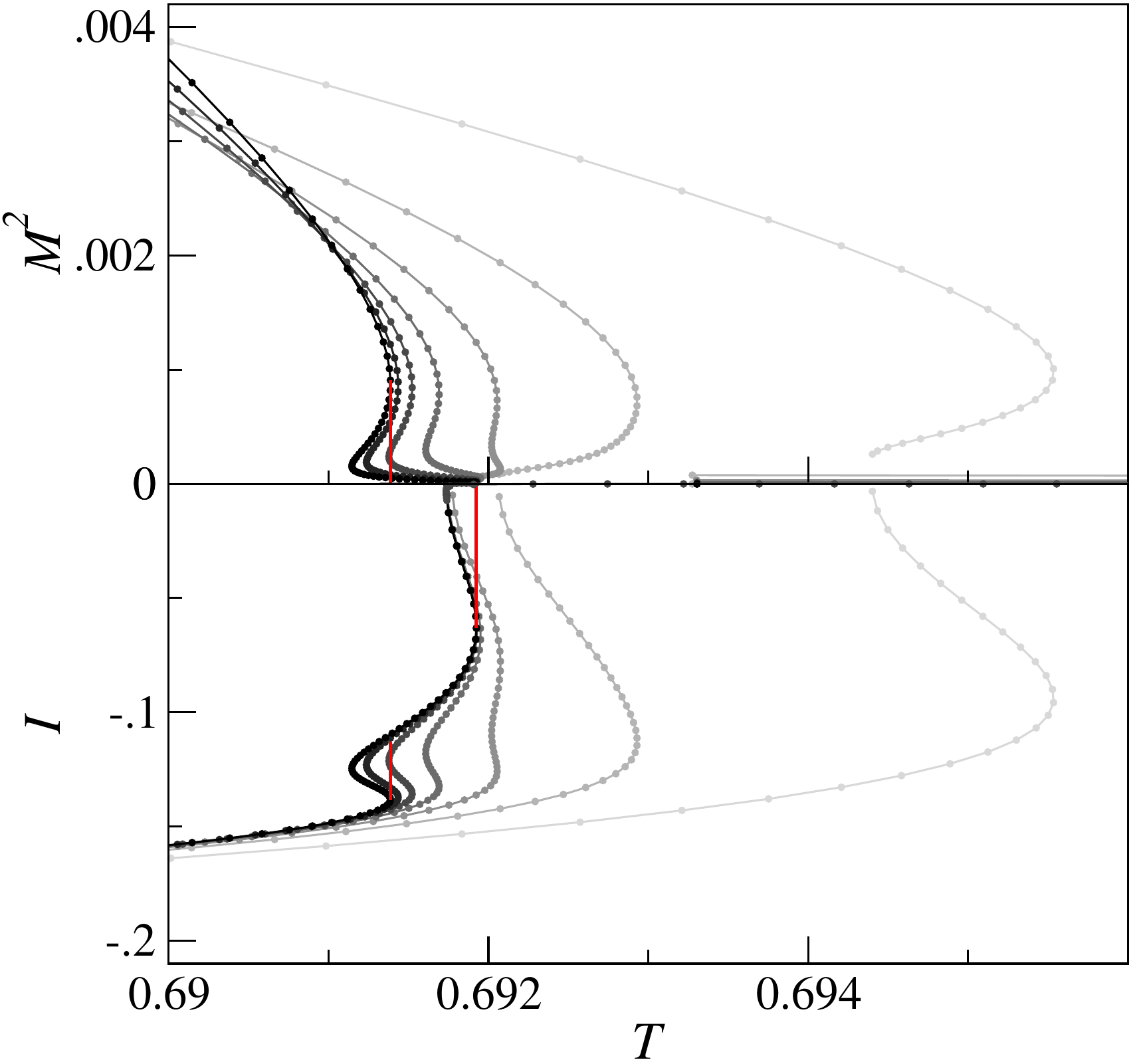}};
\node[] at (3.87,3.63) {(e)};
\end{scope}
\begin{scope}[xshift =0.24\textwidth]
\node[anchor=south west,inner sep=0] (image2) at (0\textwidth,0){\includegraphics[width=0.23\textwidth]{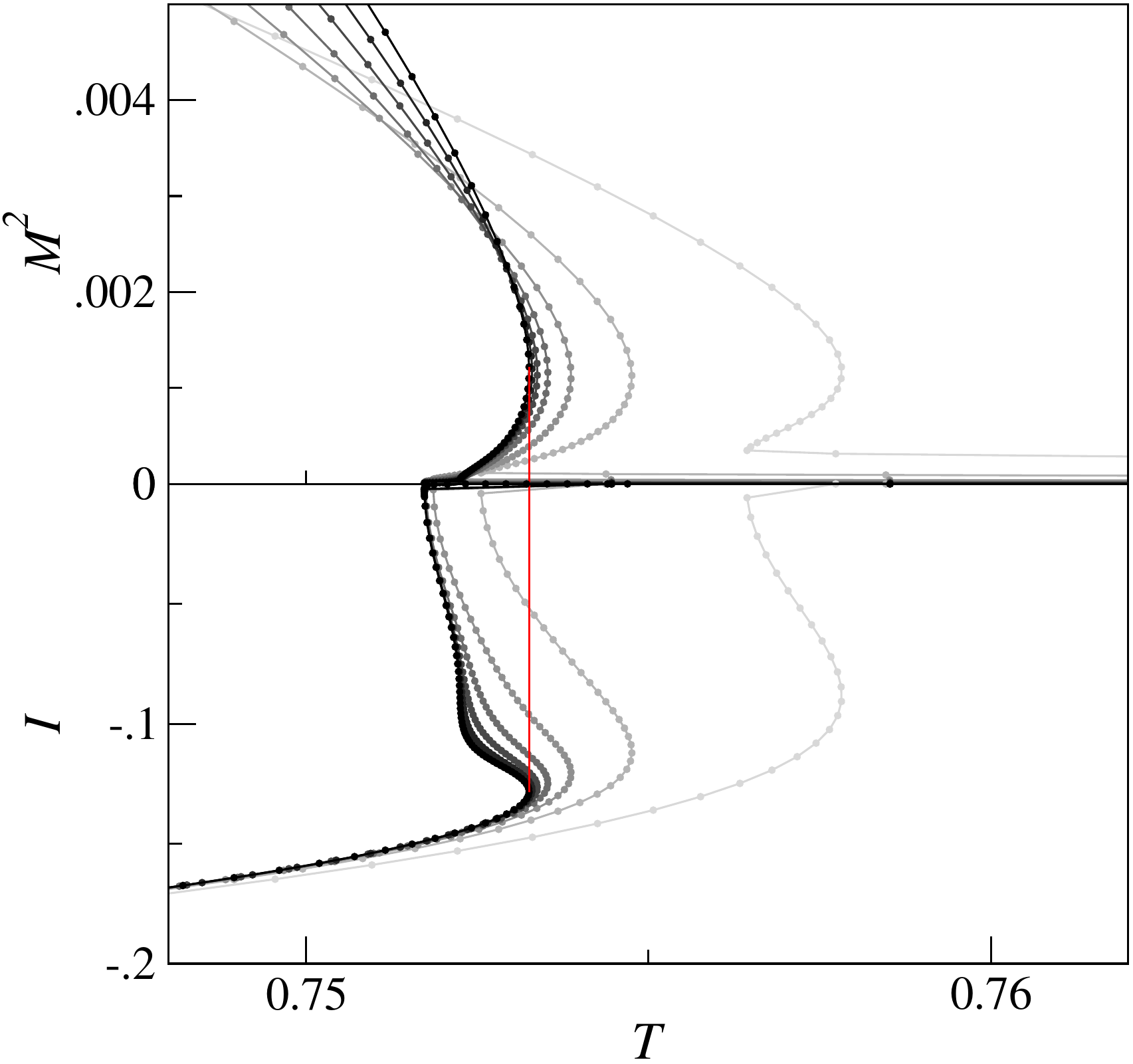}};
\node[] at (3.87,3.63) {(f)};
\end{scope}
\end{tikzpicture}

\caption{Order parameters for $J_z=-0.002$, (a) $J_2=0.505$, (b) $J_2=0.51$, (c) $J_z=0.55$, (d) $J_z=0.9$, (e) $J_z=1.1$, and (f) $J_2=1.2$ for various system sizes.\label{JZ-0.002}}
\end{center}
\end{figure}
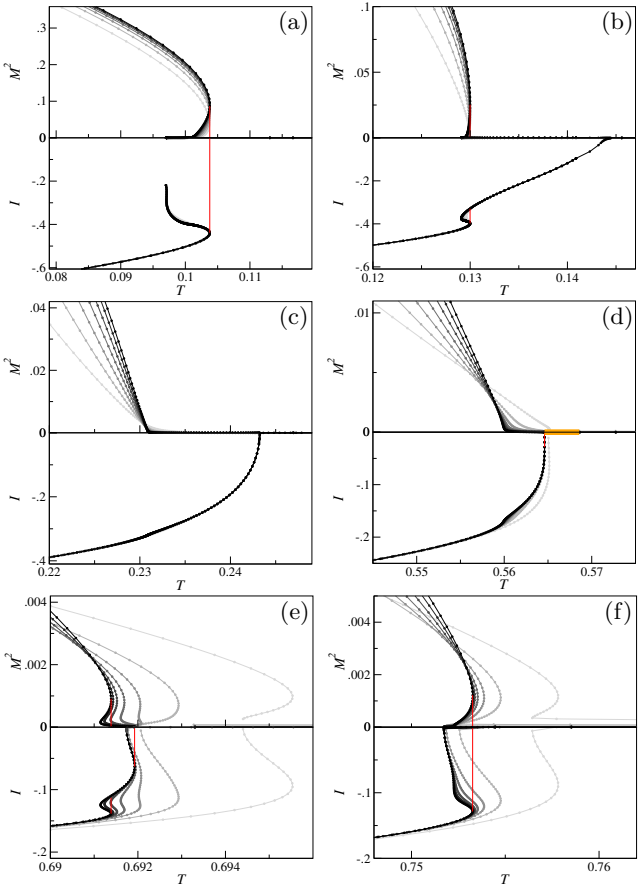

\begin{figure}[t]
\begin{center}
\begin{tikzpicture}
\begin{scope}[]
\node[anchor=south west,inner sep=0] (image1) at (0\textwidth,0) {\includegraphics[width=0.45\textwidth]{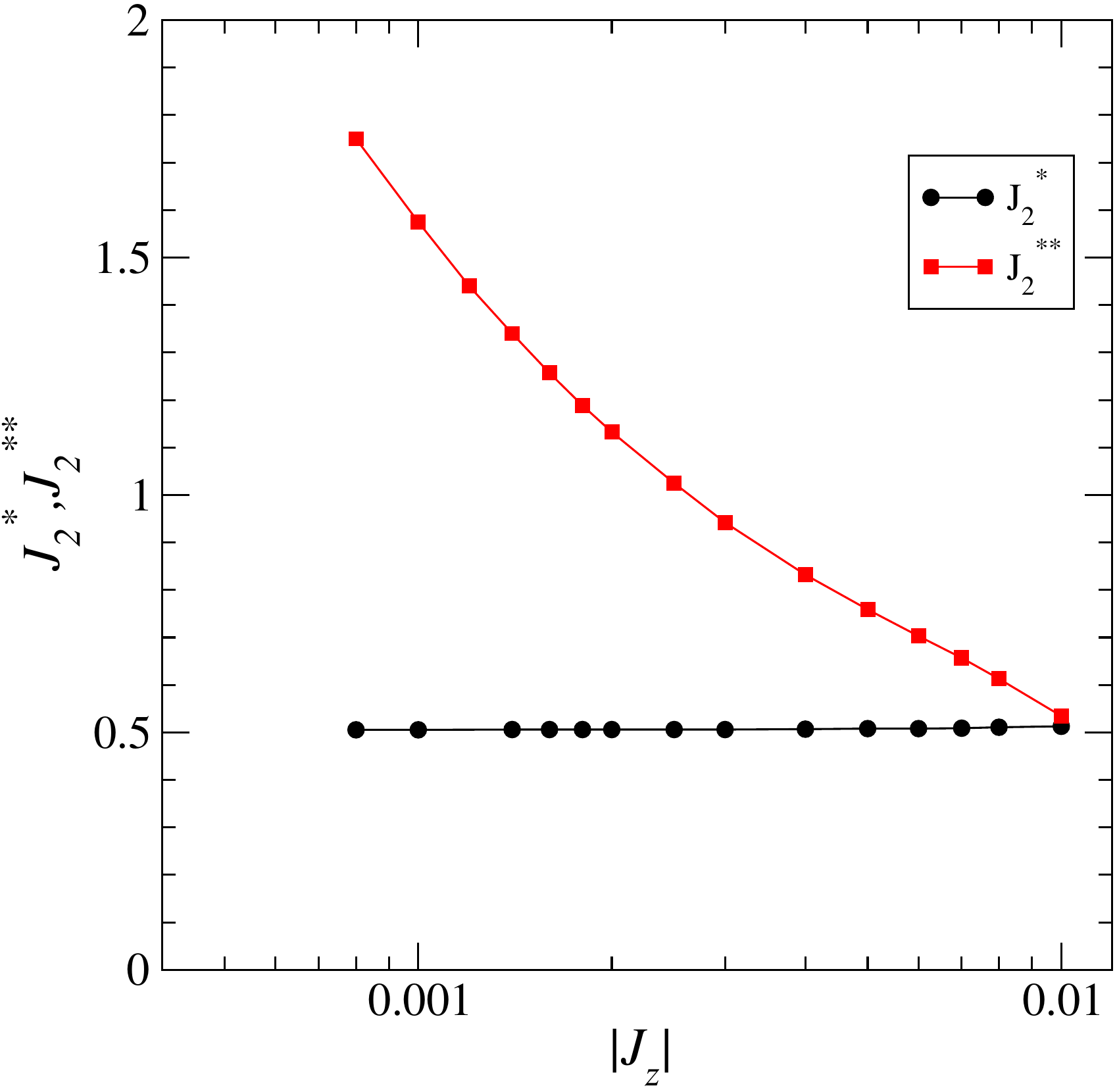}};
\end{scope}
\end{tikzpicture}
\caption{$J_2^*$ and $J_2^{**}$ as functions of interplane coupling $|J_z|$. \label{JSTAR}}
\end{center}
\end{figure}

A wider region of split behavior exists when $|J_z|$ is further lowered to $J_z=-0.002$. What is especially interesting is how the {\em nature} of the split phase transitions appears to change as $J_2$ is varied. For $J_z=-0.002$ this is illustrated in Fig.~\ref{JZ-0.002} and goes as follows: A simultaneous {\IdMsd} exists up to $J_2^* \approx 0.505$, Fig.~\ref{JZ-0.002}(a). For $J_2=0.51$, Fig.~\ref{JZ-0.002}(b), the nematic order parameter experiences a {\Ic} before there is a {\Msd} concomittant with a discontinuous jump in the nematic order parameter between two finite values, i.e. a ({\Ic},{\IfdMsd}). The discontinuous character of this lower temperature metanematic {\IfdMsd} transition weakens rapidly as $J_2$ is increased, see Fig.~\ref{JZ-0.002}(c). At $J_2=0.55$ the magnetic transition appears continuous ({\Mc}), and the jump in the nematic order parameter at the magnetic transition is changed into a weak kink-like feature signalling a regime of two distinct continuous phase transitions with separate $T_c$ values,({\Ic},{\Mc}), similar to that shown in Fig.~\ref{cubic_m2x_JZ-0.0001_J20.8}(a). The relative difference in the $T_c$ values is largest for the smallest values of $J_2$ after the splitting occurs and changes only slightly for an extended range of $J_2$. However, for $J_2$ large enough the $T_c$ values approach each other again and the nature of the phase transitions changes. At $J_2=0.9$, Fig.~\ref{JZ-0.002}(d), the nematic order parameter bends slightly back for the biggest system sizes, thus the highest temperature transition becomes {\Id}, while the lower temperature phase transition still appears to be continuous {\Msc}; ({\Id},{\Msc}). For even larger $J_2$ the two phase transitions come even closer in temperature, and now also the magnetic phase transition becomes discontinuous, so that at $J_2=1.1$, Fig.~\ref{JZ-0.002}(e) there are two slightly separated discontinuous phase transitions; a ({\Id},{\IfdMsd}). At even bigger values of $J_2$, Fig.~\ref{JZ-0.002}(f), the upper discontinuous transition {\Id} moves below the simultaneous transition,  resulting in a single simultaneous {\IdMsd} again. The boundary value, for $J_z=-0.002$,  where the split transitions merge again is $J_2^{**} = 1.13$. The value of $J_2^{**}$ increases rapidly as $|J_z|$ is further lowered, and exceeds the largest value considered here ($J_2=2$) for $J_z=-0.0001$. A plot of $J_2^*$ and $J_2^{**}$ vs. $|J_z|$ is shown in Fig.~\ref{JSTAR}.

\section{Spin correlations \label{sec:correlations}}

\begin{figure}[t]
\begin{center}
\begin{tikzpicture}
\begin{scope}[]
\node[anchor=south west,inner sep=0] (image1) at (0\textwidth,0){\includegraphics[width=0.45\textwidth]{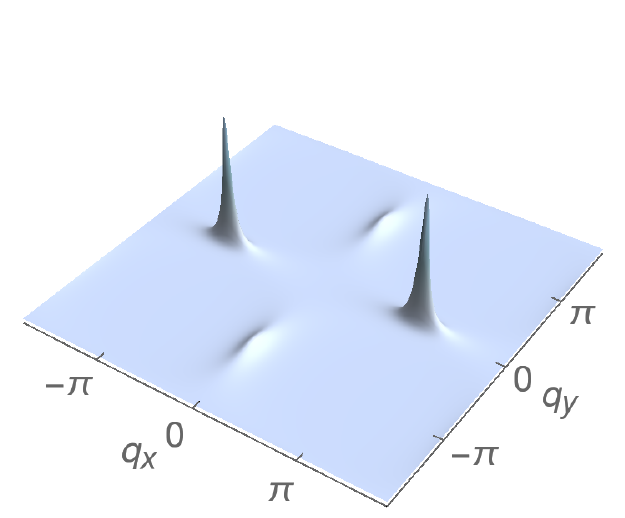}};
\end{scope}
\end{tikzpicture}
\caption{$\chi_{\qv}$ in the $(q_x,q_y)$--plane for $J_2=0.6$, $J_z=-0.0001$ at a temperature $T=0.31$ just below the nematic phase transition.
\label{contourplot}}
\end{center}
\end{figure}

For $J_2>0.5$, the Fourier transform of the spin-spin correlation function $\chi_{\rv} \equiv \langle \vec{S}_{\rv} \cdot \vec{S}_{0} \rangle$ is dominated by peaks around $(\pm \pi,0,0)$ and $(0,\pm \pi,0)$.
These peaks generally have different shapes along $q_x$, $q_y$ and $q_z$ directions, and in the nematic phase the peaks related by a $\pi/2$ lattice rotation about the $z-$axis also have different heights, see Fig.~\ref{contourplot}, while peaks related by a $\pi$ rotation about the $z-$axis have equal heights.

Approximating these peaks by Lorentzians, we can write the real-space spin-spin correlation function as
\begin{align}\label{eq:cor}
\chi_{\rv} & \propto \sum_{\qv}
\left( \f{A}{\xi_2^2 (q_x-\pi)^2+ \xi_1^2 q_y^2+ \xi_z^2 q_z^2+1}
\nonumber \right. \\
&+
\left.
\f{A^\prime}{\xi^{\prime 2}_1 q_x^2+ \xi^{\prime 2}_2 (q_y-\pi)^2+ \xi^{\prime 2}_z q_z^2+1} \right) e^{i \qv \cdot \rv}
\end{align}
where the lattice spacing has been set to unity. Unprimed(primed) symbols refer to the $(\pm \pi,0,0)$($(0,\pm \pi,0)$) peaks, and $\xi_i^{-1}$ is the half width half maximum of the peak in the direction along the real space stripes ($i=1$), in-plane perpendicular to the stripes ($i=2$) and perpendicular to the stripes in the z-direction ($i=z$).
Carrying out the summations one obtains
\begin{align}
\chi_{\rv} &\propto
\f{A\cos(\pi x)}{\xi_1 \xi_2 \xi_z} f\left(|\vec{r}^{\,\prime}|\right)
+
\f{A^\prime \cos(\pi y)}{\xi^\prime_1 \xi^\prime_2 \xi^\prime_z} f\left(|\vec{r}^{\,\prime\prime}|\right)
\end{align}
where $\vec{r}^{\,\prime}=\left(\f{x}{\xi_2},\f{y}{\xi_1},\f{z}{\xi_z}\right)$, $\vec{r}^{\,\prime\prime}=\left(\f{x}{\xi_2^\prime},\f{y}{\xi_1^\prime},\f{z}{\xi_z^\prime}\right)$ and $f(r)=\f{2 \pi^2}{r} e^{-r}$.

For temperatures above the nematic phase transition the peaks are related by a $\pi/2$ rotation about the $z-$axis, i.e. $\xi_1=\xi_1^\prime$, $\xi_2=\xi_2^\prime$, $\xi_z=\xi_z^\prime$ and $A=A^\prime$. This means that the in-plane correlation function along one of the axes, here the $x-$axis, is
\be
\label{eq:corr1}
\chi_{(x,0,0)} \propto
f\left(\f{x}{\xi_2}\right) \cos(\pi x) + f\left(\f{x}{\xi_1}\right).
\ee
The spin correlation function, Eq.~\eqref{eq:corr1}, is thus governed by two components: an oscillating component that decays with correlation length $\xi_2$ and a uniform component that decays with correlation length $\xi_1$. The difference between $\xi_1$ and $\xi_2$ measures the ellipticity of the peaks of the susceptibility in momentum space. For weak interlayer couplings the correlation length $\xi_z$ is much smaller than $\xi_1$ and $\xi_2$.

For temperatures below the nematic phase transition, one set of peaks will become higher and narrower than the other set. In the case of a negative nematic order parameter, the peaks at $(\pm \pi,0,0)$ will dominate, leading to $A \gg A^\prime$ and $\xi_1 \gg \xi_1^\prime$ etc. Therefore, in the nematic phase with negative nematic order parameter the spin-spin correlation function along the stripes ($y-$direction) is given at long distances by
\be
\chi_{(0,y,0)} \propto
f\left(\f{y}{\xi_1}\right).
\ee
Similarly the in-plane correlation function perpendicular to the stripes becomes
\be
\chi_{(x,0,0)} \propto
f\left(\f{x}{\xi_2}\right) \cos(\pi x).
\ee

\begin{figure}[t]
\begin{center}
\begin{tikzpicture}
\begin{scope}[]
\node[anchor=south west,inner sep=0] (image1) at (0\textwidth,0){\includegraphics[width=0.45\textwidth]{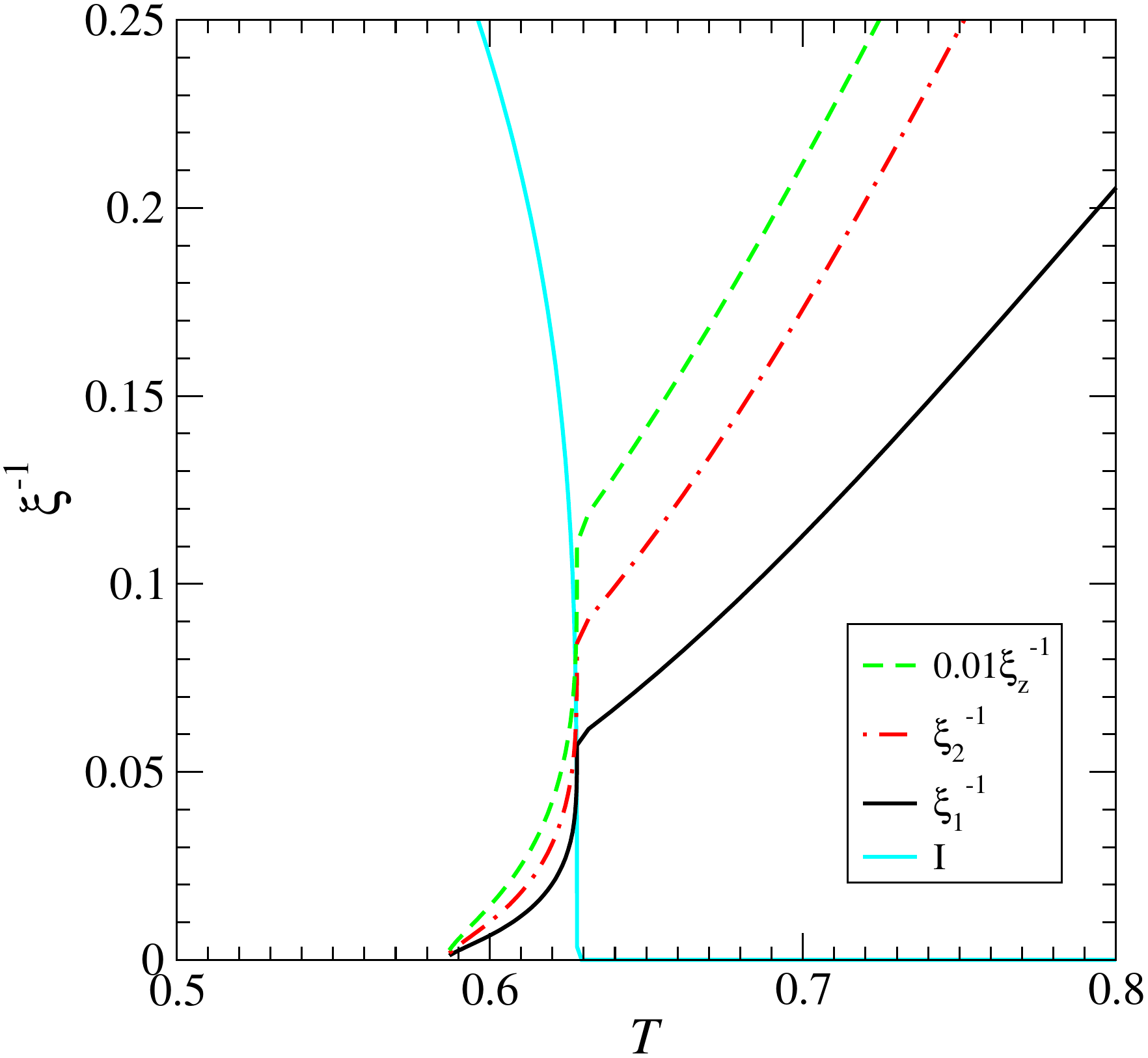}};
\end{scope}
\end{tikzpicture}
\caption{Inverse correlation lengths vs. $T$ for $J_z=-0.0001$ and $J_2=1$ for a lattice with $2400 \times 2400 \times 48$ sites.
Shown are
$0.01\xi_z^{-1}$ (green dashed),
$\xi_2^{-1}$ (red dot-dashed),
$\xi_1^{-1}$ (black solid), and the nematic order parameter I (light blue solid).
\label{inversecorrlengths}}
\end{center}
\end{figure}

For our finite lattice system with periodic boundary conditions we extract the width of the peak in the $i$-direction by fitting the values of $\chi_{\qv}$ along a line in direction $i$ in $\qv$-space through the point $\Qv$ using the functional form
\be
\f{A}{2 \left(1-\cos(q_i-Q_i) \right) \xi_i^2 + 1 }
\ee
where $A$ is independent of $\qv$ and the cosine takes care of the $q$-space periodicity. Here we have $\Qv = (\pi,0,0)$. The inverse correlation lengths in the three directions so obtained are shown as a function of $T$ for $J_2=1$ and $J_z=-0.0001$ in Fig.~\ref{inversecorrlengths}. We note that the correlation lengths are in general different, also above the nematic phase transition. Due to the very weak coupling $J_z$ between the layers, the correlation length in the $z$-direction is correspondingly small (note that we have plotted $0.01 \xi_z^{-1}$ in Fig.~\ref{inversecorrlengths}). Especially noteworthy is the fact that the correlation lengths increase much more rapidly with lowering temperature below the nematic phase transition than above. In fact this difference in behavior can be taken as an observational signature of the nematic phase transition alone.

\begin{figure}[t]
\begin{center}
\begin{tikzpicture}
\begin{scope}[]
\node[anchor=south west,inner sep=0] (image1) at (0\textwidth,0){\includegraphics[width=0.45\textwidth]{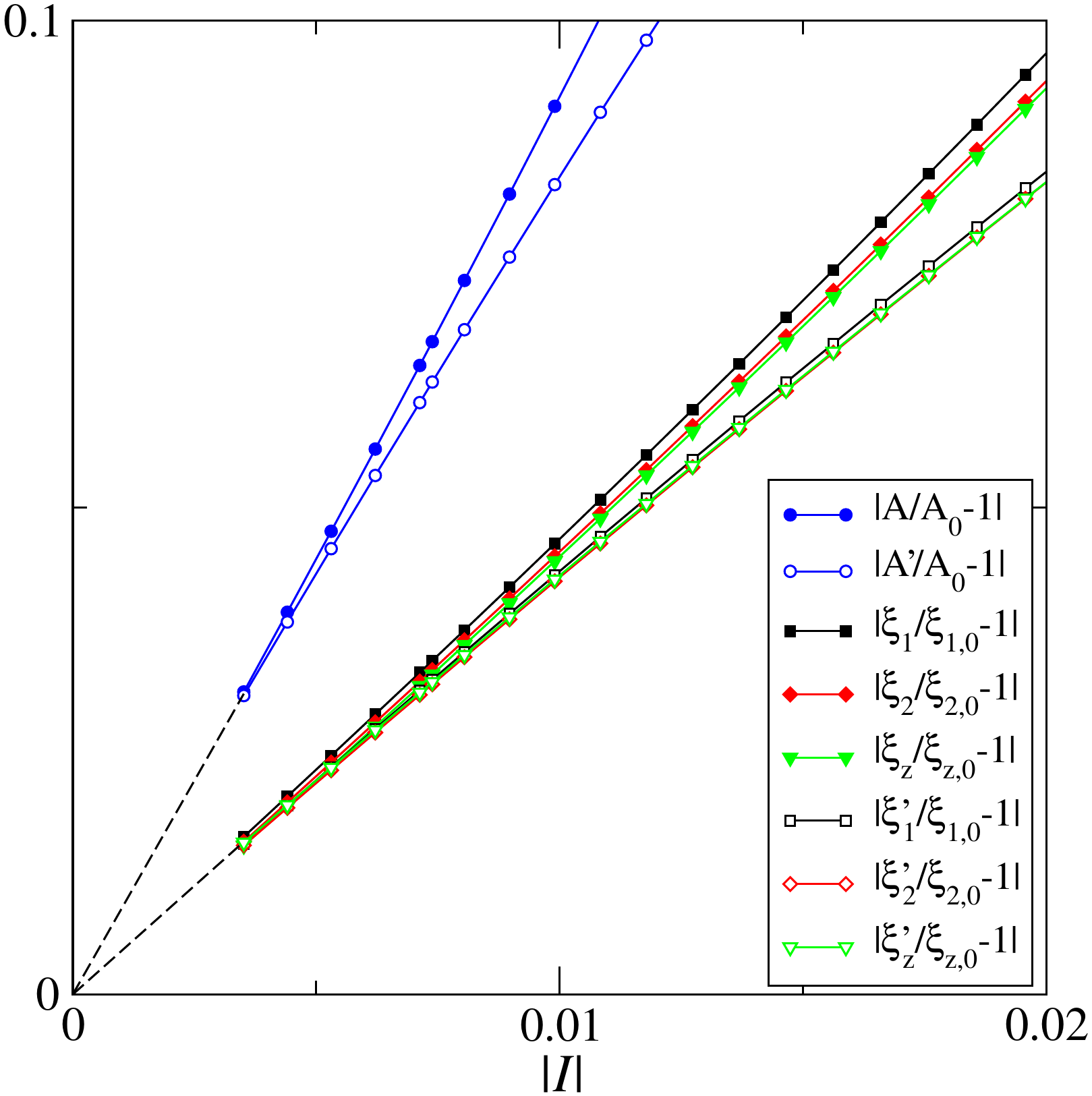}};
\end{scope}
\end{tikzpicture}
\caption{Scaled spin correlation lengths and susceptibility peak heights vs. $I$ close to the nematic phase transition for $J_z=-0.0001$ and $J_2=1$. The quantities specified in the legends are plotted as the y values. The dashed lines have slopes $4.46$ and $8.79$. For $J_z=-0.0001$ and $J_2=1$, Eq.~(\ref{eq:alpha}) gives $\alpha=8.55$.
\label{alphabeta2}}
\end{center}
\end{figure}

To derive the relation between the magnetic correlation length and the Ising-nematic order parameter near $T_c$, we expand the self-energy in Eqs.~\eqref{Keffinv}, \eqref{chi_eq} to linear order in the order parameter $I$. Next we expand near the peaks of $\chi_{\vec q}$ as in Eq.~\eqref{eq:cor}, and substitute this expression into the equation for the order parameter, Eq.~\eqref{eq:I}. Since $\chi_{\vec{q}}$ is peaked near $\vec{Q}$ and its symmetry-related points, we may perform a calculation similar to the one yielding Eq.~\eqref{eq:corr1} to write the order parameter as
\begin{align}
\label{eq:Icorr}
I &= \f{N_s T}{2V}
\left[
\sum_{\qv \sim (0,\pi,0)} \f{1}{J_{{\rm eff}\,\qv}+\Delta_{0} (1-\alpha I)}  \right.
\nonumber \\
& \left.  -\sum_{\qv \sim (\pi,0,0)} \f{1}{J_{{\rm eff}\,\qv}+\Delta_{0} (1+\alpha I)}
\right],
\end{align}
where the first[second] sum over $\qv$ is restricted to the region around the peak at $(0,\pi,0)[(\pi,0,0)]$, and $J_{{\rm eff}\,\vec{q}}=J_{\vec{q}}-\Sigma_{\vec{q}}^\prime$ may be expanded near its minima (as in Eq.~\eqref{eq:cor}), $\Sigma_{\vec{q}}^\prime$ is the point-group symmetric part of the self-energy and $\alpha$ is a parameter determined below. $\Delta_0$ is the value of $\Delta$ at $T_c$. Expanding Eq.~\eqref{eq:Icorr} to linear order in $I$ on its right side, we obtain the relation

\begin{equation}
\alpha=\left(\frac{N_s T \Delta_0}{2V}\sum_{\vec{q}}K_{{\rm eff}\,\vec{q}}^{-2}\right)^{-1}\Bigg\rvert_{T=T_c}. \label{eq:alpha}
\end{equation}
Upon inspection of Eqs.~\eqref{eq:cor}, \eqref{eq:Icorr} one sees that the susceptibility peak heights and widths may be expanded in powers of the order parameter $I$. Keeping only the linear term (valid near $T_c$ where $I\ll1$), one finds that
\begin{eqnarray}\label{eq:Icorr3}
A(A^\prime)&=&A_0(1\mp \alpha I),
\\
\xi_i(\xi_i^\prime)&=&\xi_{i,0}(1\mp \alpha I/2),\,\,\,\, (i=1,2,z)
\end{eqnarray}
where the 0 subscript indicates the value at $T=T_c$ (i.e. $I=0$). As a result, both the peak height ($A$) and width $(\xi^{-1})$ exhibit a nonanalytic temperature variation across the nematic phase transition, in an amount directly proportional to $I$. This scaling behavior for the correlation lengths and the amplitudes is shown in Fig.~\ref{alphabeta2}.

\begin{figure}[t]
\begin{center}
\begin{tikzpicture}
\begin{scope}[]
\node[anchor=south west,inner sep=0] (image1) at (0\textwidth,0) {\includegraphics[width=0.41\textwidth]{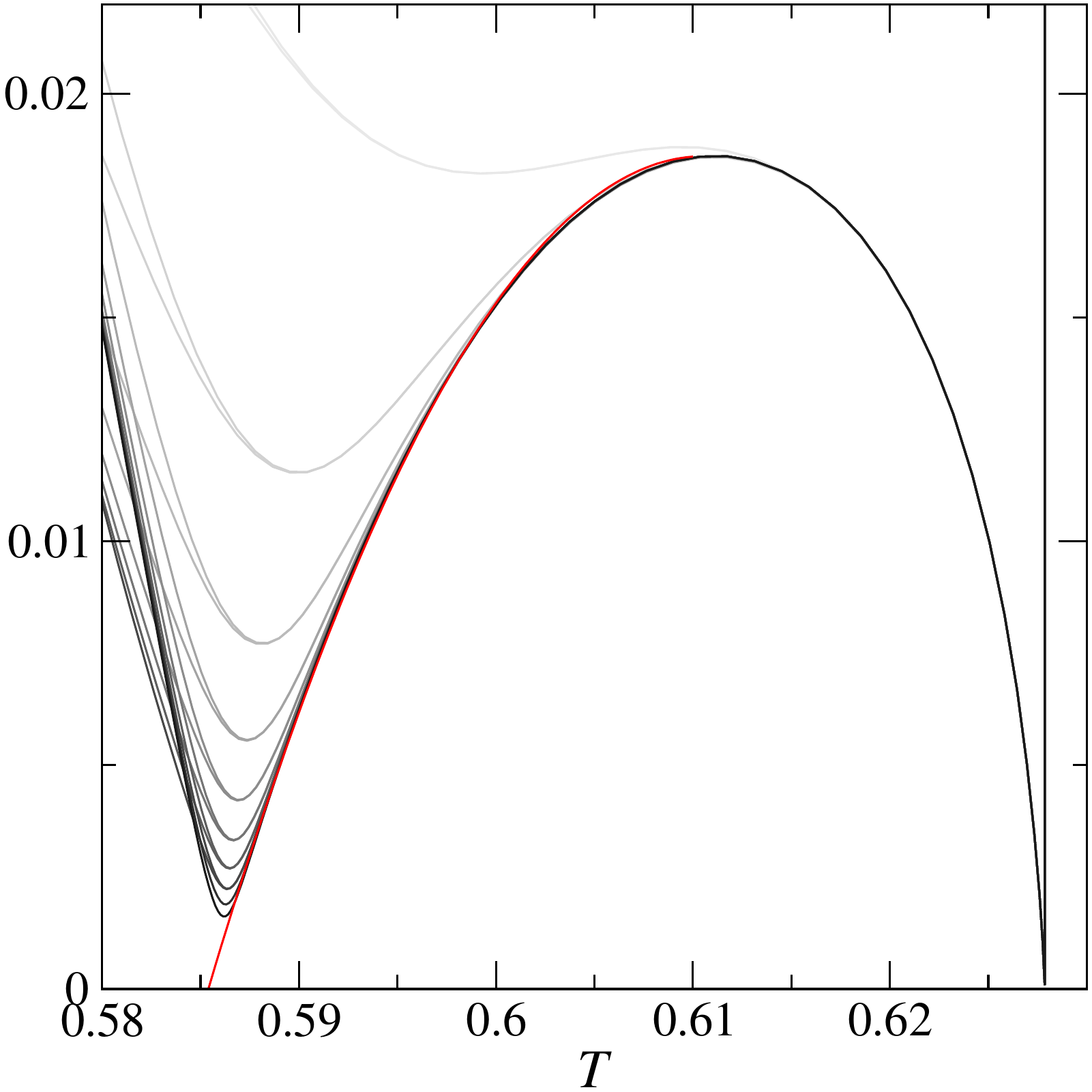}};
\node[rotate=90] at (-0.3,3.) {$-[\f{\partial \ln{\xi}}{\partial T}]^{-1}$};
\end{scope}
\end{tikzpicture}
\caption{Kouvel-Fisher plot showing $K(T) \equiv -[\f{\partial \ln{\xi_1}}{\partial T}]^{-1}$ vs. $T$.
$J_z = -0.0001$ and $J_2=1$. The different curves are for different system sizes and aspect ratios with grey scale coding such that the darkest curves correspond to the largest system sizes. Two families of system sizes are used $L \times L \times L/50$ with $L = 400-2400$, and $L \times L \times L/20$ with $L = 400-1800$.
The red curve is the best fit to a quadratic polynomial for the largest system size $2400 \times 2400 \times 48$ restricted to the temperature region $T \in [0.587,0.61]$. \label{KF}}
\end{center}
\end{figure}

To investigate the diverging behavior of the correlation length associated to the magnetic phase transition which occurs below the nematic phase transition we have made a Kouvel-Fisher analysis in Fig.~\ref{KF} showing $K(T) \equiv -[\f{\partial \ln{\xi_1}}{\partial T}]^{-1}$ vs. $T$ for the same parameters as used in Fig.~\ref{inversecorrlengths}. Taking into account also the leading irrelevant operator with a scaling dimension $y_1 <0$ we expect that close to $T_c$ the Kouvel-Fisher function behaves as $K(T)=\f{1}{\nu} \left( T-T_c \right) + B \left( T-T_c \right)^{y+1}$ where $y=-y_1 \nu >0$ and $B$ is a constant. In Fig.~\ref{KF} we have plotted $K(T)$ for different system sizes.
We see that finite-size effects are visible as low-temperature upturns for all system sizes, but that the infinite size behavior can be inferred by extrapolating the largest system size results for temperatures above the finite size upturn. We find that a value $y=1$ gives a good fit to the behavior of the largest system. Fixing $y=1$ we find a best fit, shown as the red curve in Fig.~\ref{KF}, of $K(T) = \f{1}{0.678} (T-T_c)-29.316 (T-T_c)^2$ with $T_c=0.5854$. The value of $\nu=0.678$ so obtained is reasonably close to the value $\nu=0.7112$ for the 3D Heisenberg universality class \cite{Campostrini2002}.

\section{Discussion \label{sec:discussion}}
Our results show that the phase diagram of the layered \Jonetwo-model is remarkably rich. In particular, in the frustrated regime, $J_2>0.5$, it has separate nematic and magnetic phase transitions as the temperature is lowered, if the interplane coupling $|J_z|$ is small enough. For weak interplane couplings, $|J_z| \lesssim 0.0001$, both of these phase transitions are continuous for most values of $J_2 (> 0.5)$. At $|J_z| =0.0001$ the relative difference in critical temperatures of these transitions is rather small, $\sim 7\%$, but increases as $|J_z|$ is further reduced. This is because the critical temperature of the nematic phase transition stays almost unchanged below $|J_z|=0.0001$, while the critical temperature of the magnetic phase transition goes to zero.

For values of $J_2$ very close to $0.5$, the nature of the phase transitions change, and eventually both become simultaneous first order phase transitions for $J_2 \to 0.5$. Exactly how this change happens is complicated as evidenced by Fig.~\ref{cubic_m2x_JZ-0.0001_J20.8}. In particular it involves a metanematic phase transition where the already finite nematic order parameter exhibits a jump to another value when the magnetic phase transition occurs, Fig.~\ref{cubic_m2x_JZ-0.0001_J20.8}(d).

In contrast, the regime of split transitions does not exist for strong interplane couplings $|J_z| >0.01$. There the phase transition into the magnetic phase from the disordered side is a discontinuous transition where the nematic and stripe magnetic order sets in simultaneously for all values of $J_2>0.5$.

For intermediate interplane couplings, $0.0001 < |J_z| < 0.01$, the $J_2$ regime of split phase transitions opens up from a value $J_2^* \approx 0.5$ which is very weakly dependent on $J_z$ and continues up to a value $J_2^{**}$ where it closes and reverts to simultaneous first order transitions. The value $J_2^{**}$ is very close to $J_2^*$ for $|J_z|=0.01$ and increases rapidly as $|J_z|$ is lowered.

Our results show also that the spin-spin correlation length, or more accurately its temperature derivative, can serve as a probe to detect the nematic phase transition. This is so because the correlation length depends directly on the nematic order parameter, thus it exhibits a sharp increase with decreasing temperature exactly at the nematic phase transition. For split transitions this is separate from the normal critical divergence of the correlation length that happens at a lower temperature where the stripe magnetic phase transition occurs.

Our results are strictly only valid for the classical layered {\Jonetwo} model, so we can only speculate on how our results carry over to the corresponding quantum model. What seems plausible on general grounds is that quantum effects will enhance fluctuations and be most dramatic in the regions of the phase diagram where there are adjacent phases and the critical temperatures are the lowest, especially the region close to $J_2=0.5$. Evidences of a gapless spin liquid phase in the two dimensional quantum spin-1/2 \Jonetwo-model near $J_2=0.5$ was recently reported in Ref.~\onlinecite{WangSandvik2018} but this region is notoriusly difficult to address. For higher spins with less quantum fluctuations we do expect that the high temperature features discussed here, such as the splitting of the nematic and the stripe magnetic phase transitions, carry over to the quantum case.

When it comes to the applicability of our results to the iron-based superconductors, it will depend on how well iron-based superconductors are based on models of localized magnetic moments. We do not address this question. Nevertheless, our results show that virtually all of the rich phenomena predicted to occur in effective classical models of itinerant magnetic moments\cite{Fernandes2012} appear also in the classical \Jonetwo--model for weak interplane couplings. Split continuous phase transitions is a generic feature for weak interplane couplings.
More exotic phase transitions, such as split metanematic transitions, were found only in narrow regions of the phase diagram.

In addition to obtaining the results for the \Jonetwo--model we have in this paper also taken the opportunity to outline the details of the nematic bond theory introduced in Ref.~\onlinecite{Schecter2017}.
As demonstrated here, this is an efficient and versatile method for dealing with frustrated magnetism, even for very large system sizes, and can serve as a supplement to computationally demanding Monte Carlo techniques.
\\
\acknowledgments{
The Center for Quantum Devices is funded by the Danish National Research Foundation. We acknowledge support from the Laboratory for Physical Sciences, and Microsoft (M.S.). The computations were performed on resources provided by UNINETT Sigma2 - the National Infrastructure for High Performance Computing and Data Storage in Norway.}

\bibliography{lay}

\begin{thebibliography}{22}%
\makeatletter
\providecommand \@ifxundefined [1]{%
 \@ifx{#1\undefined}
}%
\providecommand \@ifnum [1]{%
 \ifnum #1\expandafter \@firstoftwo
 \else \expandafter \@secondoftwo
 \fi
}%
\providecommand \@ifx [1]{%
 \ifx #1\expandafter \@firstoftwo
 \else \expandafter \@secondoftwo
 \fi
}%
\providecommand \natexlab [1]{#1}%
\providecommand \enquote  [1]{``#1''}%
\providecommand \bibnamefont  [1]{#1}%
\providecommand \bibfnamefont [1]{#1}%
\providecommand \citenamefont [1]{#1}%
\providecommand \href@noop [0]{\@secondoftwo}%
\providecommand \href [0]{\begingroup \@sanitize@url \@href}%
\providecommand \@href[1]{\@@startlink{#1}\@@href}%
\providecommand \@@href[1]{\endgroup#1\@@endlink}%
\providecommand \@sanitize@url [0]{\catcode `\\12\catcode `\$12\catcode
  `\&12\catcode `\#12\catcode `\^12\catcode `\_12\catcode `\%12\relax}%
\providecommand \@@startlink[1]{}%
\providecommand \@@endlink[0]{}%
\providecommand \url  [0]{\begingroup\@sanitize@url \@url }%
\providecommand \@url [1]{\endgroup\@href {#1}{\urlprefix }}%
\providecommand \urlprefix  [0]{URL }%
\providecommand \Eprint [0]{\href }%
\providecommand \doibase [0]{http://dx.doi.org/}%
\providecommand \selectlanguage [0]{\@gobble}%
\providecommand \bibinfo  [0]{\@secondoftwo}%
\providecommand \bibfield  [0]{\@secondoftwo}%
\providecommand \translation [1]{[#1]}%
\providecommand \BibitemOpen [0]{}%
\providecommand \bibitemStop [0]{}%
\providecommand \bibitemNoStop [0]{.\EOS\space}%
\providecommand \EOS [0]{\spacefactor3000\relax}%
\providecommand \BibitemShut  [1]{\csname bibitem#1\endcsname}%
\let\auto@bib@innerbib\@empty
\bibitem [{\citenamefont {Paglione}\ and\ \citenamefont
  {Greene}(2010)}]{PaglioneGreene2010}%
  \BibitemOpen
  \bibfield  {author} {\bibinfo {author} {\bibfnamefont {J.}~\bibnamefont
  {Paglione}}\ and\ \bibinfo {author} {\bibfnamefont {R.~L.}\ \bibnamefont
  {Greene}},\ }\href {\doibase 10.1038/nphys1759} {\bibfield  {journal}
  {\bibinfo  {journal} {Nature Physics}\ }\textbf {\bibinfo {volume} {6}},\
  \bibinfo {pages} {645} (\bibinfo {year} {2010})}\BibitemShut {NoStop}%
\bibitem [{\citenamefont {Stewart}(2011)}]{Stewart2011}%
  \BibitemOpen
  \bibfield  {author} {\bibinfo {author} {\bibfnamefont {G.~R.}\ \bibnamefont
  {Stewart}},\ }\href {\doibase 10.1103/RevModPhys.83.1589} {\bibfield
  {journal} {\bibinfo  {journal} {Rev. Mod. Phys.}\ }\textbf {\bibinfo {volume}
  {83}},\ \bibinfo {pages} {1589} (\bibinfo {year} {2011})}\BibitemShut
  {NoStop}%
\bibitem [{\citenamefont {Dai}\ \emph {et~al.}(2012)\citenamefont {Dai},
  \citenamefont {Hu},\ and\ \citenamefont {Dagotto}}]{Dai2012}%
  \BibitemOpen
  \bibfield  {author} {\bibinfo {author} {\bibfnamefont {P.}~\bibnamefont
  {Dai}}, \bibinfo {author} {\bibfnamefont {J.}~\bibnamefont {Hu}}, \ and\
  \bibinfo {author} {\bibfnamefont {E.}~\bibnamefont {Dagotto}},\ }\href
  {\doibase 10.1038/nphys2438} {\bibfield  {journal} {\bibinfo  {journal}
  {Nature Physics}\ }\textbf {\bibinfo {volume} {8}},\ \bibinfo {pages} {709}
  (\bibinfo {year} {2012})}\BibitemShut {NoStop}%
\bibitem [{\citenamefont {Fang}\ \emph {et~al.}(2008)\citenamefont {Fang},
  \citenamefont {Yao}, \citenamefont {Tsai}, \citenamefont {Hu},\ and\
  \citenamefont {Kivelson}}]{Fang2008}%
  \BibitemOpen
  \bibfield  {author} {\bibinfo {author} {\bibfnamefont {C.}~\bibnamefont
  {Fang}}, \bibinfo {author} {\bibfnamefont {H.}~\bibnamefont {Yao}}, \bibinfo
  {author} {\bibfnamefont {W.-F.}\ \bibnamefont {Tsai}}, \bibinfo {author}
  {\bibfnamefont {J.~P.}\ \bibnamefont {Hu}}, \ and\ \bibinfo {author}
  {\bibfnamefont {S.~A.}\ \bibnamefont {Kivelson}},\ }\href {\doibase
  10.1103/PhysRevB.77.224509} {\bibfield  {journal} {\bibinfo  {journal} {Phys.
  Rev. B}\ }\textbf {\bibinfo {volume} {77}},\ \bibinfo {pages} {224509}
  (\bibinfo {year} {2008})}\BibitemShut {NoStop}%
\bibitem [{\citenamefont {Xu}\ \emph {et~al.}(2008)\citenamefont {Xu},
  \citenamefont {M\"uller},\ and\ \citenamefont {Sachdev}}]{CenkeXu2008}%
  \BibitemOpen
  \bibfield  {author} {\bibinfo {author} {\bibfnamefont {C.}~\bibnamefont
  {Xu}}, \bibinfo {author} {\bibfnamefont {M.}~\bibnamefont {M\"uller}}, \ and\
  \bibinfo {author} {\bibfnamefont {S.}~\bibnamefont {Sachdev}},\ }\href
  {\doibase 10.1103/PhysRevB.78.020501} {\bibfield  {journal} {\bibinfo
  {journal} {Phys. Rev. B}\ }\textbf {\bibinfo {volume} {78}},\ \bibinfo
  {pages} {020501} (\bibinfo {year} {2008})}\BibitemShut {NoStop}%
\bibitem [{\citenamefont {Fernandes}\ \emph {et~al.}(2014)\citenamefont
  {Fernandes}, \citenamefont {Chubukov},\ and\ \citenamefont
  {Schmalian}}]{Fernandes2014}%
  \BibitemOpen
  \bibfield  {author} {\bibinfo {author} {\bibfnamefont {R.~M.}\ \bibnamefont
  {Fernandes}}, \bibinfo {author} {\bibfnamefont {A.~V.}\ \bibnamefont
  {Chubukov}}, \ and\ \bibinfo {author} {\bibfnamefont {J.}~\bibnamefont
  {Schmalian}},\ }\href {\doibase 10.1038/nphys2877} {\bibfield  {journal}
  {\bibinfo  {journal} {Nature Physics}\ }\textbf {\bibinfo {volume} {10}},\
  \bibinfo {pages} {97} (\bibinfo {year} {2014})}\BibitemShut {NoStop}%
\bibitem [{\citenamefont {Yildirim}(2008)}]{Yildrim2008}%
  \BibitemOpen
  \bibfield  {author} {\bibinfo {author} {\bibfnamefont {T.}~\bibnamefont
  {Yildirim}},\ }\href {\doibase 10.1103/PhysRevLett.101.057010} {\bibfield
  {journal} {\bibinfo  {journal} {Phys. Rev. Lett.}\ }\textbf {\bibinfo
  {volume} {101}},\ \bibinfo {pages} {057010} (\bibinfo {year}
  {2008})}\BibitemShut {NoStop}%
\bibitem [{\citenamefont {Si}\ and\ \citenamefont
  {Abrahams}(2008)}]{SiAbrahams2008}%
  \BibitemOpen
  \bibfield  {author} {\bibinfo {author} {\bibfnamefont {Q.}~\bibnamefont
  {Si}}\ and\ \bibinfo {author} {\bibfnamefont {E.}~\bibnamefont {Abrahams}},\
  }\href {\doibase 10.1103/PhysRevLett.101.076401} {\bibfield  {journal}
  {\bibinfo  {journal} {Phys. Rev. Lett.}\ }\textbf {\bibinfo {volume} {101}},\
  \bibinfo {pages} {076401} (\bibinfo {year} {2008})}\BibitemShut {NoStop}%
\bibitem [{\citenamefont {Chandra}\ \emph {et~al.}(1990)\citenamefont
  {Chandra}, \citenamefont {Coleman},\ and\ \citenamefont {Larkin}}]{CCL1990}%
  \BibitemOpen
  \bibfield  {author} {\bibinfo {author} {\bibfnamefont {P.}~\bibnamefont
  {Chandra}}, \bibinfo {author} {\bibfnamefont {P.}~\bibnamefont {Coleman}}, \
  and\ \bibinfo {author} {\bibfnamefont {A.~I.}\ \bibnamefont {Larkin}},\
  }\href {\doibase 10.1103/PhysRevLett.64.88} {\bibfield  {journal} {\bibinfo
  {journal} {Phys. Rev. Lett.}\ }\textbf {\bibinfo {volume} {64}},\ \bibinfo
  {pages} {88} (\bibinfo {year} {1990})}\BibitemShut {NoStop}%
\bibitem [{\citenamefont {Weber}\ \emph {et~al.}(2003)\citenamefont {Weber},
  \citenamefont {Capriotti}, \citenamefont {Misguich}, \citenamefont {Becca},
  \citenamefont {Elhajal},\ and\ \citenamefont {Mila}}]{Weber2003}%
  \BibitemOpen
  \bibfield  {author} {\bibinfo {author} {\bibfnamefont {C.}~\bibnamefont
  {Weber}}, \bibinfo {author} {\bibfnamefont {L.}~\bibnamefont {Capriotti}},
  \bibinfo {author} {\bibfnamefont {G.}~\bibnamefont {Misguich}}, \bibinfo
  {author} {\bibfnamefont {F.}~\bibnamefont {Becca}}, \bibinfo {author}
  {\bibfnamefont {M.}~\bibnamefont {Elhajal}}, \ and\ \bibinfo {author}
  {\bibfnamefont {F.}~\bibnamefont {Mila}},\ }\href {\doibase
  10.1103/PhysRevLett.91.177202} {\bibfield  {journal} {\bibinfo  {journal}
  {Phys. Rev. Lett.}\ }\textbf {\bibinfo {volume} {91}},\ \bibinfo {pages}
  {177202} (\bibinfo {year} {2003})}\BibitemShut {NoStop}%
\bibitem [{\citenamefont {Mermin}\ and\ \citenamefont
  {Wagner}(1966)}]{MerminWagner1966}%
  \BibitemOpen
  \bibfield  {author} {\bibinfo {author} {\bibfnamefont {N.~D.}\ \bibnamefont
  {Mermin}}\ and\ \bibinfo {author} {\bibfnamefont {H.}~\bibnamefont
  {Wagner}},\ }\href {\doibase 10.1103/PhysRevLett.17.1133} {\bibfield
  {journal} {\bibinfo  {journal} {Phys. Rev. Lett.}\ }\textbf {\bibinfo
  {volume} {17}},\ \bibinfo {pages} {1133} (\bibinfo {year}
  {1966})}\BibitemShut {NoStop}%
\bibitem [{\citenamefont {Singh}(2009)}]{Singh2009}%
  \BibitemOpen
  \bibfield  {author} {\bibinfo {author} {\bibfnamefont {D.}~\bibnamefont
  {Singh}},\ }\href {\doibase https://doi.org/10.1016/j.physc.2009.03.035}
  {\bibfield  {journal} {\bibinfo  {journal} {Physica C: Superconductivity}\
  }\textbf {\bibinfo {volume} {469}},\ \bibinfo {pages} {418 } (\bibinfo {year}
  {2009})}\BibitemShut {NoStop}%
\bibitem [{\citenamefont {Fernandes}\ \emph {et~al.}(2012)\citenamefont
  {Fernandes}, \citenamefont {Chubukov}, \citenamefont {Knolle}, \citenamefont
  {Eremin},\ and\ \citenamefont {Schmalian}}]{Fernandes2012}%
  \BibitemOpen
  \bibfield  {author} {\bibinfo {author} {\bibfnamefont {R.~M.}\ \bibnamefont
  {Fernandes}}, \bibinfo {author} {\bibfnamefont {A.~V.}\ \bibnamefont
  {Chubukov}}, \bibinfo {author} {\bibfnamefont {J.}~\bibnamefont {Knolle}},
  \bibinfo {author} {\bibfnamefont {I.}~\bibnamefont {Eremin}}, \ and\ \bibinfo
  {author} {\bibfnamefont {J.}~\bibnamefont {Schmalian}},\ }\href {\doibase
  10.1103/PhysRevB.85.024534} {\bibfield  {journal} {\bibinfo  {journal} {Phys.
  Rev. B}\ }\textbf {\bibinfo {volume} {85}},\ \bibinfo {pages} {024534}
  (\bibinfo {year} {2012})}\BibitemShut {NoStop}%
\bibitem [{\citenamefont {Schecter}\ \emph {et~al.}(2017)\citenamefont
  {Schecter}, \citenamefont {Sylju{\aa}sen},\ and\ \citenamefont
  {Paaske}}]{Schecter2017}%
  \BibitemOpen
  \bibfield  {author} {\bibinfo {author} {\bibfnamefont {M.}~\bibnamefont
  {Schecter}}, \bibinfo {author} {\bibfnamefont {O.~F.}\ \bibnamefont
  {Sylju{\aa}sen}}, \ and\ \bibinfo {author} {\bibfnamefont {J.}~\bibnamefont
  {Paaske}},\ }\href {\doibase 10.1103/PhysRevLett.119.157202} {\bibfield
  {journal} {\bibinfo  {journal} {Phys. Rev. Lett.}\ }\textbf {\bibinfo
  {volume} {119}},\ \bibinfo {pages} {157202} (\bibinfo {year}
  {2017})}\BibitemShut {NoStop}%
\bibitem [{\citenamefont {Kamiya}\ \emph {et~al.}(2011)\citenamefont {Kamiya},
  \citenamefont {Kawashima},\ and\ \citenamefont {Batista}}]{Kamiya2011}%
  \BibitemOpen
  \bibfield  {author} {\bibinfo {author} {\bibfnamefont {Y.}~\bibnamefont
  {Kamiya}}, \bibinfo {author} {\bibfnamefont {N.}~\bibnamefont {Kawashima}}, \
  and\ \bibinfo {author} {\bibfnamefont {C.~D.}\ \bibnamefont {Batista}},\
  }\href {\doibase 10.1103/PhysRevB.84.214429} {\bibfield  {journal} {\bibinfo
  {journal} {Phys. Rev. B}\ }\textbf {\bibinfo {volume} {84}},\ \bibinfo
  {pages} {214429} (\bibinfo {year} {2011})}\BibitemShut {NoStop}%
\bibitem [{\citenamefont {Henley}(1989)}]{Henley1989}%
  \BibitemOpen
  \bibfield  {author} {\bibinfo {author} {\bibfnamefont {C.~L.}\ \bibnamefont
  {Henley}},\ }\href {\doibase 10.1103/PhysRevLett.62.2056} {\bibfield
  {journal} {\bibinfo  {journal} {Phys. Rev. Lett.}\ }\textbf {\bibinfo
  {volume} {62}},\ \bibinfo {pages} {2056} (\bibinfo {year}
  {1989})}\BibitemShut {NoStop}%
\bibitem [{\citenamefont {{Villain, J.}}\ \emph {et~al.}(1980)\citenamefont
  {{Villain, J.}}, \citenamefont {{Bidaux, R.}}, \citenamefont {{Carton,
  J.-P.}},\ and\ \citenamefont {{Conte, R.}}}]{Villain1980}%
  \BibitemOpen
  \bibfield  {author} {\bibinfo {author} {\bibnamefont {{Villain, J.}}},
  \bibinfo {author} {\bibnamefont {{Bidaux, R.}}}, \bibinfo {author}
  {\bibnamefont {{Carton, J.-P.}}}, \ and\ \bibinfo {author} {\bibnamefont
  {{Conte, R.}}},\ }\href {\doibase 10.1051/jphys:0198000410110126300}
  {\bibfield  {journal} {\bibinfo  {journal} {J. Phys. France}\ }\textbf
  {\bibinfo {volume} {41}},\ \bibinfo {pages} {1263} (\bibinfo {year}
  {1980})}\BibitemShut {NoStop}%
\bibitem [{\citenamefont {Kouvel}\ and\ \citenamefont
  {Fisher}(1964)}]{KouvelFisher1964}%
  \BibitemOpen
  \bibfield  {author} {\bibinfo {author} {\bibfnamefont {J.~S.}\ \bibnamefont
  {Kouvel}}\ and\ \bibinfo {author} {\bibfnamefont {M.~E.}\ \bibnamefont
  {Fisher}},\ }\href {\doibase 10.1103/PhysRev.136.A1626} {\bibfield  {journal}
  {\bibinfo  {journal} {Phys. Rev.}\ }\textbf {\bibinfo {volume} {136}},\
  \bibinfo {pages} {A1626} (\bibinfo {year} {1964})}\BibitemShut {NoStop}%
\bibitem [{\citenamefont {Chen}\ \emph {et~al.}(1993)\citenamefont {Chen},
  \citenamefont {Ferrenberg},\ and\ \citenamefont
  {Landau}}]{ChenFerrenbergLandau1993}%
  \BibitemOpen
  \bibfield  {author} {\bibinfo {author} {\bibfnamefont {K.}~\bibnamefont
  {Chen}}, \bibinfo {author} {\bibfnamefont {A.~M.}\ \bibnamefont
  {Ferrenberg}}, \ and\ \bibinfo {author} {\bibfnamefont {D.~P.}\ \bibnamefont
  {Landau}},\ }\href {\doibase 10.1103/PhysRevB.48.3249} {\bibfield  {journal}
  {\bibinfo  {journal} {Phys. Rev. B}\ }\textbf {\bibinfo {volume} {48}},\
  \bibinfo {pages} {3249} (\bibinfo {year} {1993})}\BibitemShut {NoStop}%
\bibitem [{\citenamefont {Affleck}\ and\ \citenamefont
  {Halperin}(1996)}]{AffleckHalperin1996}%
  \BibitemOpen
  \bibfield  {author} {\bibinfo {author} {\bibfnamefont {I.}~\bibnamefont
  {Affleck}}\ and\ \bibinfo {author} {\bibfnamefont {B.~I.}\ \bibnamefont
  {Halperin}},\ }\href {http://stacks.iop.org/0305-4470/29/i=11/a=003}
  {\bibfield  {journal} {\bibinfo  {journal} {Journal of Physics A:
  Mathematical and General}\ }\textbf {\bibinfo {volume} {29}},\ \bibinfo
  {pages} {2627} (\bibinfo {year} {1996})}\BibitemShut {NoStop}%
\bibitem [{\citenamefont {Campostrini}\ \emph {et~al.}(2002)\citenamefont
  {Campostrini}, \citenamefont {Hasenbusch}, \citenamefont {Pelissetto},
  \citenamefont {Rossi},\ and\ \citenamefont {Vicari}}]{Campostrini2002}%
  \BibitemOpen
  \bibfield  {author} {\bibinfo {author} {\bibfnamefont {M.}~\bibnamefont
  {Campostrini}}, \bibinfo {author} {\bibfnamefont {M.}~\bibnamefont
  {Hasenbusch}}, \bibinfo {author} {\bibfnamefont {A.}~\bibnamefont
  {Pelissetto}}, \bibinfo {author} {\bibfnamefont {P.}~\bibnamefont {Rossi}}, \
  and\ \bibinfo {author} {\bibfnamefont {E.}~\bibnamefont {Vicari}},\ }\href
  {\doibase 10.1103/PhysRevB.65.144520} {\bibfield  {journal} {\bibinfo
  {journal} {Phys. Rev. B}\ }\textbf {\bibinfo {volume} {65}},\ \bibinfo
  {pages} {144520} (\bibinfo {year} {2002})}\BibitemShut {NoStop}%
\bibitem [{\citenamefont {Wang}\ and\ \citenamefont
  {Sandvik}(2018)}]{WangSandvik2018}%
  \BibitemOpen
  \bibfield  {author} {\bibinfo {author} {\bibfnamefont {L.}~\bibnamefont
  {Wang}}\ and\ \bibinfo {author} {\bibfnamefont {A.~W.}\ \bibnamefont
  {Sandvik}},\ }\href {\doibase 10.1103/PhysRevLett.121.107202} {\bibfield
  {journal} {\bibinfo  {journal} {Phys. Rev. Lett.}\ }\textbf {\bibinfo
  {volume} {121}},\ \bibinfo {pages} {107202} (\bibinfo {year}
  {2018})}\BibitemShut {NoStop}%
\end{thebibliography}%


\end{document}